\documentclass[reprint,amsmath,amssymb,aps]{revtex4-2}
\usepackage{graphicx}
\usepackage{dcolumn}
\usepackage{bm}
\usepackage{color}
\usepackage{subfigure}
\usepackage{xspace}
\usepackage{textpos}
\usepackage[english]{babel}
\newcommand*\diff{\mathop{}\!\mathrm{d}}
\usepackage[switch]{lineno}
\usepackage{hyperref}
\usepackage{afterpage}

\begin{document}

\title{Measurement of the strong-phase difference between \texorpdfstring{$D^0$}{D0} and \texorpdfstring{$\bar{D^0}\to K^+K^-\pi^+\pi^-$}{D0bar2KKpipi} in bins of phase space}

\begin{abstract}
    A first determination of the strong-phase difference between $D^0$ and $\bar{D^0}\to K^+K^-\pi^+\pi^-$  is performed using $e^+e^-\to\psi(3770)\to D\bar{D}$ data collected by the BESIII detector corresponding to an integrated luminosity of $20.3$~fb$^{-1}$.  The measurements are made in four pairs of bins in phase space, which are chosen to provide optimal sensitivity to the angle $\gamma$ of the Unitarity Triangle in $B^\pm \to DK^\pm$ decays. From these  measurements, it follows that the $C\!P$-even fraction of the decay is $F_+ = 0.754 \pm 0.010_{\rm \, stat.} \pm 0.008_{\rm \,  syst.}$.  In addition, the branching fraction of $D^0 \to K^+K^-\pi^+\pi^-$  is measured to be $(2.863 \pm 0.028_{\rm \, stat.} \pm 0.045_{\rm \, syst.})\times10^{-3}$, which is twice as precise as previous results obtained at other experiments.
\end{abstract}

\date{18th Februrary 2025}

\author{
M.~Ablikim$^{1}$, M.~N.~Achasov$^{4,c}$, P.~Adlarson$^{76}$, O.~Afedulidis$^{3}$, X.~C.~Ai$^{81}$, R.~Aliberti$^{35}$, A.~Amoroso$^{75A,75C}$, Q.~An$^{72,58,a}$, Y.~Bai$^{57}$, O.~Bakina$^{36}$
% ...
\\
\vspace{0.2cm}
(BESIII Collaboration)\\
\vspace{0.2cm} {\it
$^{1}$ Institute of High Energy Physics, Beijing 100049, People's Republic of China\\
% ...
$^{3}$ Bochum  Ruhr-University, D-44780 Bochum, Germany\\
$^{4}$ Budker Institute of Nuclear Physics SB RAS (BINP), Novosibirsk 630090, Russia\\
% ...
$^{35}$ Johannes Gutenberg University of Mainz, Johann-Joachim-Becher-Weg 45, D-55099 Mainz, Germany\\
$^{36}$ Joint Institute for Nuclear Research, 141980 Dubna, Moscow region, Russia\\
% ...
$^{57}$ Southeast University, Nanjing 211100, People's Republic of China\\
$^{58}$ State Key Laboratory of Particle Detection and Electronics, Beijing 100049, Hefei 230026, People's Republic of China\\
% ...
$^{72}$ University of Science and Technology of China, Hefei 230026, People's Republic of China\\
% ...
$^{75}$ University of Turin and INFN, (A)University of Turin, I-10125, Turin, Italy; (B)University of Eastern Piedmont, I-15121, Alessandria, Italy; (C)INFN, I-10125, Turin, Italy\\
$^{76}$ Uppsala University, Box 516, SE-75120 Uppsala, Sweden\\
% ...
$^{81}$ Zhengzhou University, Zhengzhou 450001, People's Republic of China\\
\vspace{0.2cm}
$^{a}$ Deceased\\
% ...
$^{c}$ Also at the Novosibirsk State University, Novosibirsk, 630090, Russia\\
% ...
~\\}
}

\maketitle

\section{Introduction}
\noindent In the Standard Model (SM), all $C\!P$-violation phenomena in the quark sector can be described by the Cabibbo-Kobayashi-Maskawa (CKM) matrix~\cite{Cabibbo:1963yz,Kobayashi:1973fv}. The unitary nature of this matrix leads to a set of relations that may be represented as triangles in the complex plane.   One of these representations, the so-called Unitarity Triangle (UT), has particular importance in flavor-physics studies as all its parameters may be conveniently measured in the decays of $b$~hadrons~\cite{pdg}.    

In order to test the SM description of $C\!P$ violation, it is important to verify that all measurements of the sides and angles of the UT are self consistent.  In this task, measurements of the angle $\gamma = \arg(-V_{us}{V_{ub}}^\ast / V_{cs} {V_{cb}}^*)$, sometimes denoted as $\phi_3$, are of particular importance, as they involve only tree-level processes, which are assumed to be dominated by SM contributions and have negligible theoretical uncertainties~\cite{cite:BrodZupan}. Hence, these measurements provide an important benchmark of the SM, which may be compared to indirect determinations of $\gamma$ arising from measurements of other parameters of the UT that are more susceptible to any New Physics effects lying beyond our present knowledge. 

The angle $\gamma$ can be measured using $B^\pm\to DK^\pm$ decays, where $D$ is a superposition of $D^0$ and $\bar{D^0}$ mesons. When these $D$ mesons decay to a final state common to both $D^0$ and $\bar{D^0}$, the interference effects have a dependence on $\gamma$. A powerful implementation of this approach is to consider $D$ decays to a mixed-$CP$ final state~\cite{BondarPoluektov2006,BondarPoluektov2008,GiriGrossmanSofferZupan}. The strong-phase difference $\delta_D$, defined as the $C\!P$-invariant phase difference between a $D^0$ decay and its charge-conjugated $\bar{D^0}$ decay, has a non-trivial variation across phase space due to the intermediate resonances that contribute to these decays. If the strong-phase difference is known, measurements of the different interference effects across phase space can be interpreted in terms of $\gamma$, with excellent precision.

The $D$ strong-phase difference may be predicted using an amplitude model, but then any $B^\pm\to DK^\pm$ analysis using this as input itself becomes model dependent. This introduces systematic uncertainties that are very difficult to quantify. Therefore, it is desirable to make direct measurements of the strong-phase difference, which can be achieved using quantum-correlated $D\bar{D}$ pairs produced in $e^+e^-$ collisions at charm threshold. Parameters related to this strong-phase difference have been measured in localised regions of phase space for the three-body decays $D\to K_S^0\pi^+\pi^-$ and $D\to K_S^0K^+K^-$, using data collected by CLEO-c and BESIII~\cite{cite:KSpipiStrongPhase,cite:CLEOcisiKSpipi,cite:cisiKSKK}. These measurements have been used subsequently in direct determinations of $\gamma$ by the LHCb~\cite{LHCb-PAPER-2020-019,LHCb-PAPER-2016-007} and Belle/Belle II collaborations~\cite{cite:Belle_gamma}. This method has been generalised to the four-body decays $D\to\pi^+\pi^-\pi^+\pi^-$\cite{cite:cisi4pi,cite:cisi4pi_BESIII} and $D\to K_S^0\pi^+\pi^-\pi^0$~\cite{cite:KSpipipi0_CLEOc,Belle:2019uav}.

The decay $D \to K^+K^-\pi^+\pi^-$ has been proposed as an additional mode for the measurement of $\gamma$~\cite{cite:Rademacker_Wilkinson_KKpipi}, and a recent model-dependent analysis of LHCb data has confirmed that it provides interesting sensitivity~\cite{LHCb-PAPER-2022-037}. Therefore, there is a strong motivation to perform direct measurements of the $D$-decay strong-phase parameters in this mode in order to allow for a model-independent determination of $\gamma$.  A previous study was performed with BESIII data corresponding to an integrated luminosity of $(2.932 \pm 0.014)$~fb$^{-1}$~\cite{cite:IntegratedLuminosity_2010_2011,cite:NDD_corr}, which considered the strong-phase properties integrated over all phase space~\cite{cite:KKpipi_FPlus}.  This analysis determined the $C\!P$-even fraction of the decay to be $F_+ = 0.73 \pm 0.04$,  where $(2F_+ - 1)$ is the amplitude-averaged cosine of the  strong-phase difference~\cite{cite:Firstpipipi0}.  However, the most sensitive approach to the measurement of $\gamma$, as pursued in Ref.~\cite{LHCb-PAPER-2022-037}, requires that the strong-phase parameters are known in localised regions of phase space.   Such knowledge is also potentially useful for model-independent studies of $D$-mixing and $C\!P$ violation in the charm sector~\cite{cite:BinFlip}.

This paper reports measurements of the strong-phase parameters of the decay $D \to K^+K^-\pi^+\pi^-$ performed in bins of phase space, as defined in Ref.~\cite{LHCb-PAPER-2022-037}. The measurement uses a data set of $e^+e^- \to \psi(3770) \to D \bar{D}$ events collected by the BESIII detector, corresponding to an integrated luminosity of $(20.28 \pm 0.04)$~fb$^{-1}$~\cite{cite:NDD,cite:NDD_corr,cite:NewLuminosity}. A value of the $C\!P$-even fraction of the inclusive decay is also obtained, which supersedes that presented in Ref.~\cite{cite:KKpipi_FPlus}.

\section{Measurement strategy}
\subsection{Binning of phase space}
\noindent Binning schemes are presented in Ref.~\cite{LHCb-PAPER-2022-037} that partition the phase space of $D \to K^+K^-\pi^+\pi^-$ decays in a manner that is designed to yield optimal sensitivity to $\gamma$ in a $B^\pm \to D K^\pm$ analysis. The schemes were developed assuming that the variation of the magnitude and phase of the decay amplitude follows the behavior predicted by an amplitude model developed by LHCb~\cite{LHCb-PAPER-2018-041,LHCb-PAPER-2022-037}. If the true variation in amplitude differs from that predicted by the model, the sensitivity to $\gamma$ will be reduced with respect to expectation. However, no associated bias will enter the analysis, as the strong-phase parameters within each bin are measured from data. 

Using the LHCb model, the amplitudes of the $D^0$ and $\bar{D^0}\to K^+K^-\pi^+\pi^-$ decays, $\mathcal{A}(D^0)$ and $\mathcal{A}(\bar{D^0})$, respectively, are predicted for each $D$ candidate. From the predicted strong-phase difference between $\mathcal{A}(D^0)$ and $\mathcal{A}(\bar{D^0})$, candidates with similar strong-phase differences are placed in the same bin, labelled $i = 1, 2, ..., \mathcal{N}$. Furthermore, each bin is split in two, such that candidates with $\lvert\mathcal{A}(D^0)\rvert$ smaller (larger) than $\lvert\mathcal{A}(\bar{D^0})\rvert$ are assigned a positive (negative) bin number. In summary, the binning scheme has $2\times\mathcal{N}$ bins, labelled $-\mathcal{N}, ..., -2, -1, +1, +2, ..., \mathcal{N}$. The analysis presented in this paper uses the binning scheme with $\mathcal{N} = 4$, which is well-matched to the size of the available data set.  Definitions of the binning scheme are available in Ref.~\cite{LHCb-PAPER-2022-037}. Since four-body decays have a five-dimensional phase space, the binning schemes cannot be visualised in a convenient manner.

\subsection{Categories of tag and event-yield formalism}
\noindent The strong decay $\psi(3770)\to D\bar{D}$ conserves the $C = -1$ quantum number of the initial state, which results in a $D$-meson pair with an entangled, anti-symmetric wave function. The quantum correlation allows for a direct measurement of the strong-phase difference between the $D^0$ and $\bar{D^0}$ mesons through a double-tag (DT) analysis~\cite{cite:DT_method}.

If one of the $D$ decays is reconstructed in a decay of known $C\!P$ content, referred to as a tag mode, the other $D$ decay, referred to as the signal decay, then has a decay rate that depends on the $C\!P$ content of the tag mode. The strong-phase parameters of the signal mode $D\to K^+K^-\pi^+\pi^-$ may be inferred from the suppression or enhancement of the DT yield with respect to the case where no quantum correlations are present. Furthermore, for each tag mode, the corresponding single-tag (ST) events are also considered. These are events where one of the $D$ mesons decays into the tag mode, and the other $D$ meson is not reconstructed. The ST yields are used to normalise the DT yields.

A useful parameter is the fraction of $D^0\to K^+K^-\pi^+\pi^-$ decays in bin $i$, defined as
\begin{equation}
    K_i = \frac{\int_i\diff\Phi \, \lvert\mathcal{A}(D^0)\rvert^2}{\int\diff\Phi \, \lvert\mathcal{A}(D^0)\rvert^2},
\end{equation}
where $\Phi$ represents the five-dimensional phase-space. At the current level of precision, it is assumed that $C\!P$ in the charm system is conserved, and therefore the fractional bin yields of $\bar{D^0}$ in bin $i$, $\bar{K}_i$, can be shown to be equal to $K_{-i}$. By definition, the sum of $K_i$ over all bins is unity. The $K_i$ parameters correspond to the fractional yield of $D^0$ decays in each bin in the absence of quantum correlations. The parameters of interest for the measurement of $\gamma$ are the amplitude-averaged cosine of the strong-phase difference,
\begin{equation}
    c_i = \frac{1}{\sqrt{K_i\bar{K_i}}}\int_i\diff\Phi \, \lvert\mathcal{A}(D^0)\rvert\lvert\mathcal{A}^*(\bar{D^0})\rvert\cos(\delta_D),
    \label{equation:ci}
\end{equation}
and $s_i$, the amplitude-averaged sine of the strong-phase difference, which is defined analogously. The parameter $\delta_D \equiv \arg\big(\mathcal{A}(D^0)\mathcal{A}^\ast(\bar{D^0})\big)$ is the strong-phase difference between the $D^0$ and $\bar{D^0}$ decays at the phase-space point $\Phi$, and the effect of averaging over the phase-space bin $i$ is to dilute the coherence of the contributing amplitudes such that $c_i^2 + s_i^2 < 1$. In the limit of an infinite number of bins, $c_i^2 + s_i^2\to1$. From $C\!P$ conservation, it can be shown that $c_{-i} = c_i$ and $s_{-i}$ = $-s_i$.

\begin{table}[htb]
    \centering
    \caption{Tag modes used in this analysis. The three tag modes with a $\dagger$ are also considered with a partially reconstructed $D\to K^+K^-\pi^+\pi^-$ decay, where a charged kaon is missing.}
    \vspace*{0.15cm}
    \label{table:Tag_modes}
    \begin{tabular}{cc}
        \hline
        Category     & Tag mode \\
        \hline
        Flavor       & $K^-\pi^+$, $K^-\pi^+\pi^0$, $K^-\pi^+\pi^-\pi^+$, $K^-e^+\nu_e$ \\
        $C\!P$ even  & $K^+K^{-\dagger}$, $\pi^+\pi^-$, $K_S^0\pi^0\pi^0$, $\pi^+\pi^-\pi^0$, $K_L^0\pi^0$ \\
        $C\!P$ odd   & $K_S^0\pi^{0\dagger}$, $K_S^0\eta$, $K_S^0\eta^\prime(\pi\pi\eta)$, $K_S^0\eta^\prime(\rho^0\gamma)$, $K_S^0\pi^+\pi^-\pi^0$ \\
        Mixed $C\!P$ & $K_S^0\pi^+\pi^{-\dagger}$, $K_L^0\pi^+\pi^-$ \\
        \hline
    \end{tabular}
\end{table}

In order to measure $K_i$, $c_i$ and $s_i$, three different categories of tag modes, listed in Table~\ref{table:Tag_modes}, are used. In the first category, the signal mode $D\to K^+K^-\pi^+\pi^-$ is tagged with modes that have a non-zero strangeness, which are referred to as flavor tags. In the case of $D^0\to K^-e^+\nu_e$, which proceeds through a single weak amplitude, there is an unambiguous correlation between the flavor of the final and initial states, and therefore the kaon charge of this tag indicates the flavor of the parent meson of the signal decay in DT events (note that charge conjugation is implicit throughout this paper). Therefore, for such a $D^0\to K^-e^+\nu_e$ tag, if the $\bar{D^0}\to K^+K^-\pi^+\pi^-$ decay is reconstructed in bin $i$, the predicted DT yield is
\begin{linenomath}
    \begin{equation*}
        \hat{N}_i^{\rm DT}\propto K_{-i}.
    \end{equation*}
\end{linenomath}
Here and subsequently the `hat' accent denotes the predicted value, whereas the measured DT yield carries no accent. The flavor tags are therefore sensitive to the $K_i$ parameters. The proportionality constant, in the absence of reconstruction effects, is $2N_{D\bar{D}}\mathcal{B}_{\rm tag}\mathcal{B}$, where $N_{D\bar{D}} = (7.33 \pm 0.08) \times 10^7$~\cite{cite:NDD,cite:NDD_corr,cite:NewLuminosity} is the total number of $D\bar{D}$ pairs, $\mathcal{B}_{\rm tag}$ is the branching fraction of the tag and $\mathcal{B}$ is the branching fraction of $D^0\to K^+K^-\pi^+\pi^-$. The same proportionality constant applies to all subsequent expressions in this section.

For other flavor tags, such as $D^0\to K^-\pi^+$, which is Cabibbo favored (CF), there exists a corresponding doubly Cabibbo-suppressed decay (DCS) $\bar{D^0}\to K^-\pi^+$, which can interfere with the CF decay. Thus, the generalised DT yield expression is
\begin{equation}
    \hat{N}_i^{\rm DT}\propto K_{-i} + r_D^2K_i - 2Rr_D\sqrt{K_iK_{-i}}\big(c_i\cos(\delta_D) + s_i\sin(\delta_D)\big),
    \label{eq:flavourDT}
\end{equation}
where $r_D$ is the magnitude of the ratio between the DCS and CF decay amplitudes and $\delta_D = \delta_D^{\rm CF} - \delta_D^{\rm DCS}$ is the strong-phase difference between the CF and DCS decays.

The coherence parameter $R$ describes the dilution of interference effects when the multi-body flavor-tag mode is integrated over phase space, taking a value of $0\leq R \leq 1$ for these decays, and $R = 1$ for $D^0\to K^-\pi^+$ tags. In the multi-body case the values of $r_D$ and $\delta_D$ are averaged over the full phase space. The parameters $r_D$, $R$ and $\delta_D$ can either be obtained from direct measurements at charm threshold or from charm-mixing measurements~\cite{Atwood:2003mj,Harnew:2013wea}. For this analysis, the values of $r_D$, $R$ and $\delta_D$ for the $D^0\to K^-\pi^+\pi^0$ and $D^0\to K^-\pi^+\pi^-\pi^+$ tags are taken from Ref.~\cite{cite:K3piStrongPhase}.  The values of $r_D$ and $\delta_D$ for $D^0\to K^-\pi^+$ taken from Ref. \cite{cite:deltaKpi} are used as constraints, as discussed later.  

In the second category, the tag modes are  $C\!P$ eigenstates. The mode $D \to K^0_S\eta'$ is listed twice, as it may be conveniently reconstructed using both the decays $\eta' \to \pi^+\pi^-\eta$ and $\eta' \to \rho^0 \gamma$. Also included are the quasi-$C\!P$-eigenstates $D\to\pi^+\pi^-\pi^0$ and $D\to K_S^0\pi^+\pi^-\pi^0$, which are known to be predominantly $C\!P$ even and $C\!P$ odd, respectively~\cite{cite:pipipi0_CPfraction,cite:pipipi0_BESIII,cite:KSpipipi0_BESIII}. Note that in Ref.~\cite{cite:KKpipi_FPlus} only the part of the $D\to K_S^0\pi^+\pi^-\pi^0$ phase space containing the $C\!P$-odd resonance $D\to K_S^0\omega$ was selected. In the current analysis, the full phase space, which contains a mixture of $C\!P$-even and $C\!P$-odd contributions, is included. The DT yield is described by
\begin{equation}
    \hat{N}_i^{\rm DT}\propto K_i + K_{-i} - 2\sqrt{K_iK_{-i}}c_i(2F_+^{f} - 1),
\end{equation}
where $F_+^{f}$ is the $C\!P$-even fraction of the tag mode $f$. For pure $C\!P$-even (odd) tags, $F_+^{f} = 1$ ($0$). This class of tag modes is particularly important for measuring $c_i$.

Finally, in the last category of tag modes, the mixed-$CP$ decays $D\to K_{S, L}^0\pi^+\pi^-$ are used. These tag modes are also binned on the tag side, which provides sensitivity to $s_i$, as well as to $c_i$,
\begin{equation}
    \hat{N}_{ij}^{\rm DT}\propto K_iK_{-j}^{f} + K_{-i}K_j^{f} - 2\sqrt{K_iK_{-i}K_j^{f}K_{-j}^{f}}\big(c_ic_j^{f} + s_is_j^{f}\big),
\end{equation}
where $j$ denotes the bin number on the tag side. The parameters $K_j^{f}$, $c_j^{f}$ and $s_j^{f}$ for the mode $f$ are defined in an analogous manner to those of the signal decay, and have been measured previously in quantum-correlated $D\bar{D}$ decays~\cite{cite:CLEOcisiKSpipi,cite:KSpipiStrongPhase}. For this analysis, the `equal $\Delta\delta_D$' binning scheme is used. In terms of the two-body invariant mass $m(K_S^0\pi^\pm)\equiv m_\pm$, bins with $m_+^2 > m_-^2$ are defined to be positive, while the region where $m_+^2 < m_-^2$ have bins labelled by a negative index.

To convert the DT yield expressions into normalised equations, the DT 
yields must first be corrected for reconstruction efficiencies and migrations between bins. A DT efficiency matrix $\epsilon_{ia}^{\rm DT}$, which is defined as the fraction of events generated in bin $a$ that are reconstructed in bin $i$, is introduced. Second, the DT yields must be normalised by the efficiency-corrected ST yields. The expected ST yield of a tag mode $f$ with $C\!P$-even fraction $F_+^{f}$ is given by~\cite{cite:AsnerSun}
\begin{equation}
    N^{\rm ST}(f) = 2N_{D\bar{D}}\mathcal{B}(f)\epsilon_{\rm ST}(f)\big(1 - (2F_+^{f} - 1)y\big),
    \label{equation:ST_yield}
\end{equation}
where $\epsilon_{\rm ST}$ is the reconstruction efficiency of the ST mode, and $y = (0.615^{+0.056}_{-0.055})\times 10^{-2}$ is the charm-mixing parameter~\cite{cite:HFLAV2021}. The equivalent expression for flavor tags is
\begin{equation}
    N^{\rm ST}(f) = 2N_{D\bar{D}}\mathcal{B}(f)\epsilon_{\rm ST}(f)\big(1 + r_D^2 + 2yr_D\cos(\delta_D)\big).
    \label{equation:ST_yield_flavour}
\end{equation}
In Eqs.~\eqref{equation:ST_yield}, \eqref{equation:ST_yield_flavour} and subsequent expressions, $\mathcal{O}(y^2)$ terms are neglected.

By taking a ratio of DT and ST yields, most of the proportionality constants cancel. The predicted DT yield of flavor-tagged events in bin $i$, now accounting for reconstruction effects, is shown in Eq.~\eqref{equation:Yield_equation_flavor}. For the purely CF $D^0\to K^-e^+\nu_e$ flavor tag , $r_D = 0$, and only the first term in Eq.~\eqref{equation:Yield_equation_flavor} contributes.

\begin{widetext}
    \begin{align}
        \hat{N}^{\rm DT}_i =& \frac{N^{\rm ST}}{\epsilon^{\rm ST}\big(1 + r_D^2 + 2yr_D\cos(\delta_D)\big)}\mathcal{B}\epsilon_{ia}^{\rm DT}\Big[K_{-a} + r_D^2K_a - 2Rr_D\sqrt{K_aK_{-a}}\big(c_a\cos(\delta_D) + s_a\sin(\delta_D)\big)\Big].
        \label{equation:Yield_equation_flavor} \\
        \hat{N}^{\rm DT}_i =& \frac{N^{\rm ST}}{\epsilon^{\rm ST}\big(1 - (2F_+^{f} - 1)y\big)}\mathcal{B}\epsilon_{ia}^{\rm DT}\Big[K_a + K_{-a} - 2\sqrt{K_aK_{-a}}c_a(2F_+^{f} - 1)\Big],
        \label{equation:Yield_equation_CP} \\
        \hat{N}^{\rm DT}_{ij} =& \frac{N^{\rm ST}}{\epsilon^{\rm ST}\big(1 - (2F_+^{f} - 1)y\big)}\mathcal{B}\epsilon_{ijab}^{\rm DT}\Big[K_{a}K_{-b}^{f} + K_{-a}K_{b}^{f} - 2\sqrt{K_{a}K_{-a}K_{b}^{f} K_{-b}^{f}}(c_{a}c_{b}^{f} + s_{a}s_{b}^{f})\Big].
        \label{equation:Yield_equation_SCMB}
    \end{align}
\end{widetext}

Similarly, Eq.~\eqref{equation:Yield_equation_CP} is the predicted DT yield for $C\!P$-tagged events and Eq.~\eqref{equation:Yield_equation_SCMB} describes the DT yield of mixed-$CP$ tags. The index $a$ is implicitly summed over in Eqs.~\eqref{equation:Yield_equation_flavor} and \eqref{equation:Yield_equation_CP}. In Eq.~\eqref{equation:Yield_equation_SCMB}, both $a$ and $b$ indices are summed over.

The strategy for measuring $K_i$, $c_i$ and $s_i$ is therefore to count the DT yield in each bin for all tag modes, as well as the ST yields, and use Eqs.~\eqref{equation:Yield_equation_flavor}-\eqref{equation:Yield_equation_SCMB} to perform a maximum-likelihood fit with $K_i$, $c_i$ and $s_i$ as free parameters.

\section{Description of BEPCII and the BESIII detector}
\noindent The BESIII detector~\cite{Ablikim:2009aa} records symmetric $e^+e^-$ collisions provided by the BEPCII storage ring~\cite{Yu:2016cof}, which operates with a centre-of-mass energy range from $\sqrt{s} = 2.00$~GeV to $4.95$~GeV, with a peak luminosity of $1.1\times10^{33}$~cm$^{-2}$s$^{-1}$ achieved at $\sqrt{s} = 3.773$~GeV. \mbox{BESIII} has collected large data samples in this energy region~\cite{Ablikim:2019hff,EcmsMea,EventFilter}. The cylindrical core of the BESIII detector covers 93\% of the full solid angle and consists of a helium-based multilayer drift chamber~(MDC), a plastic scintillator time-of-flight system~(TOF), and a CsI(Tl) electromagnetic calorimeter~(EMC), which are all enclosed in a superconducting solenoidal magnet providing a 1.0~T magnetic field. The solenoid is supported by an octagonal flux-return yoke with resistive plate counter muon-identification modules interleaved with steel. The charged-particle momentum resolution at $1~{\rm GeV}/c$ is $0.5\%$, and the resolution of the rate of energy loss, ${\rm d}E/{\rm d}x$, is $6\%$ for electrons from Bhabha scattering. The EMC measures photon energies with a resolution of $2.5\%$ ($5\%$) at $1$~GeV in the barrel (end-cap) region. The time resolution in the TOF barrel region is 68~ps, while that in the end-cap region is 110~ps. The end-cap TOF system was upgraded in 2015 using multigap resistive plate chamber technology, providing a time resolution of 60~ps, which benefits 86\% of the data used in this analysis~\cite{etof1,etof2,etof3}.

Simulated data samples produced with a {\sc geant4}-based~\cite{geant4} Monte Carlo (MC) package, which includes the geometric description of the BESIII detector and the detector response, are used to determine detection efficiencies and to estimate backgrounds. The simulation models the beam-energy spread and initial-state radiation in the $e^+e^-$ annihilations with the generator {\sc kkmc}~\cite{ref:kkmc}. The inclusive MC sample includes the production of $D\bar{D}$ pairs, the non-$D\bar{D}$ decays of the $\psi(3770)$, the initial-state radiation production of the $J/\psi$ and $\psi(3686)$ states, and the continuum processes incorporated in {\sc kkmc}~\cite{ref:kkmc}. All particle decays are modelled with {\sc evtgen}~\cite{Lange:2001uf,Ping:2008zz} using branching fractions either taken from the Particle Data Group~\cite{pdg}, when available, or otherwise estimated with {\sc lundcharm}~\cite{YANGRui-Ling:61301,Chen:2000tv}. Final-state radiation (FSR) from charged final-state particles is incorporated using the {\sc photos} package~\cite{RICHTERWAS1993163}. The signal decay $D \to K^+K^-\pi^+\pi^-$ is described by the amplitude model from Ref.~\cite{LHCb-PAPER-2018-041}, and quantum-correlation effects are accounted for through a re-weighting of generated events according to the nature of the tag mode.

\section{Event selection}
\noindent The selection of charged tracks and photon candidates, as well as the combination of these objects to form short-lived particles and $K^0_S$ mesons, is identical to that described in Ref.~\cite{cite:KKpipi_FPlus}. This analysis also includes the semi-leptonic tag $D^0\to K^-e^+\nu_e$, which requires the selection of electron tracks. The reconstruction of electrons follows the strategy in Ref.~\cite{cite:KeNu_selection}.

For a track to be identified as an electron, it must deposit sufficient energy in the EMC. Information about the shape of the electromagnetic shower is also used. The momentum of tracks that are identified as electrons are corrected for FSR by searching for showers within $5^\circ$ of the initial electron momentum direction. The momenta of these showers are added to the electron momentum. Additionally, any event containing showers with energy greater than $300$~MeV which are not used in the $D$ meson reconstruction are rejected. This suppresses background events with additional photons that are not from FSR, such as $D^0\to K^-\pi^+\pi^0$.

Decays of $D$ mesons into the tag modes listed in Table~\ref{table:Tag_modes} or the signal mode, here collectively referred to as tags, are reconstructed from the charged tracks, photons and composite particles. Tags can either be fully reconstructed, where all final-state particles are reconstructed, or partially reconstructed, in which the presence of a missing particle is inferred from the reconstructed momentum of all other particles in the event.

In fully reconstructed tags, the energy difference $\Delta E = E_D - \sqrt{s}/2$, where $E_D$ is the reconstructed energy of the $D$ meson, is required to be within $3\sigma$ on either side of the signal peak. This requirement removes combinatorial background. The resolutions, and hence requirements, vary from channel to channel and are slightly narrower in simulation. For example, in data (simulation), $\Delta E$ must lie within $[-16.5,13.1]$~MeV ($[-15.0,13.9]$~MeV) for $D \to K^+K^-\pi^+\pi^-$, $[-26.6,25.0]$~MeV ($[-22.4,22.1]$~MeV) for $D \to K^-\pi^+$ and $[-63.1,44.5]$~MeV ($[-55.9,42.8]$~MeV) for $D \to K^0_S \pi^0$ decays.

In addition, for fully reconstructed tags, the beam-constrained mass is defined as $M_{\rm BC} = \sqrt{E_{\rm beam}^2 - \lvert\sum_i\vec{p}_i\rvert^2}$, where $E_{\rm beam}$ is the beam energy and the sum is over all $D$-decay products. For correctly reconstructed tags, the $M_{\rm BC}$ distribution peaks at the mass of the $D^0$ meson~\cite{pdg}. This quantity is used as a fit variable to determine signal yields, as described in Sect.\ref{section:Single_and_double_tag_yield_determination}.

For the DT selection, one $D$ meson is reconstructed as the tag mode while the other $D$ meson is reconstructed as the signal mode $D\to K^+K^-\pi^+\pi^-$. For both tag and signal modes, $\Delta E$ and $M_{\rm BC}$ are calculated and the $\Delta E$ requirements are imposed on both $D$ mesons. If the DT event has multiple candidates, the combination with the average $M_{\rm BC}$ closest to the known value of the $D^0$ mass is chosen.

In the $D\to\pi^+\pi^-\pi^0$, $D\to K^-\pi^+\pi^-\pi^+$ and $D\to K^+K^-\pi^+\pi^-$ decays, there are backgrounds from $D\to K_S^0\pi^0$, $D\to K_S^0K^\mp\pi^\pm$ and $K_S^0K^+K^-$ decays, respectively. To suppress these backgrounds, all combinations of oppositely charged pions in these decays are required to originate from a point in space within twice the vertex resolution from the interaction point. Furthermore, it is found that the $D\to K_S^0K^+K^-$ events contribute as background to the $D\to K^+K^-\pi^+\pi^-$ mode.  These events have a $\pi^+\pi^-$ invariant mass distribution that is asymmetric around the $K_S^0$ mass, and events where the $\pi^+\pi^-$ mass lies within $[477, 507]$~MeV/$c^2$ are removed.  This requirement rejects over $95\%$ of the background events from $D\to K_S^0K^+K^-$, but only removes $5\%$ of the $D\to K^+K^-\pi^+\pi^-$ signal candidates.

If a tag mode contains a $K_L^0$ meson or a neutrino, a partial reconstruction technique is employed to form the DT, in which the signal mode $D\to K^+K^-\pi^+\pi^-$ is first reconstructed. The selection of $K_L^0X$ tags is similar to that in Ref.~\cite{cite:KKpipi_FPlus}, with minor improvements that are described below, while the $D^0\to K^-e^+\nu_e$ tag, which contains a neutrino, is new in this analysis.

In these modes, where the tag side is partially reconstructed, the signal mode $M_{\rm BC}$ is required to be within $[1.855, 1.875]$~GeV/$c^2$ to reduce combinatorial background on the signal side. The tag mode is reconstructed without the $K_L^0$ meson or the neutrino from the remaining tracks and showers in the event.  The four-momentum of the $K_L^0$ meson or neutrino is inferred from the missing four-momentum $p_{\rm miss}$, which is determined from the reconstructed momentum of the other charged tracks and showers, and from four-momentum conservation. Since there is a missing particle, ST events cannot be reconstructed for tag modes containing a $K_L^0$ meson or a neutrino.

For the $K_L^0$ selection, it is required that there are no additional charged tracks or $\pi^0$ candidates. Correctly reconstructed events are expected to peak at the $K_L^0$ mass~\cite{pdg} in the $M_{\rm miss}^2 = p_{\rm miss}^2$ distribution. In the selection of the $D^0\to K^-e^+\nu_e$ tag, which has a missing neutrino, the maximum shower energy in the event is required to be less than $300$~MeV to suppress backgrounds from $D^0\to K^-\pi^0e^+\nu_e$ and $D^0\to K^-\pi^+\pi^0$ decays. Any showers within $5^\circ$ of the positron direction are assumed to be from FSR and included in the calculation of the $e^+$ momentum. The variable $U_{\rm miss} = E_{\rm miss} - \lvert\vec{p}_{\rm miss}\rvert$ is determined from the missing energy and missing three-momentum, and events with a missing neutrino are  expected to peak at $U_{\rm miss} = 0$.

The statistical power of the analysis is augmented by also including events where the decay $D\to K^+K^-\pi^+\pi^-$ is missing a charged kaon. The tag modes considered for this approach are $D\to K^+K^-$, $K_S^0\pi^0$ and $K_S^0\pi^+\pi^-$, which have the largest statistical impact. The selection is similar to that of tags with a $K_L^0$, where the fully reconstructed tag mode is first selected, and the tag-side $M_{\rm BC}$ is required to be in the interval $[1.86, 1.87]$~GeV/$c^2$. It is required that there are exactly three additional charged tracks, identified as two oppositely charged pions and a kaon. The momentum of the missing kaon is inferred from the missing momentum in the event, and the distribution of $M_{\rm miss}^2$ is expected to peak at the $K^\pm$ mass~\cite{pdg}. To suppress background from $D\to K^-\pi^+\pi^-\pi^+\pi^0$ decays, events with additional $\pi^0$ candidates are rejected, and both charged kaons are required to have energies greater than $700$~MeV. From simulation, these requirements are found to reduce the background from $D\to K^-\pi^+\pi^-\pi^+\pi^0$ decays by $77\%$.

\section{Single- and double-tag yield determination}
\label{section:Single_and_double_tag_yield_determination}
\noindent The ST yield of each fully reconstructed mode is determined using a maximum-likelihood fit of the $M_{\rm BC}$ distribution. The signal shape is obtained from simulation, but is convolved with a resolution function to account for any difference between data and simulation. For the flavor tags, as well as the $D\to K_S^0\pi^+\pi^-$, $D\to K^+K^-$ and $D\to K_S^0\pi^+\pi^-\pi^0$ tags, the resolution function is a sum of two Gaussian functions, with their means and widths as free parameters in the fit. For other tag modes, which have smaller sample size, a single Gaussian function is found to be adequate in describing the difference in resolution between data and simulation. The difference in resolution is found to be a few hundred keV/$c^2$ and varies between tags.

\begin{figure*}[htb!]
    \centering
    \hspace{0.5cm}
    \includegraphics[height=3.62cm,trim={0 7.0cm 0 0},clip=true]{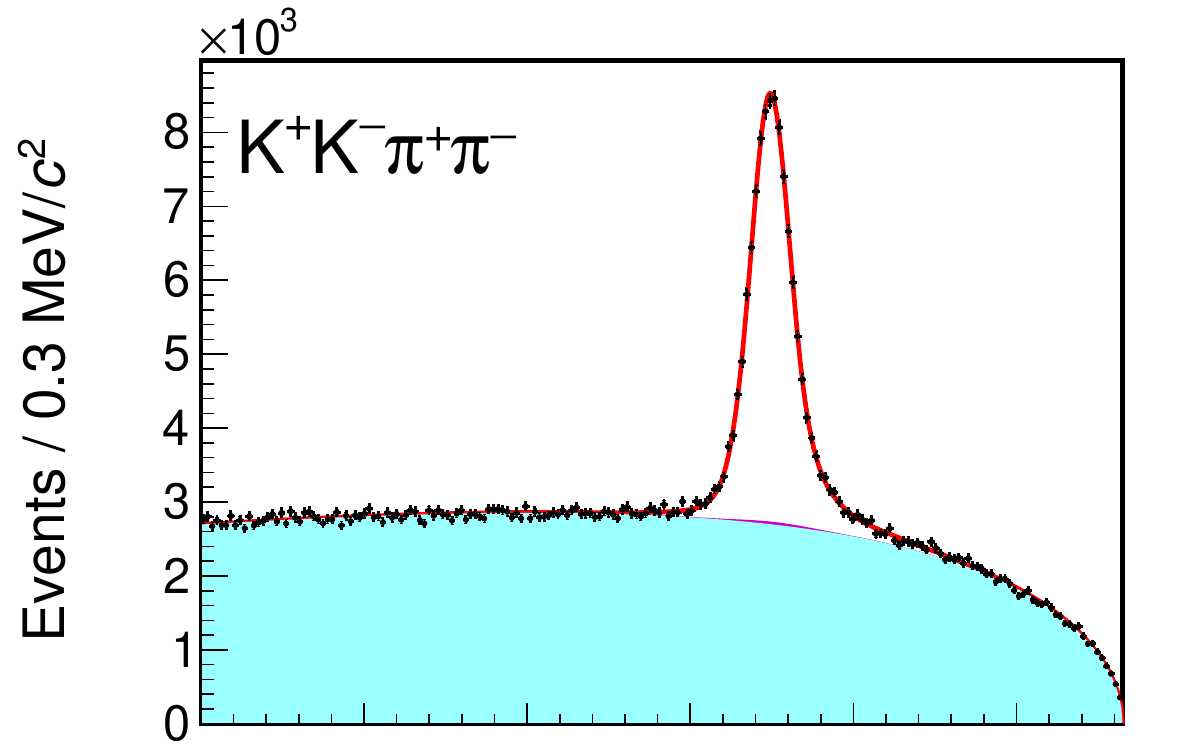}
    \includegraphics[height=3.62cm,trim={1.65cm 7.0cm -2.82cm 0},clip=true]{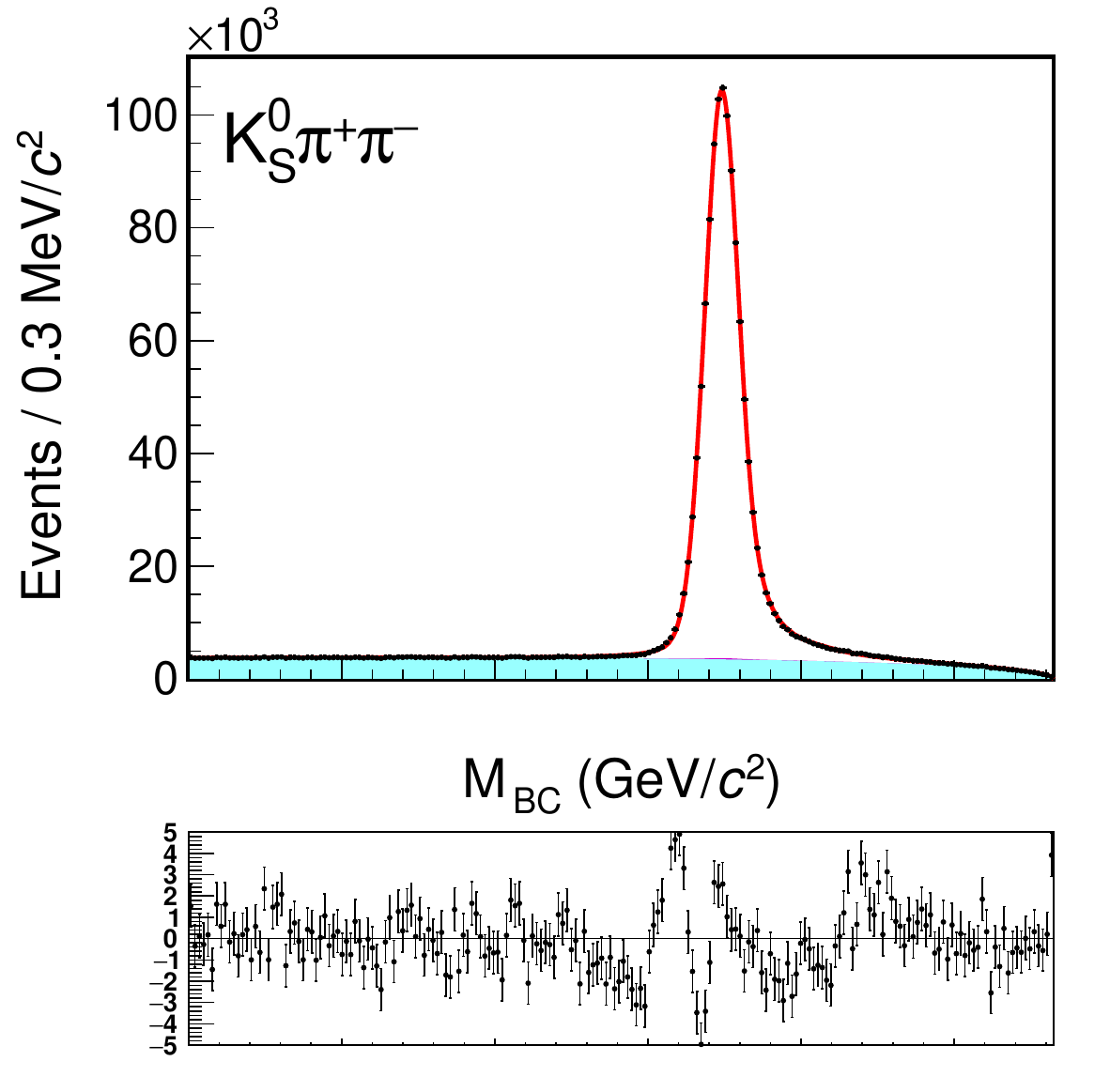}
    \hspace{0.5cm}
    \includegraphics[height=3.62cm,trim={0 7.0cm 0 0},clip=true]{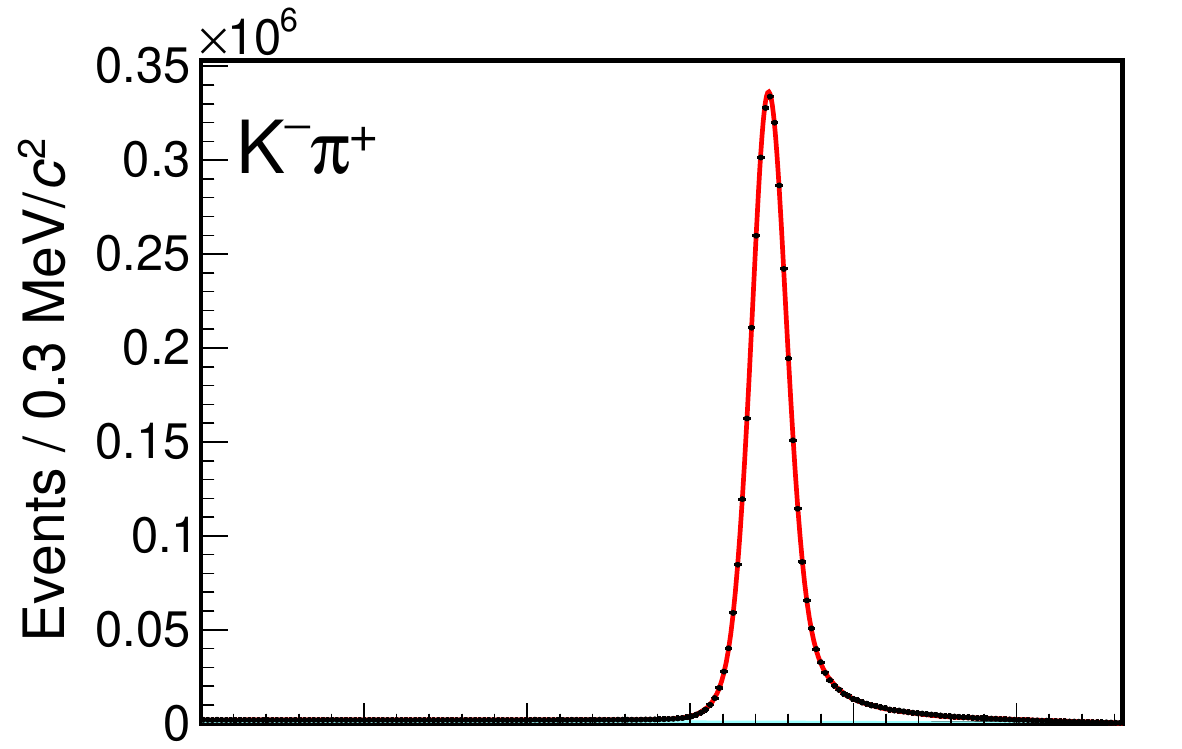}
    \includegraphics[height=3.62cm,trim={1.65cm 7.0cm 0 0},clip=true]{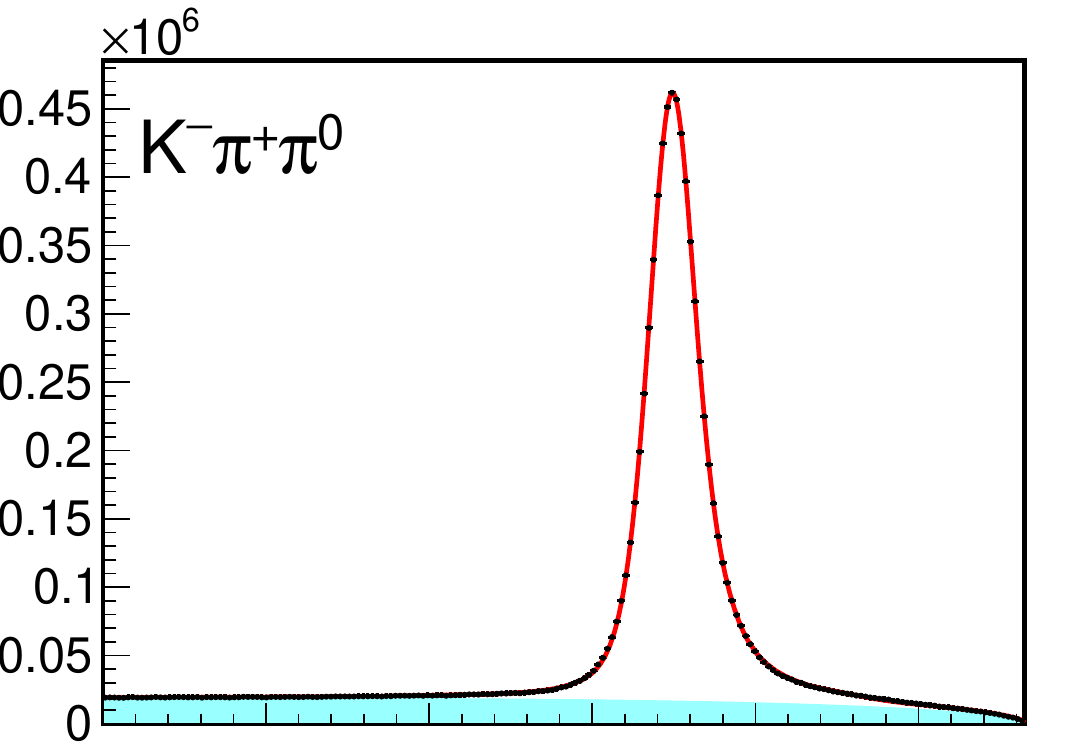}
    \includegraphics[height=3.62cm,trim={1.65cm 7.0cm 0 0},clip=true]{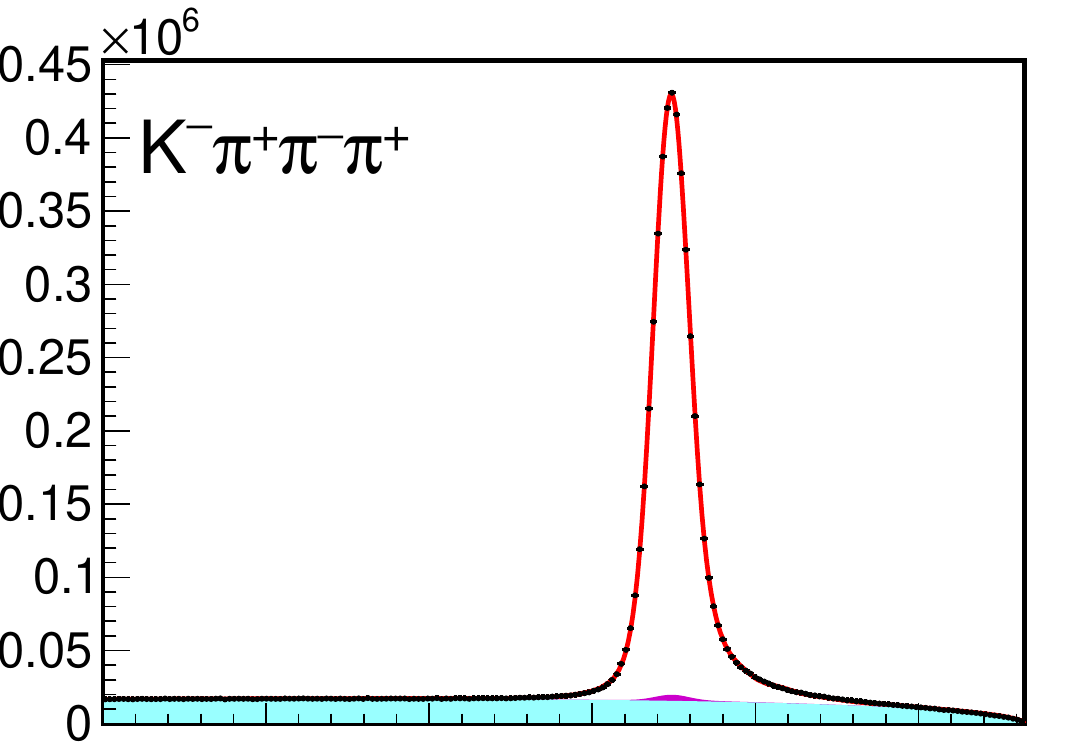}
    \includegraphics[height=3.62cm,trim={0 7.0cm 0 0},clip=true]{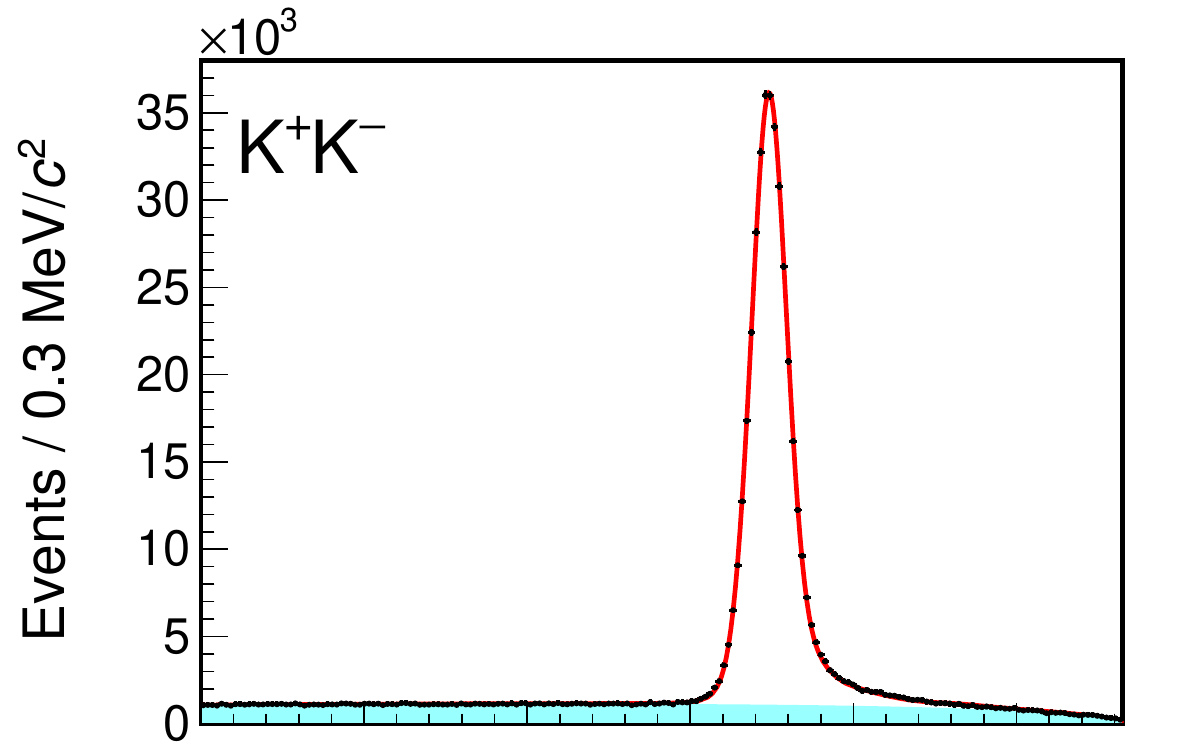}
    \includegraphics[height=3.62cm,trim={1.65cm 7.0cm 0 0},clip=true]{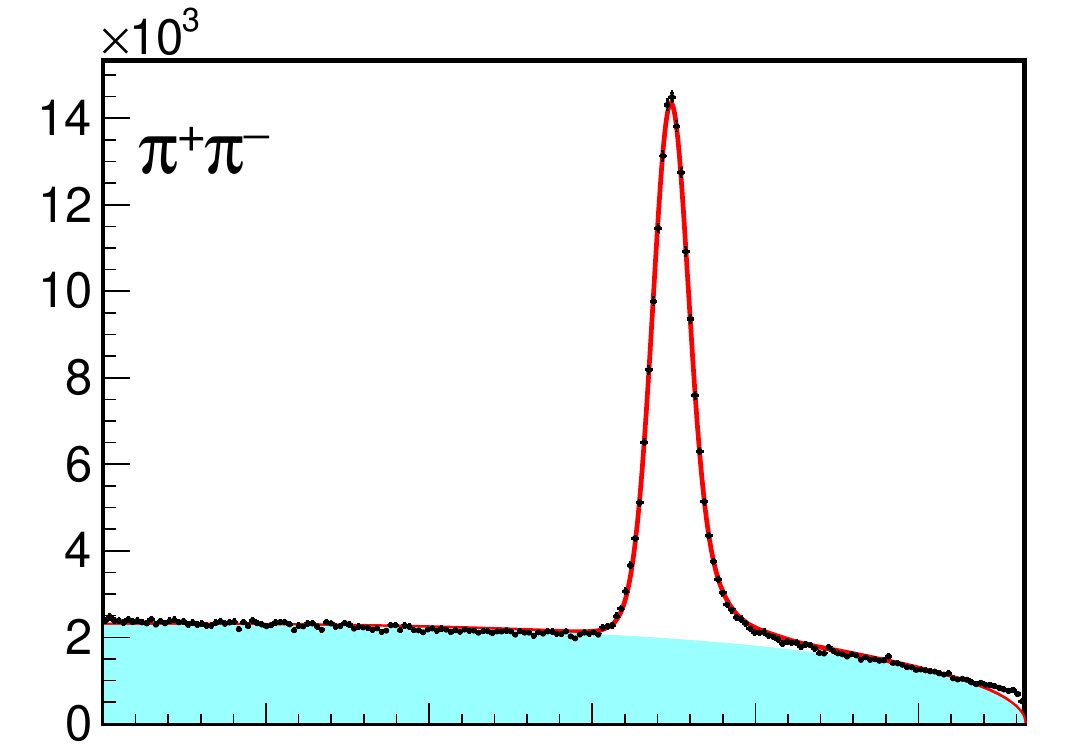}
    \includegraphics[height=3.62cm,trim={1.65cm 7.0cm 0 0},clip=true]{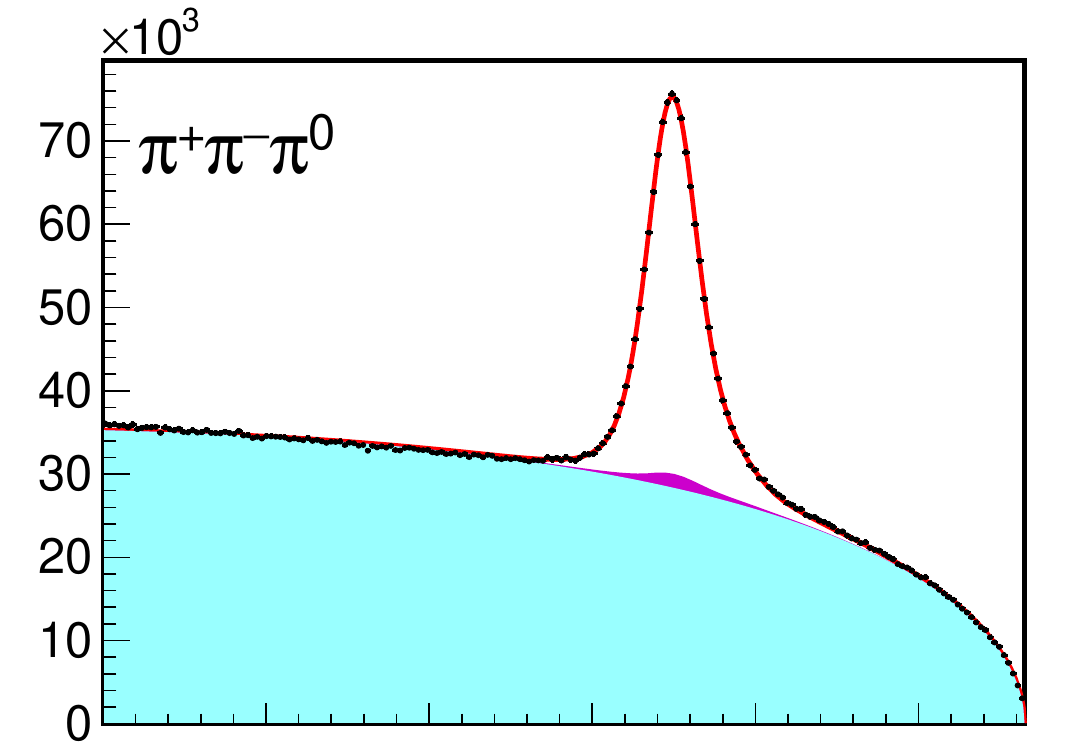}
    \includegraphics[height=3.62cm,trim={0 7.0cm 0 0},clip=true]{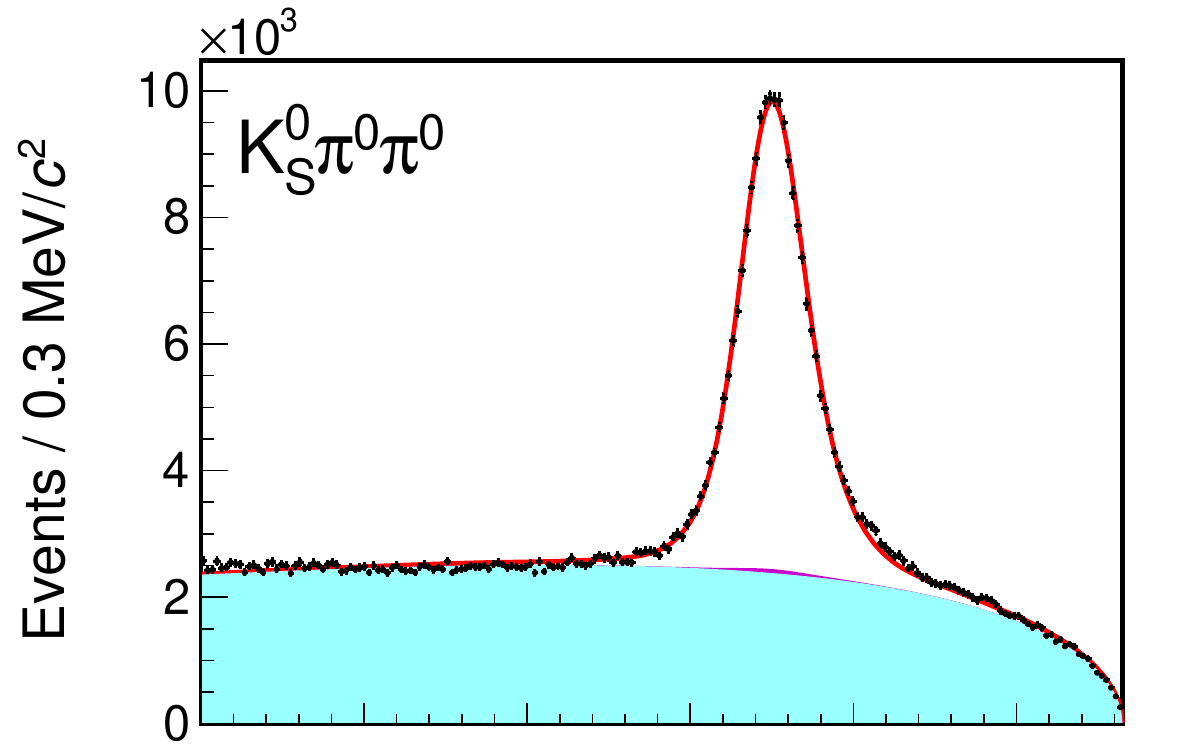}
    \includegraphics[height=3.62cm,trim={1.65cm 7.0cm 0 0},clip=true]{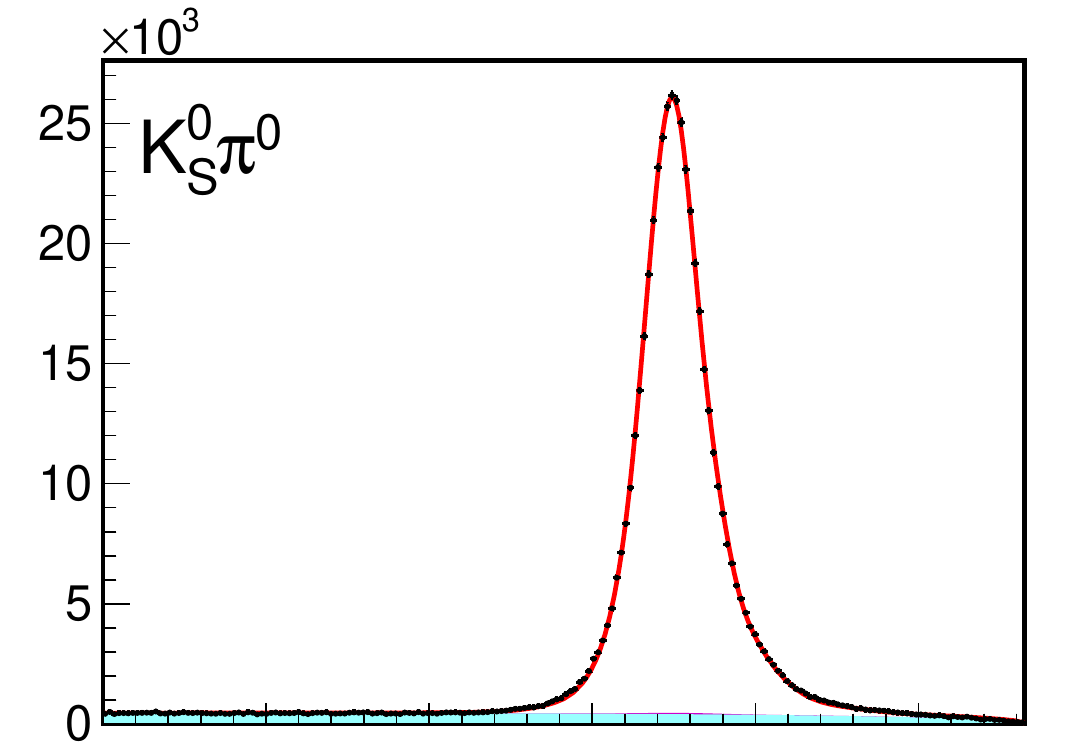}
    \includegraphics[height=3.62cm,trim={1.65cm 7.0cm 0 0},clip=true]{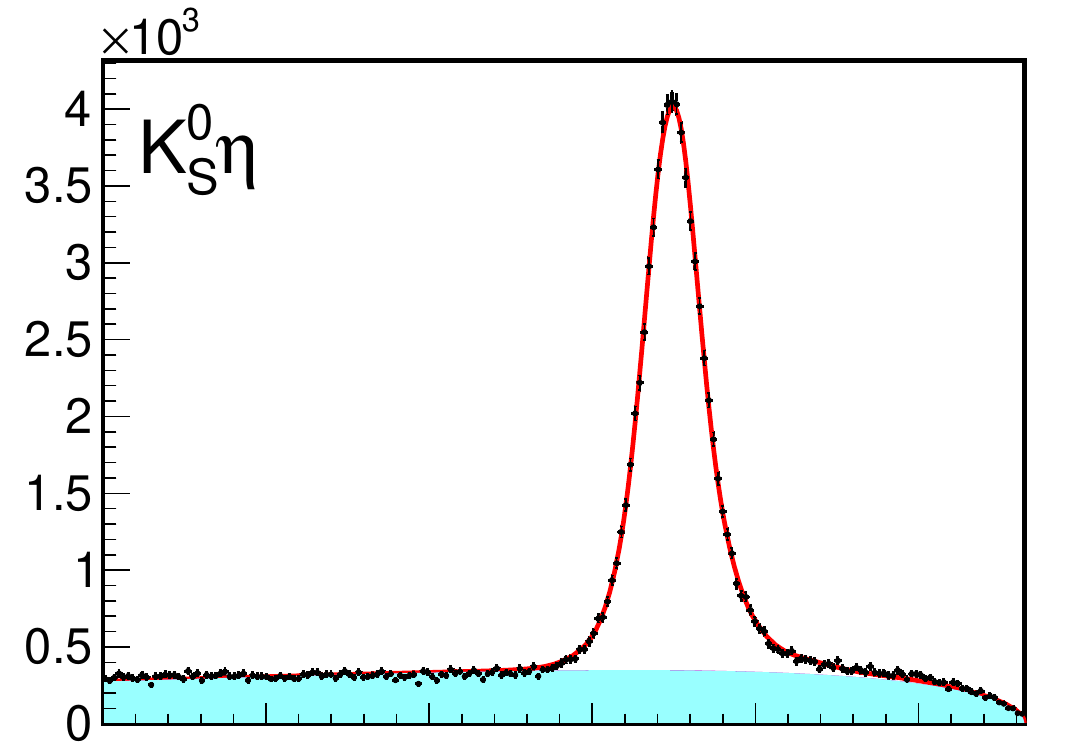}
    \includegraphics[height=4.195cm,trim={0 5.0cm 0 0},clip=true]{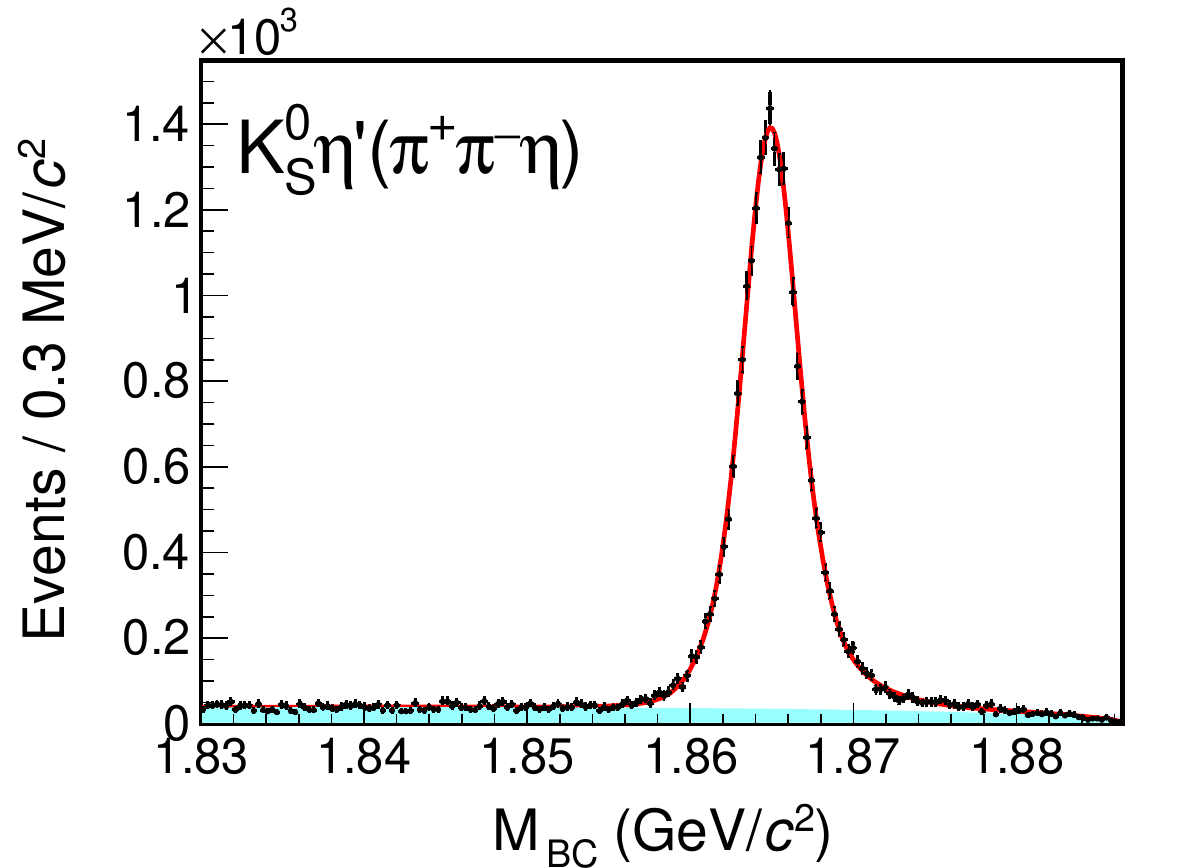}
    \includegraphics[height=4.195cm,trim={1.65cm 5.0cm 0 0},clip=true]{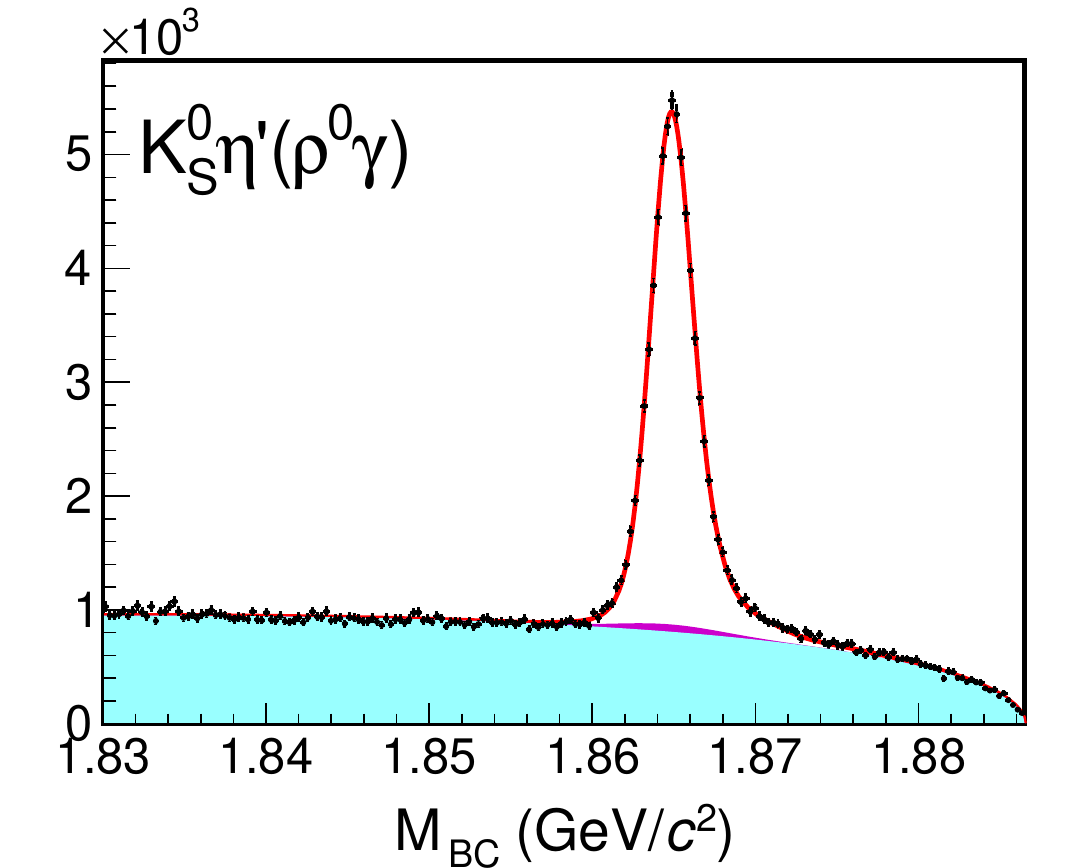}
    \includegraphics[height=4.195cm,trim={1.65cm 5.0cm 0 0},clip=true]{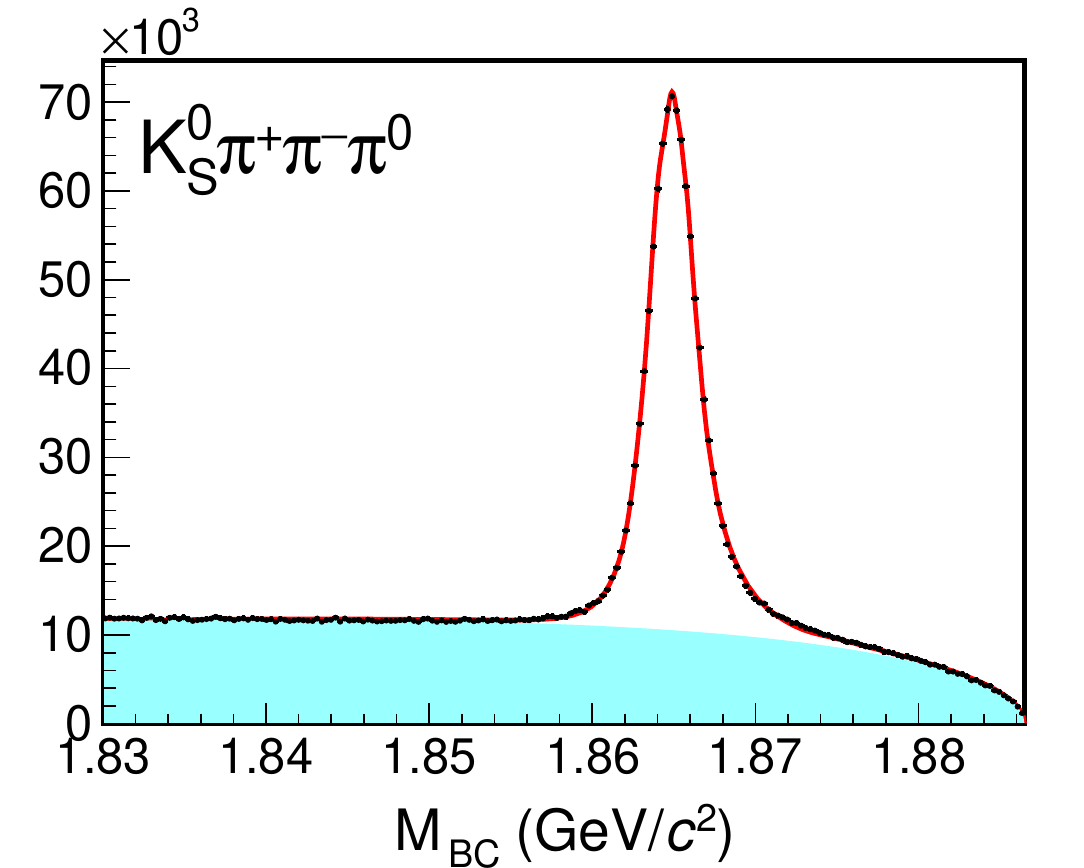}
    \caption{ST $M_{\rm BC}$ distributions. Data points are shown in black with error bars and the red curve is the fit result. The solid blue shape is the combinatorial background. The magenta, stacked on top of the blue, is the peaking background.}
    \label{figure:ST_MBC}
\end{figure*}

The non-peaking combinatorial background is described with an ARGUS~\cite{cite:Argus} function with the end point fixed at $1.8865$~GeV$/c^2$, but the shape parameter is free to vary in the fit. The yields of signal and combinatorial background are also free parameters. Where applicable, backgrounds that have peaking structures near the signal peak are included as a component with a fixed shape from simulation. The yields of peaking backgrounds are fixed relative to the signal yield, using their relative efficiencies from simulation and branching fractions from the PDG~\cite{pdg}. The level of peaking background in fully reconstructed tags is around a few percent or less. The ST yield fits of all fully reconstructed tags are shown in Fig.~\ref{figure:ST_MBC}.

For the partially reconstructed tags, the ST yield cannot be measured directly. Nonetheless, an effective ST reconstruction efficiency of the tag $f$ is calculated from $\epsilon_{\rm ST}(f) = \epsilon_{\rm DT}(KK\pi\pi|f)/\epsilon_{\rm ST}(KK\pi\pi)$. The effective ST yield is then calculated from this efficiency, the branching fraction~\cite{cite:deltaKpi}, and $N_{D\bar{D}}$, using Eq.~\eqref{equation:ST_yield}. The ST yields and their efficiencies, determined from simulation, are presented in Table~\ref{table:Single_tag_yields_efficiencies}.

\begin{table}[htb!]
    \centering
    \caption{ST yields in data and corresponding efficiencies as determined from MC simulation.}
    \label{table:Single_tag_yields_efficiencies}
    \begin{tabular}{lcc}
        \hline
        Tag mode                          & ST         yield     & ST efficiency ($\%$) \\
        \hline
        $K^+K^-\pi^+\pi^-$                & $\phantom{00}74487 \pm 426\phantom{00}$      & $18.71 \pm 0.04$     \\
        \hline
        $K^-\pi^+$                        & $3817700 \pm 2040\phantom{0}$   & $68.53 \pm 0.05$     \\
        $K^-\pi^+\pi^0$                   & $7252630 \pm 9050\phantom{0}$   & $36.90 \pm 0.05$     \\
        $K^-\pi^+\pi^-\pi^+$              & $4847930 \pm 2710\phantom{0}$   & $41.14 \pm 0.05$     \\
        $K^-e^{+}\nu_e$                   & $3030700 \pm 41200$  & $58.26 \pm 0.17$     \\
        \hline
        $K^+K^-$                          & $\phantom{0}387176 \pm 677\phantom{00}$     & $63.10 \pm 0.05$     \\
        $\pi^+\pi^-$                      & $\phantom{0}142731 \pm 475\phantom{00}$     & $66.65 \pm 0.05$     \\
        $\pi^+\pi^-\pi^0$                 & $\phantom{0}757190 \pm 1500\phantom{0}$    & $36.27 \pm 0.05$     \\
        $K_S^0\pi^0\pi^0$                 & $\phantom{0}159212 \pm 616\phantom{00}$     & $14.46 \pm 0.04$     \\
        $K_L^0\pi^0$                      & $\phantom{0}438500 \pm 15500$   & $31.19 \pm 0.12$     \\
        $K_S^0\pi^0$                      & $\phantom{0}482803 \pm 745\phantom{00}$     & $37.43 \pm 0.05$     \\
        $K_S^0\eta$                       & $\phantom{00}65382 \pm 304\phantom{00}$      & $31.34 \pm 0.05$     \\
        $K_S^0\eta^\prime_{\pi\pi\eta}$   & $\phantom{00}24059 \pm 169\phantom{00}$      & $11.78 \pm 0.03$     \\
        $K_S^0\eta^\prime_{\rho\gamma}$   & $\phantom{00}59428 \pm 310\phantom{00}$      & $18.06 \pm 0.04$     \\
        $K_S^0\pi^+\pi^-\pi^0$            & $\phantom{0}920140 \pm 1760\phantom{0}$    & $14.71 \pm 0.04$     \\
        \hline
        $K_S^0\pi^+\pi^-$                 & $1174820 \pm 1820\phantom{0}$   & $34.89 \pm 0.05$     \\
        $K_L^0\pi^+\pi^-$                 & $1839700 \pm 50900$  & $39.59 \pm 0.09$     \\
        \hline
    \end{tabular}
\end{table}

The DT yields are determined in bins of phase space, using the $2 \times 4$ binning scheme described in Ref.~\cite{LHCb-PAPER-2022-037} and provided in  Ref.~\cite{cite:KKpipi_BinningScheme}. For each fully reconstructed tag, an unbinned maximum likelihood fit of all phase-space bins is simultaneously performed on the signal-side $M_{\rm BC}$, where the shape of the signal and background contributions are shared between all bins. The yields of signal and combinatorial background in each bin are free parameters. The shapes of signal and background are obtained in an identical manner to that of ST yield fits, but only a single Gaussian function is used as a resolution function in all tag modes. In bins with low fractional bin yields or bins with large suppression due to quantum correlations, the observed DT yield may be small. Therefore, asymmetric uncertainties are assigned by scanning the profile likelihood to obtain a more accurate confidence interval.

\begin{figure*}[htb!]
    \centering
    \includegraphics[width=0.495\textwidth,trim={0 5.0cm 0 0},clip=true]{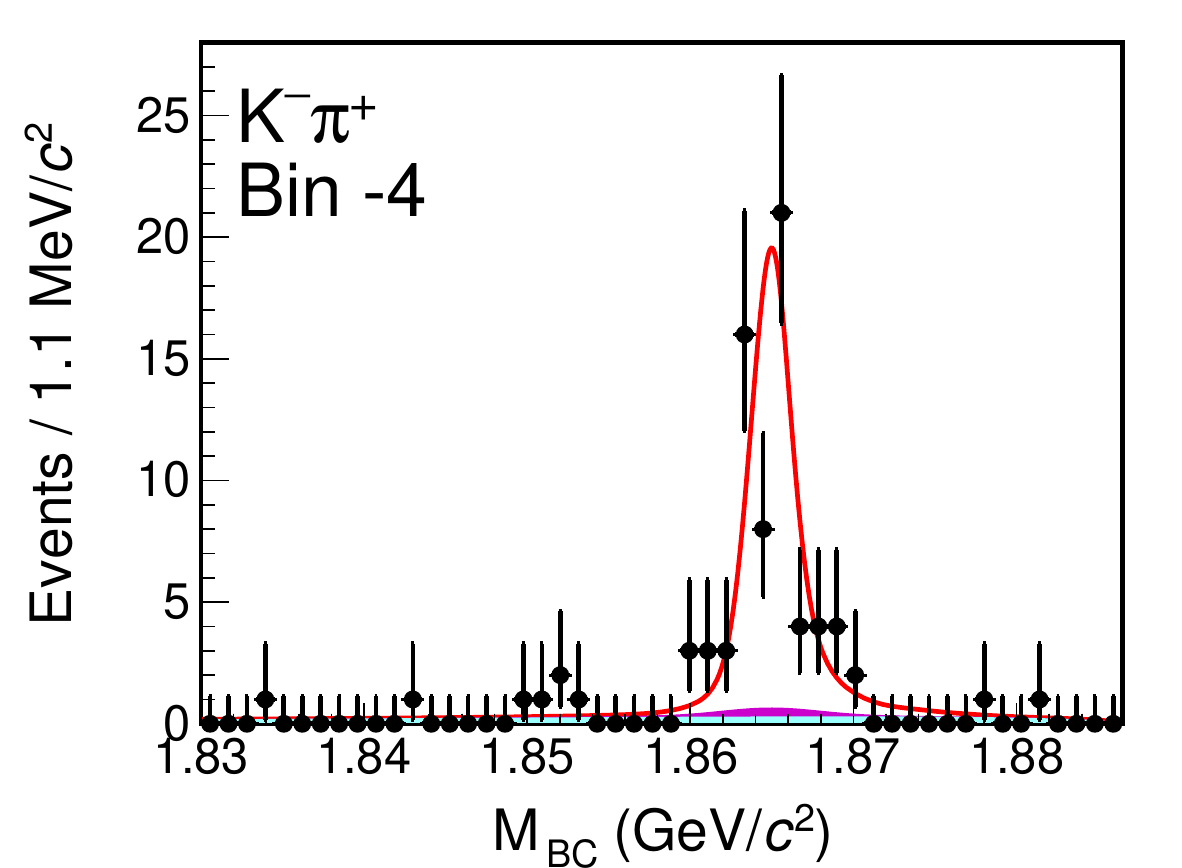}
    \includegraphics[width=0.495\textwidth,trim={0 5.0cm 0 0},clip=true]{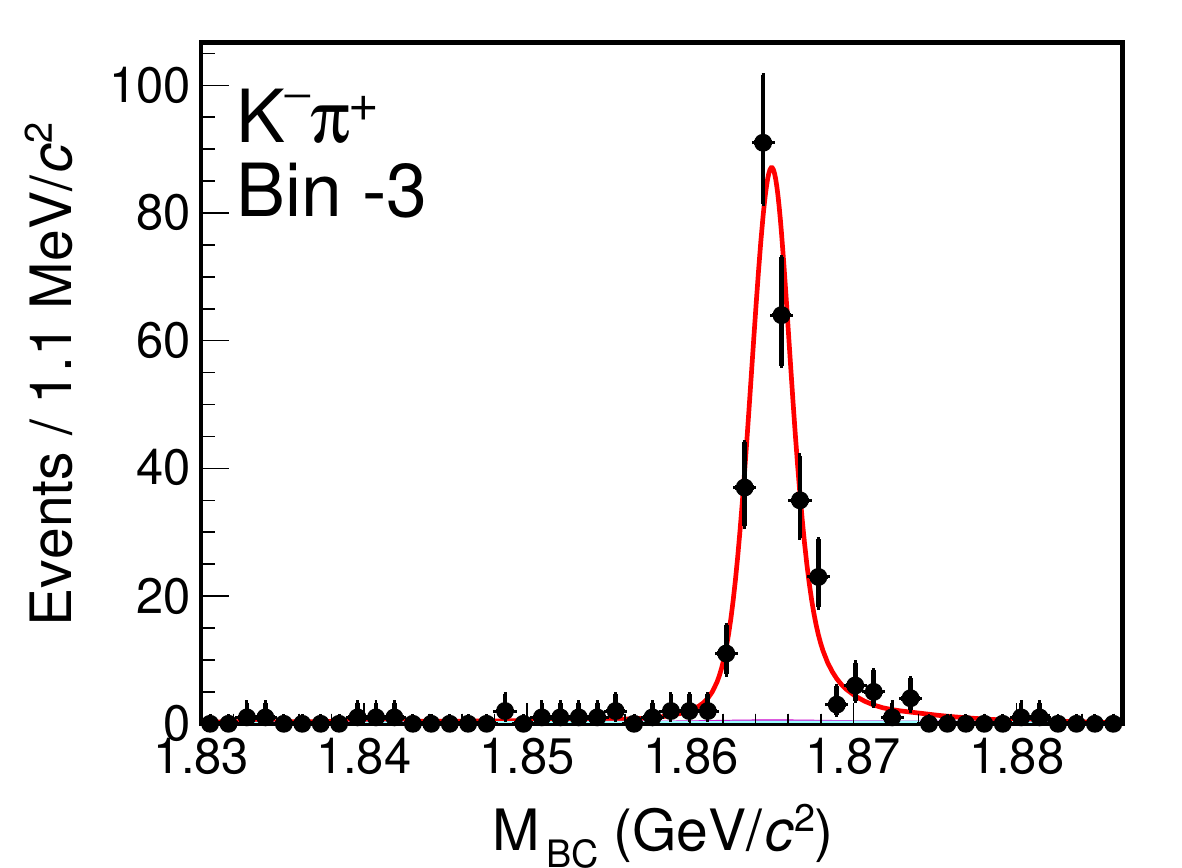}
    \includegraphics[width=0.495\textwidth,trim={0 5.0cm 0 0},clip=true]{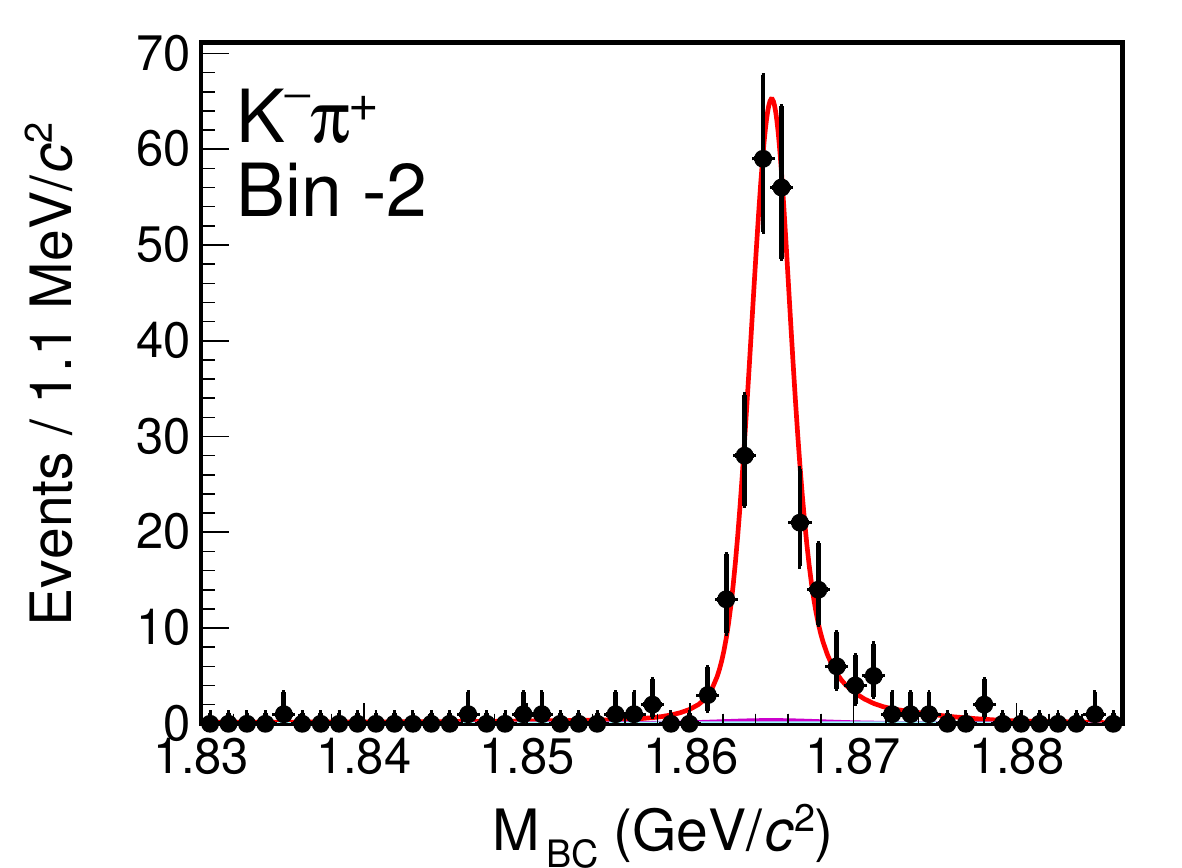}
    \includegraphics[width=0.495\textwidth,trim={0 5.0cm 0 0},clip=true]{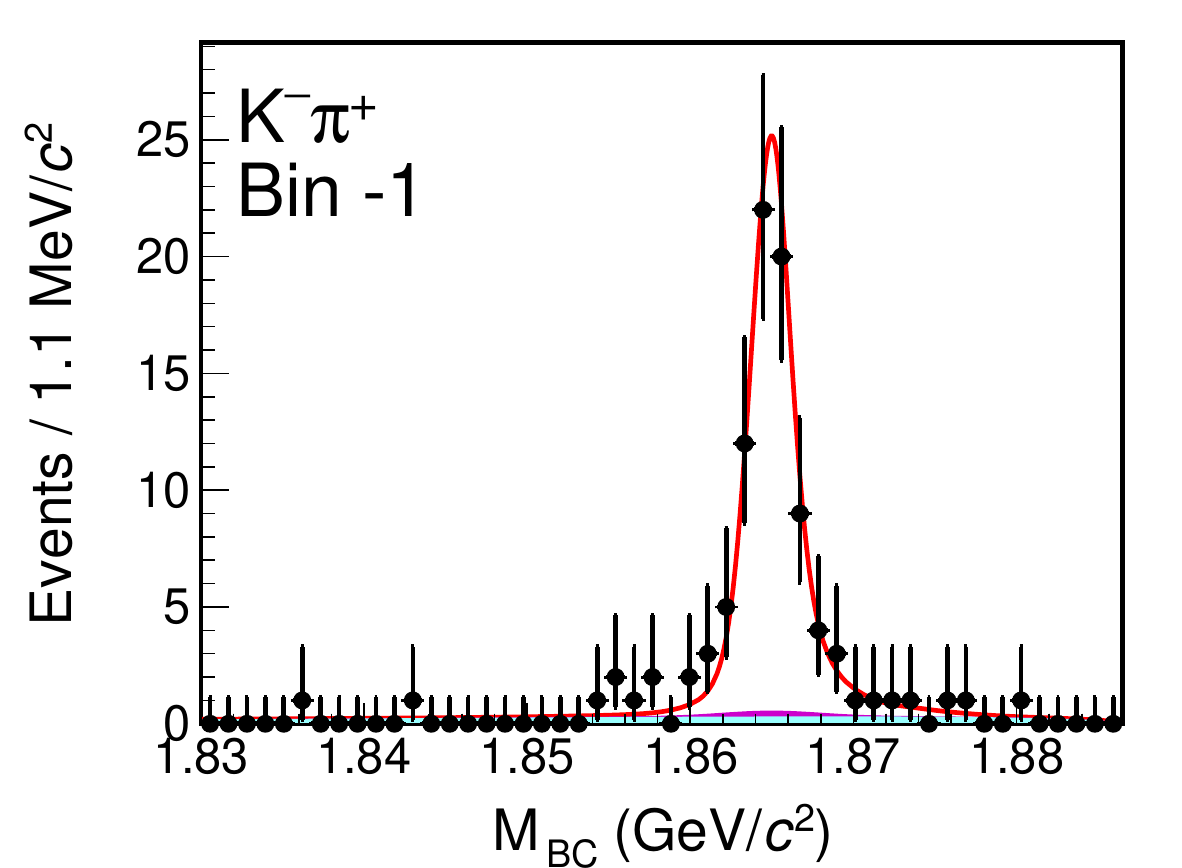}
    \caption{$M_{\rm BC}$ distributions of the signal mode in $\bar{D{}^0}\to K^+K^-\pi^+\pi^-$ versus $D^0\to K^-\pi^+$ events in bins where $i < 0$. Data points are shown in black with error bars and the red curve is the fit result. The solid blue shape is the combinatorial background. The magenta, stacked on top of the blue, is the peaking background.}
    \label{figure:DT_MBC_Kpi_minus}
\end{figure*}

\begin{figure*}[htb!]
    \centering
    \includegraphics[width=0.495\textwidth,trim={0 5.0cm 0 0},clip=true]{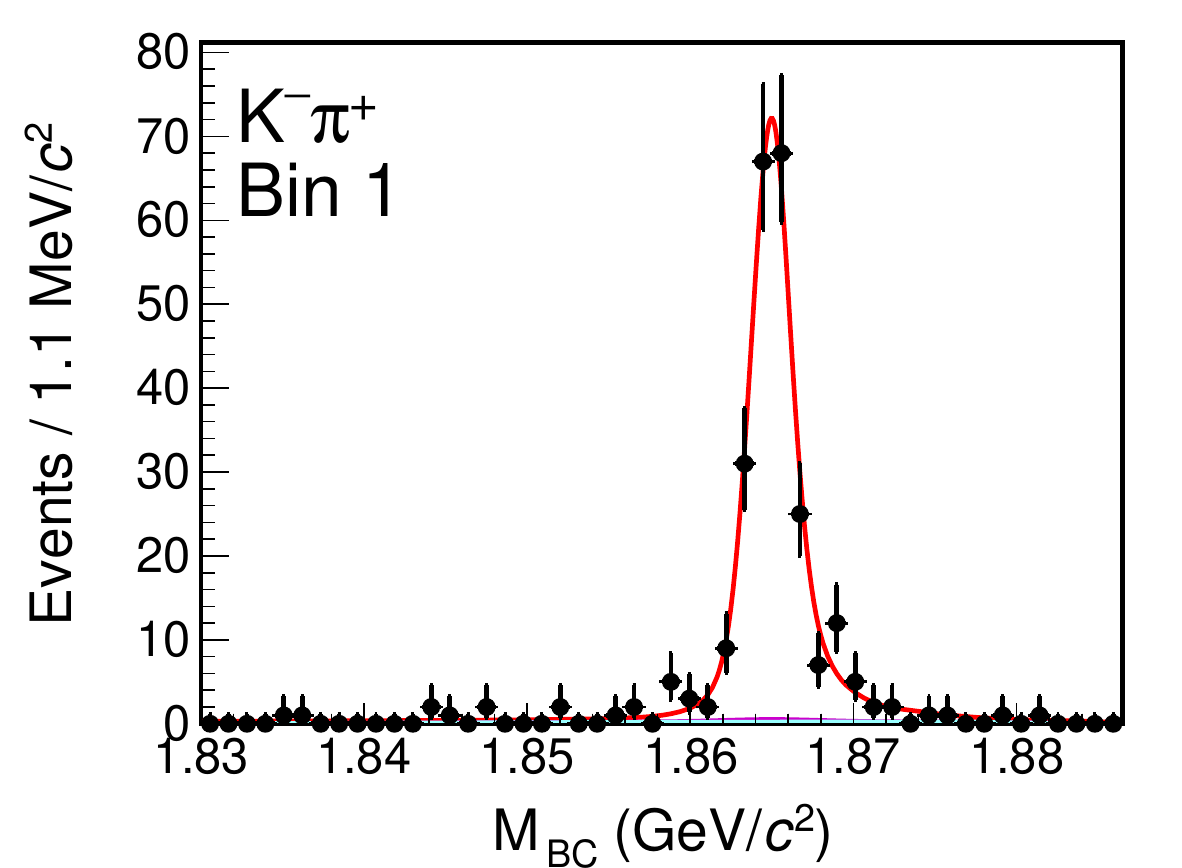}
    \includegraphics[width=0.495\textwidth,trim={0 5.0cm 0 0},clip=true]{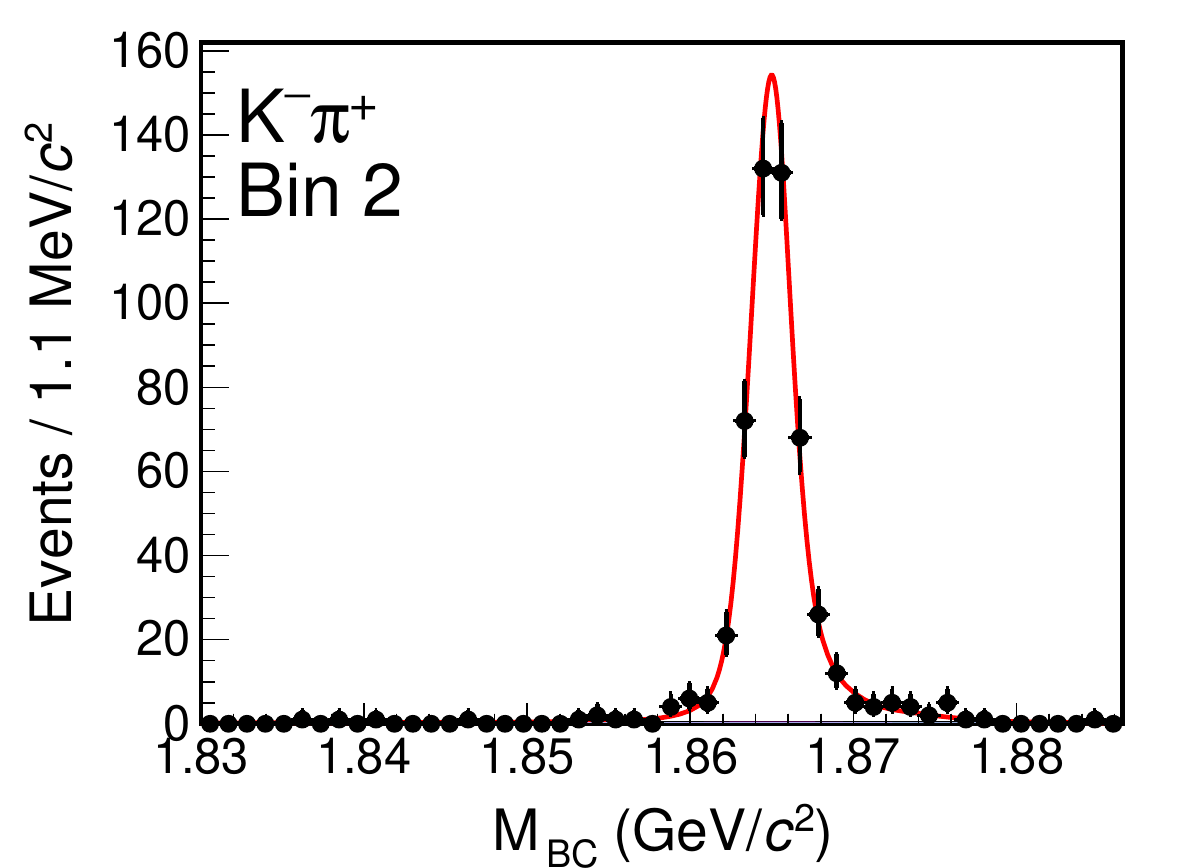}
    \includegraphics[width=0.495\textwidth,trim={0 5.0cm 0 0},clip=true]{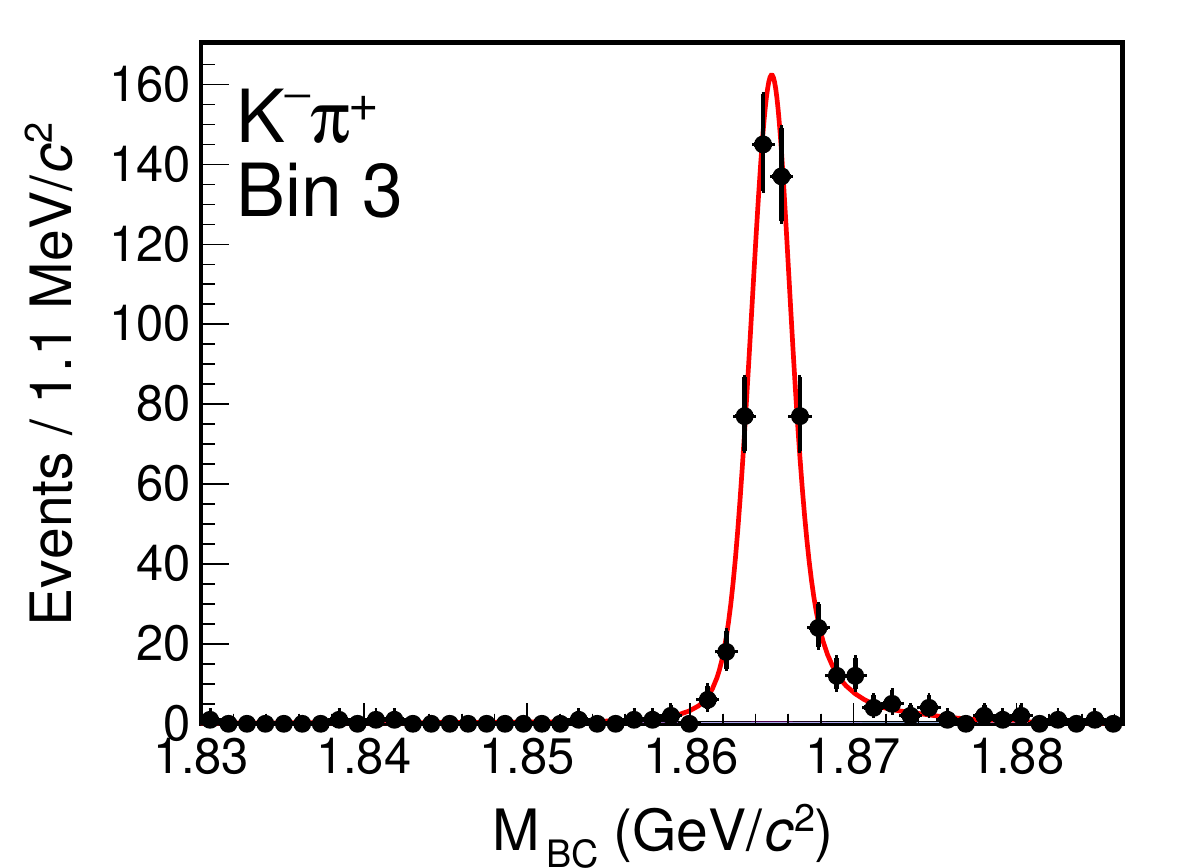}
    \includegraphics[width=0.495\textwidth,trim={0 5.0cm 0 0},clip=true]{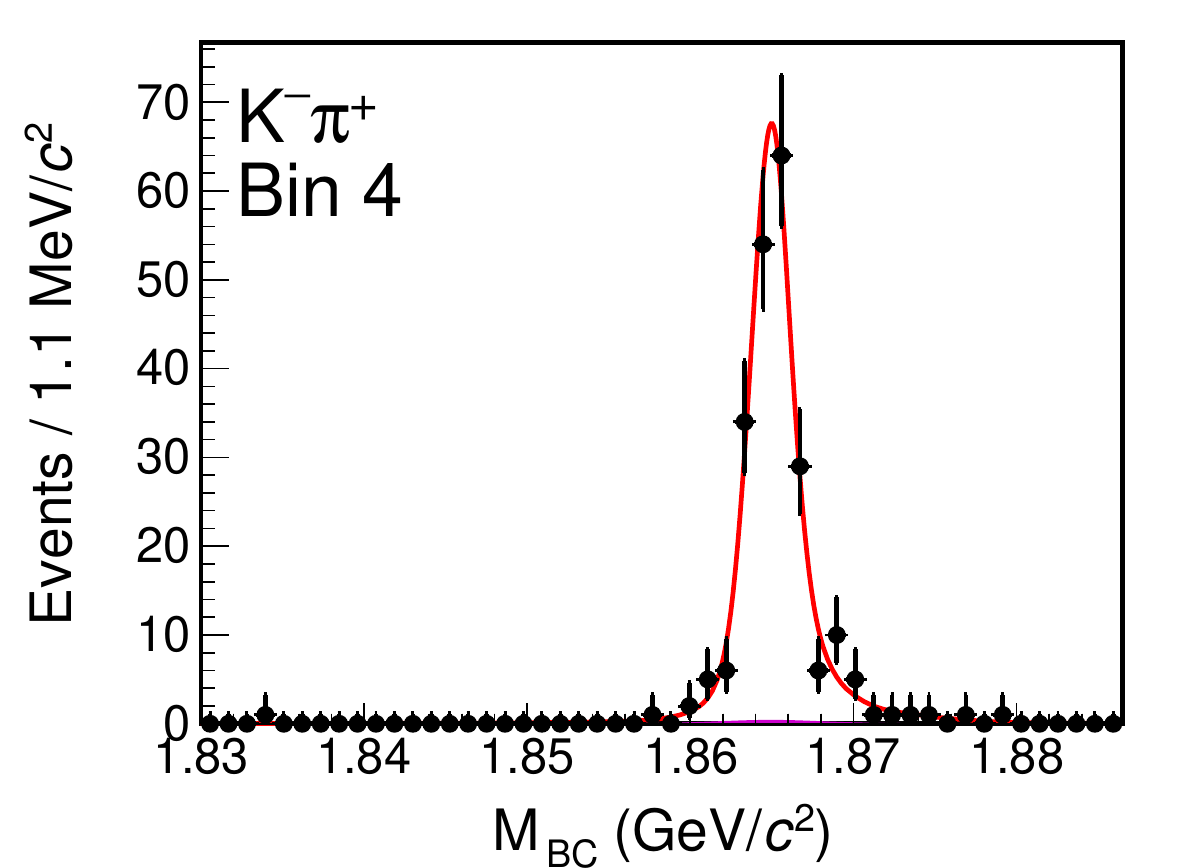}
    \caption{$M_{\rm BC}$ distributions of the signal mode in $\bar{D{}^0}\to K^+K^-\pi^+\pi^-$ versus $D^0\to K^-\pi^+$ events in bins where $i > 0$. Data points are shown in black with error bars and the red curve is the fit result. The solid blue shape is the combinatorial background. The magenta, stacked on top of the blue, is the peaking background.}
    \label{figure:DT_MBC_Kpi_plus}
\end{figure*}

In the partially reconstructed modes, where a $K_L^0$ or a charged kaon is missing, a fit of the missing-mass squared, $M_{\rm miss}^2$, is performed instead, and the flat combinatorial background is described by a first-order polynomial function. The signal shape is treated in a similar manner to that of the fully reconstructed tags. In the $D^0\to K^-e^+\nu_e$ tag, the variable $U_{\rm miss}$ is fitted with a strategy identical to that for tags with a missing kaon.  

The peaking backgrounds in the DT fits are treated similarly to those for the ST candidates, but the yields are also corrected for enhancements or suppressions due to quantum correlations. The quantum correlation corrections are calculated using knowledge of the $C\!P$ contents of the signal and tag modes. The $C\!P$ content of $D\to K^+K^-\pi^+\pi^-$ is initially obtained from the amplitude model in Ref.~\cite{LHCb-PAPER-2018-041}, but the differences in DT yields are found to be negligible when the quantum-correlation corrections are updated using the results from Sect.~\ref{section:Phase_space_binned_strong_phase_measurement}.

\begin{table*}[htb!]
    \centering
    \caption{DT yields of flavor tags. The uncertainties are statistical only.}
    \label{table:Double_tag_yields_flavor}
    {\renewcommand{\arraystretch}{1.5}
    \begin{tabular}{ccccccccc}
        \hline
        Tag mode $\rightarrow$  & $K^-\pi^+$ & $K^-\pi^+\pi^0$ & $K^-\pi^+\pi^-\pi^+$ & $K^-e^+\nu_e$ \\
        Bin number $\downarrow$ &              &                  &                      & \\
        \hline
        $-4$  & $61.2^{+8.5}_{-7.8}$      & $117.9^{+11.7}_{-11.1}$   & $84.1^{+10.1}_{-9.5}$     & $60.4^{+7.8}_{-7.2}$      \\
        $-3$  & $279.0^{+17.5}_{-16.8}$   & $442.9^{+22.6}_{-21.9}$   & $268.3^{+17.7}_{-17.0}$   & $175.9^{+13.9}_{-13.2}$   \\
        $-2$  & $209.5^{+15.1}_{-14.5}$   & $365.4^{+20.2}_{-19.6}$   & $218.5^{+15.8}_{-15.2}$   & $126.4^{+11.6}_{-11.0}$   \\
        $-1$  & $79.7^{+9.8}_{-9.1}$      & $165.2^{+14.0}_{-13.3}$   & $101.0^{+11.2}_{-10.5}$   & $67.8^{+8.5}_{-8.0}$      \\
        $\phantom{-}1$   & $231.2^{+16.0}_{-15.3}$   & $436.5^{+21.9}_{-21.2}$   & $288.1^{+17.9}_{-17.2}$   & $164.8^{+13.4}_{-12.8}$   \\
        $\phantom{-}2$   & $496.9^{+23.1}_{-22.4}$   & $931.7^{+31.8}_{-31.1}$   & $566.5^{+24.8}_{-24.1}$   & $318.1^{+18.8}_{-18.2}$   \\
        $\phantom{-}3$   & $524.3^{+23.6}_{-22.9}$   & $999.4^{+33.1}_{-32.5}$   & $576.3^{+25.2}_{-24.6}$   & $378.4^{+20.2}_{-19.5}$   \\
        $\phantom{-}4$   & $217.6^{+15.1}_{-14.4}$   & $348.9^{+19.8}_{-19.1}$   & $235.6^{+16.4}_{-15.7}$   & $162.9^{+13.0}_{-12.4}$   \\
        \hline
    \end{tabular}}
\end{table*}

For each flavor tag there are eight measured DT yields, which are listed in Table~\ref{table:Double_tag_yields_flavor}. The projections of the fit of the $D^0\to K^-\pi^+$ tag are shown in Figs.~\ref{figure:DT_MBC_Kpi_minus} and \ref{figure:DT_MBC_Kpi_plus}. The fit projections of the other flavor tags are similar.

\begin{figure*}[htb!]
    \centering
    \includegraphics[width=0.495\textwidth,trim={0 5.0cm 0 0},clip=true]{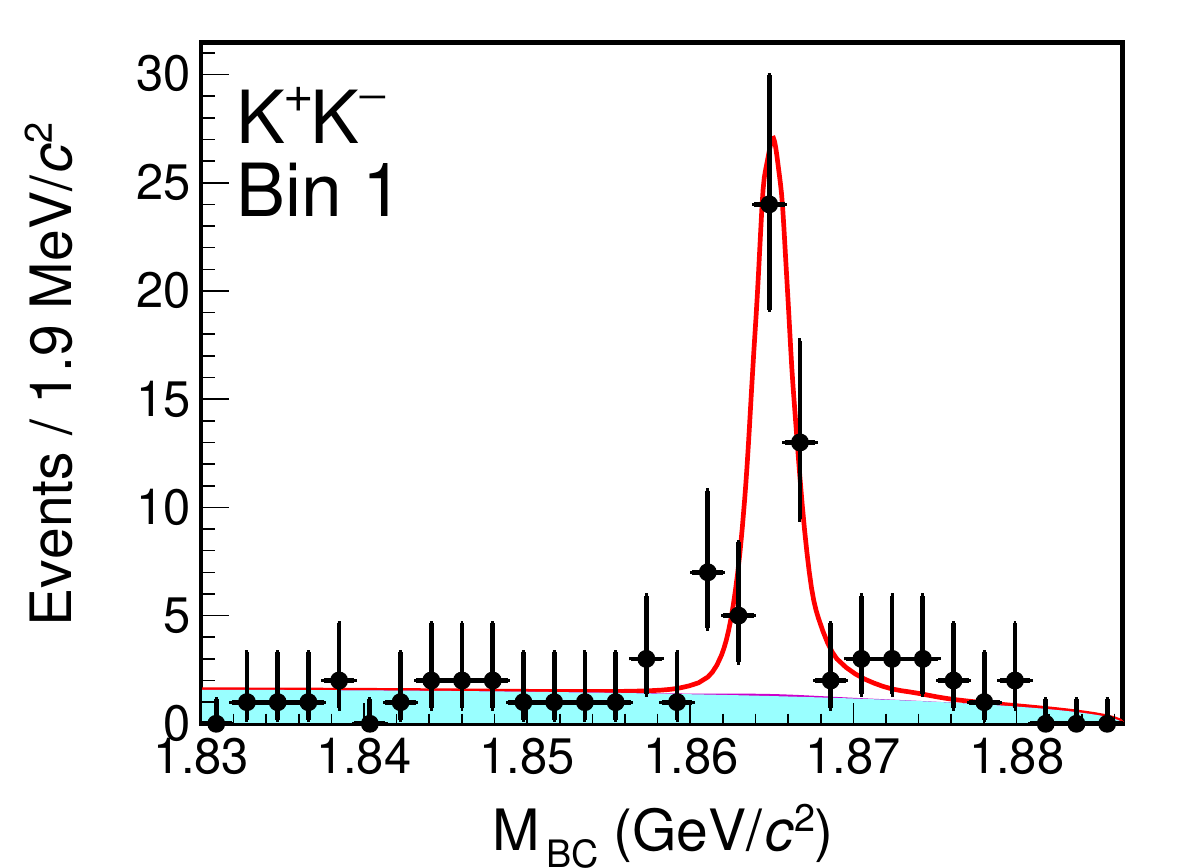}
    \includegraphics[width=0.495\textwidth,trim={0 5.0cm 0 0},clip=true]{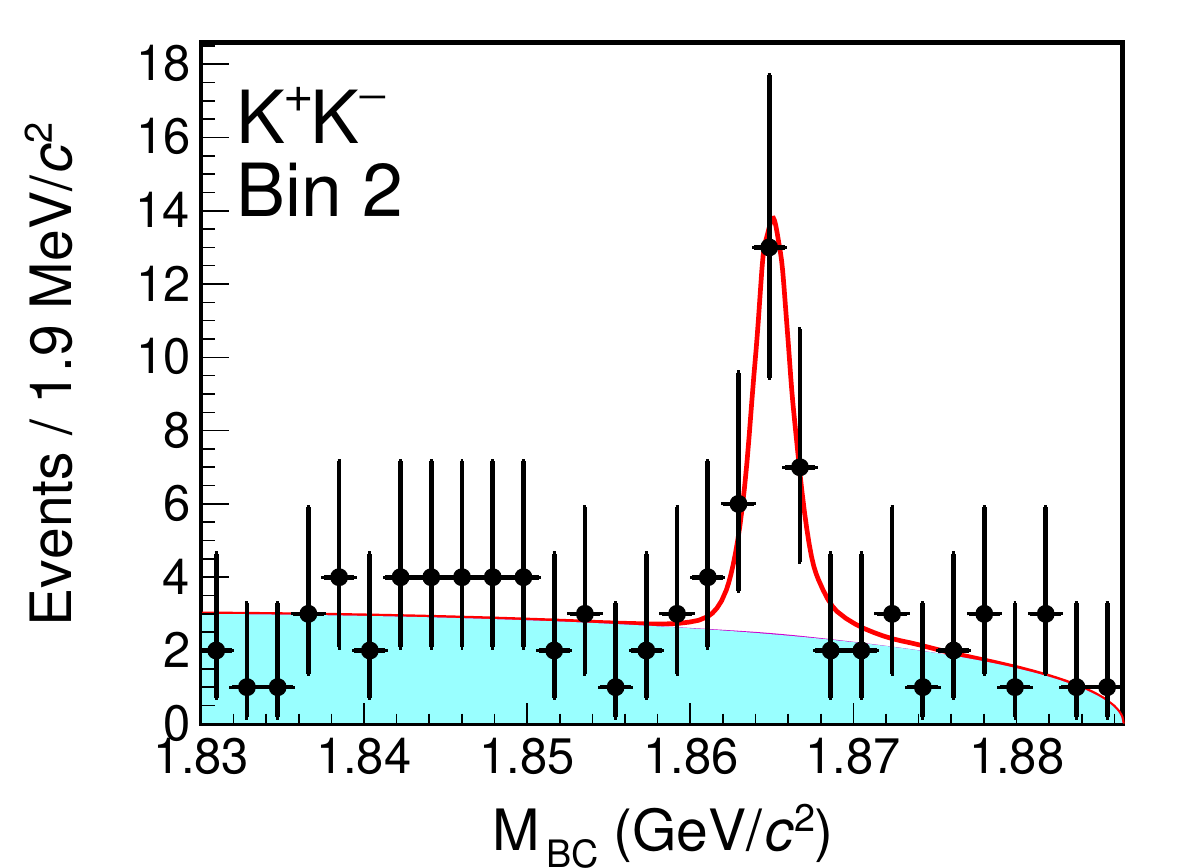}
    \includegraphics[width=0.495\textwidth,trim={0 5.0cm 0 0},clip=true]{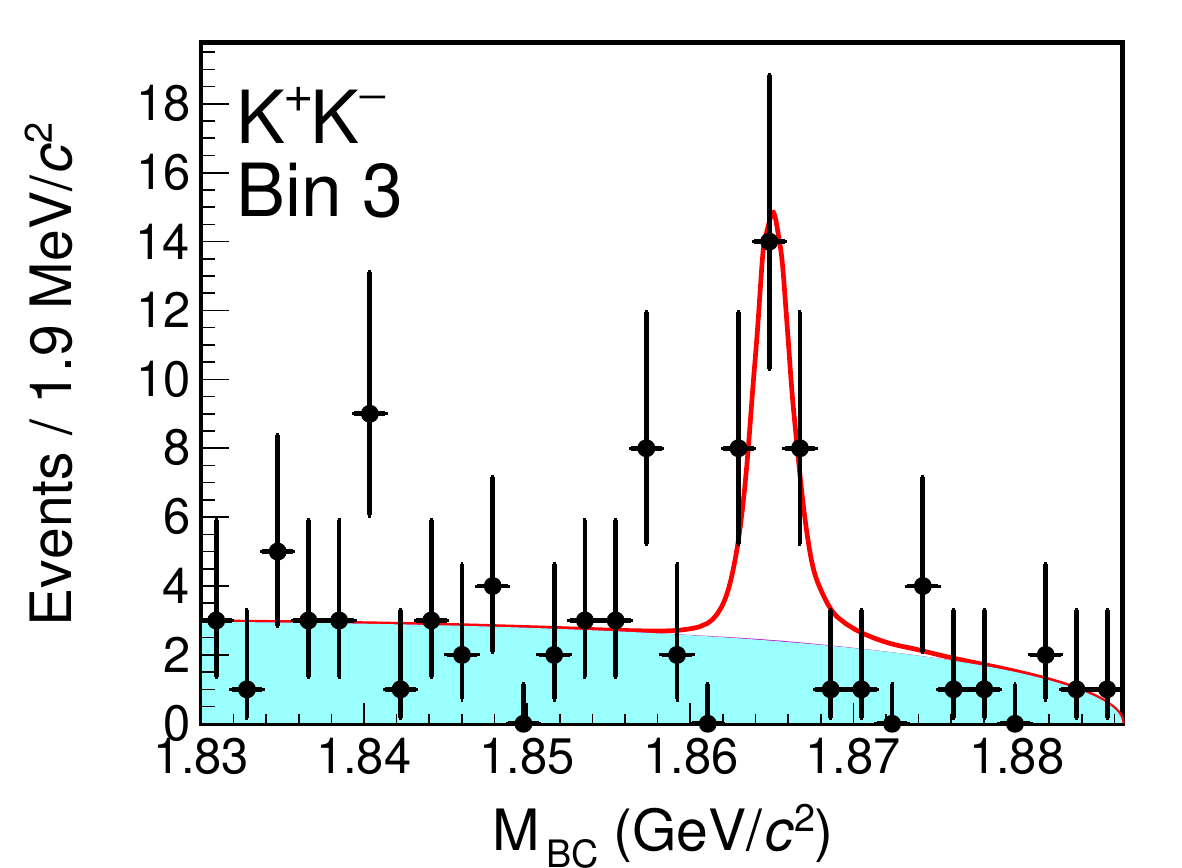}
    \includegraphics[width=0.495\textwidth,trim={0 5.0cm 0 0},clip=true]{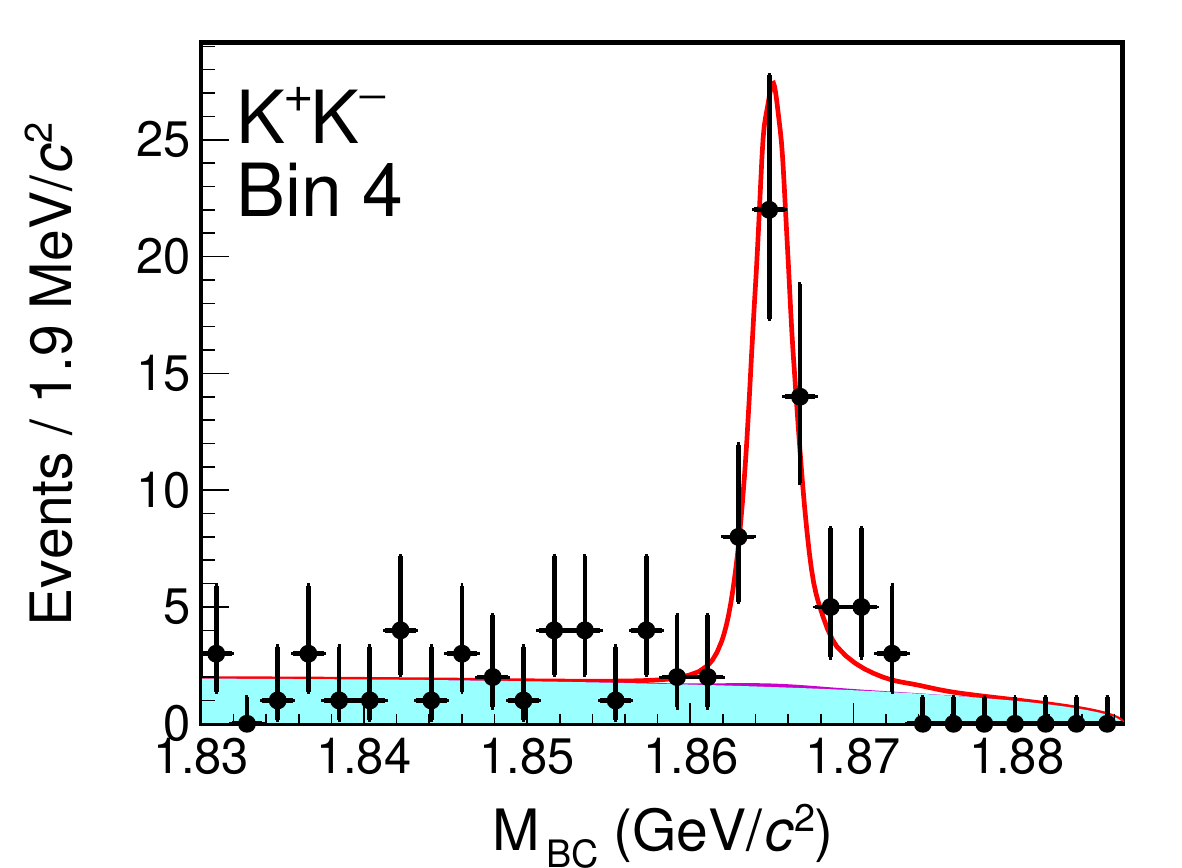}
    \caption{$M_{\rm BC}$ distributions of the signal mode in $D\to K^+K^-\pi^+\pi^-$ versus $D\to K^+K^-$ events. Data points are shown in black with error bars and the red curve is the fit result. The solid blue shape is the combinatorial background. The magenta, stacked on top of the blue, is the peaking background.}
    \label{figure:DT_MBC_KK}
\end{figure*}

In the fit of $C\!P$ tags, bin $i$ and $-i$ are equivalent and are therefore merged, resulting in four DT yields for each tag that are listed in Table~\ref{table:Double_tag_yields_CP}. The entries marked with a $\dagger$ are selections where the signal mode $D\to K^+K^-\pi^+\pi^-$ is partially reconstructed with a missing kaon. While the yields obtained from these tag modes are similar to their fully reconstructed counterparts, the uncertainties are larger due to the higher level of background.

\begin{table}[htb!]
    \centering
    \caption{DT yields of $C\!P$ tags. The uncertainties are statistical only.}
    \label{table:Double_tag_yields_CP}
    {\renewcommand{\arraystretch}{1.5}
    \begin{tabular}{ccccc}
        \hline
        Bin number $\rightarrow$ & $1$ & $2$ & $3$ & $4$ \\
        Tag mode $\downarrow$ & & & & \\
        \hline
        $K^+K^-$                            & $46.6^{+8.1}_{-7.4}$      & $20.6^{+6.0}_{-5.3}$      & $22.5^{+6.1}_{-5.4}$      & $46.7^{+8.0}_{-7.3}$      \\
        $K^+K^{-\dagger}$                   & $42.6^{+9.2}_{-8.4}$      & $34.6^{+8.9}_{-8.2}$      & $22.1^{+10.0}_{-9.0}$     & $35.1^{+7.6}_{-6.9}$      \\
        $\pi^+\pi^-$                        & $11.3^{+5.2}_{-4.5}$      & $3.0^{+4.5}_{-3.0}$       & $5.9^{+5.2}_{-4.4}$       & $11.6^{+4.9}_{-4.2}$      \\
        $\pi^+\pi^-\pi^0$                   & $74.1^{+12.7}_{-12.3}$    & $68.9^{+11.6}_{-11.1}$    & $70.6^{+11.6}_{-11.4}$    & $80.4^{+12.8}_{-12.4}$    \\
        $K_S^0\pi^0\pi^0$                   & $15.1^{+4.9}_{-4.2}$      & $6.4^{+3.2}_{-2.6}$       & $14.6^{+4.8}_{-4.2}$      & $14.0^{+4.4}_{-3.8}$      \\
        $K_L^0\pi^0$                        & $31.3^{+8.4}_{-7.8}$      & $21.9^{+4.9}_{-4.9}$      & $13.6^{+3.7}_{-3.8}$      & $19.8^{+7.2}_{-6.5}$      \\
        $K_S^0\pi^0$                        & $30.9^{+5.9}_{-5.2}$      & $125.2^{+12.0}_{-11.5}$   & $163.3^{+13.4}_{-12.6}$   & $28.7^{+5.7}_{-5.0}$      \\
        $K_S^0\pi^{0\dagger}$               & $5.6^{+6.1}_{-5.1}$       & $118.1^{+13.7}_{-13.0}$   & $144.2^{+15.0}_{-14.2}$   & $15.6^{+6.3}_{-5.5}$      \\
        $K_S^0\eta$                         & $8.7^{+3.4}_{-2.7}$       & $22.6^{+5.4}_{-4.7}$      & $24.4^{+5.7}_{-5.0}$      & $5.7^{+3.5}_{-2.8}$       \\
        $K_S^0\eta^\prime_{\pi\pi\eta}$     & $0.0^{+0.5}_{-0.0}$       & $8.7^{+3.4}_{-2.7}$       & $7.0^{+3.0}_{-2.3}$       & $2.9^{+2.0}_{-1.4}$       \\
        $K_S^0\eta^\prime_{\rho\gamma}$     & $5.6^{+2.6}_{-2.0}$       & $19.3^{+5.1}_{-4.4}$      & $18.0^{+4.7}_{-4.0}$      & $2.8^{+2.0}_{-1.4}$       \\
        $K_S^0\pi^+\pi^-\pi^0$              & $72.4^{+9.8}_{-9.1}$      & $184.5^{+15.1}_{-14.4}$   & $272.5^{+18.1}_{-17.5}$   & $56.1^{+8.6}_{-7.9}$      \\
        \hline
    \end{tabular}}
\end{table}

\begin{figure*}[htb!]
    \centering
    \includegraphics[width=0.495\textwidth,trim={0 5.0cm 0 0},clip=true]{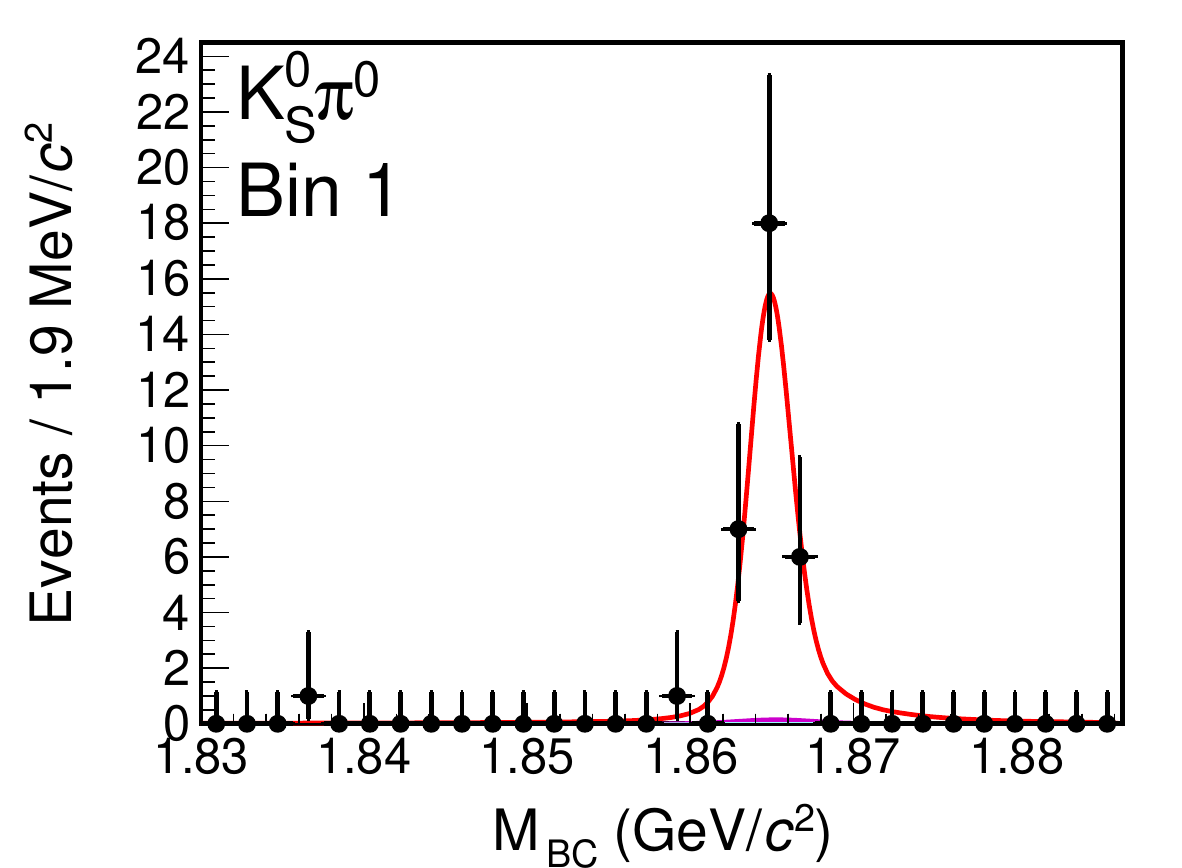}
    \includegraphics[width=0.495\textwidth,trim={0 5.0cm 0 0},clip=true]{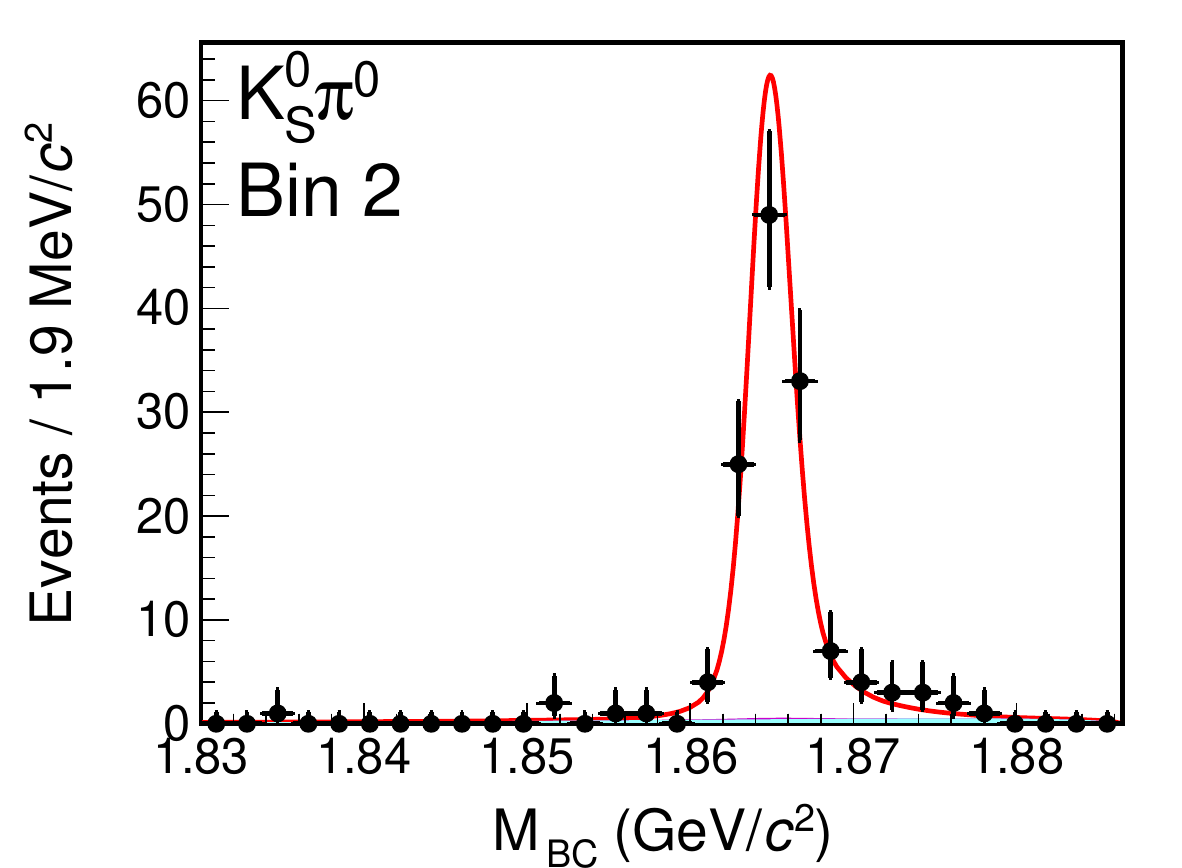}
    \includegraphics[width=0.495\textwidth,trim={0 5.0cm 0 0},clip=true]{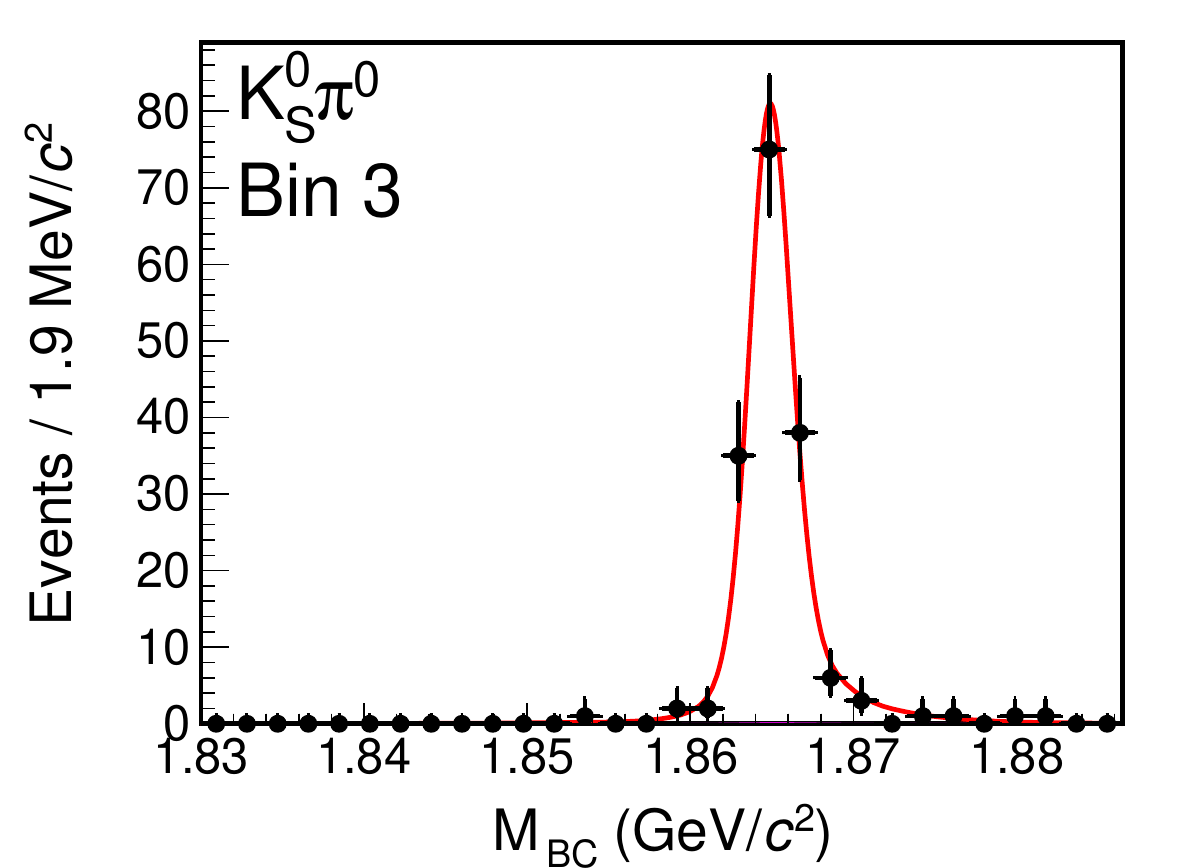}
    \includegraphics[width=0.495\textwidth,trim={0 5.0cm 0 0},clip=true]{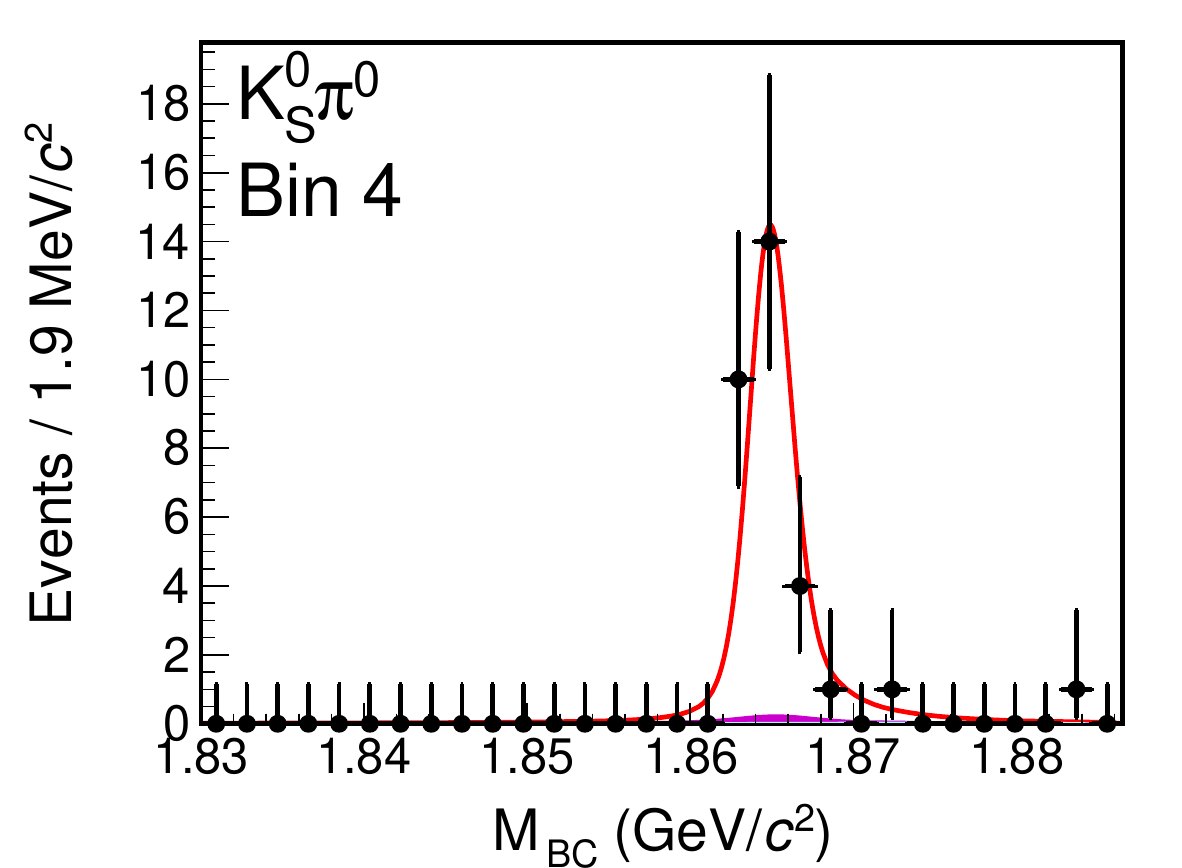}
    \caption{$M_{\rm BC}$ distributions of the signal mode in $D\to K^+K^-\pi^+\pi^-$ versus $D\to K_S^0\pi^0$ events. Data points are shown in black with error bars and the red curve is the fit result. The solid blue shape is the combinatorial background. The magenta, stacked on top of the blue, is the peaking background.}
    \label{figure:DT_MBC_KSpi0}
\end{figure*}

\begin{figure*}[htb!]
    \centering
    \includegraphics[width=0.495\textwidth,trim={0 5.0cm 0 0},clip=true]{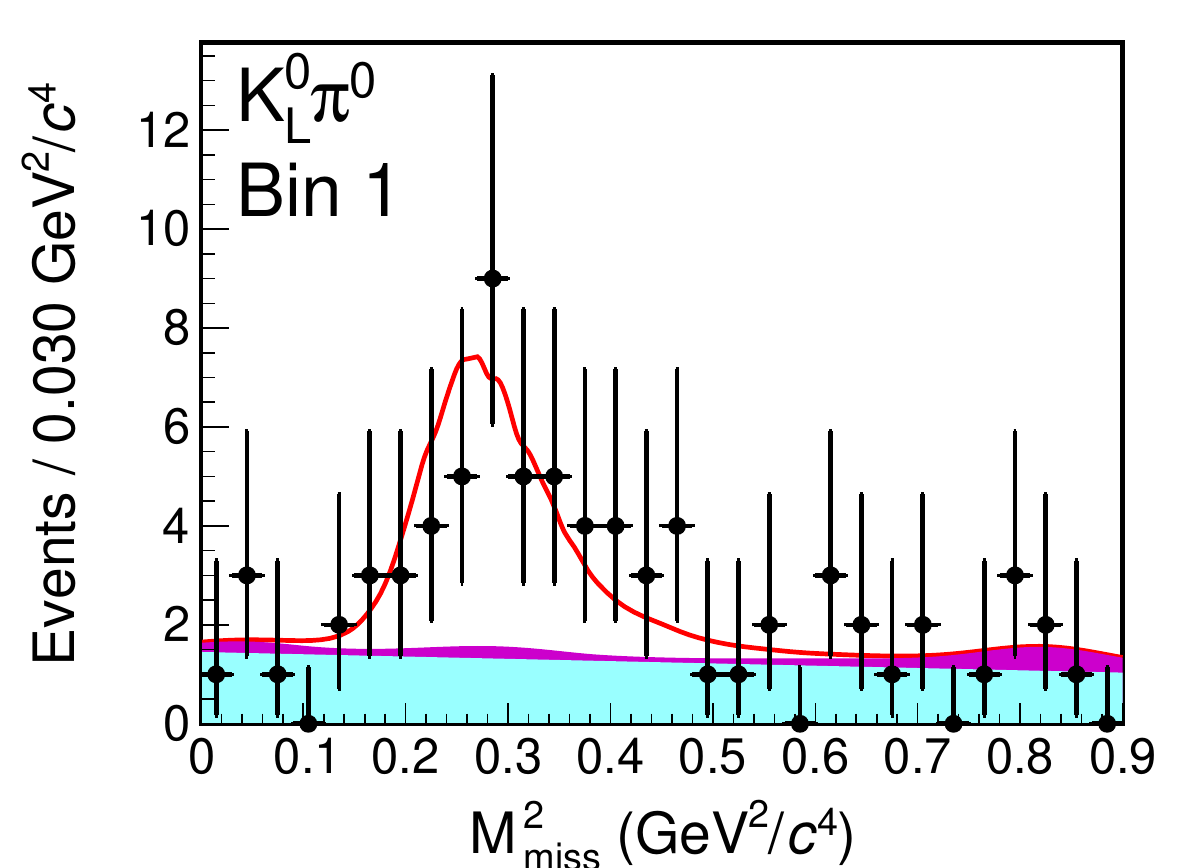}
    \includegraphics[width=0.495\textwidth,trim={0 5.0cm 0 0},clip=true]{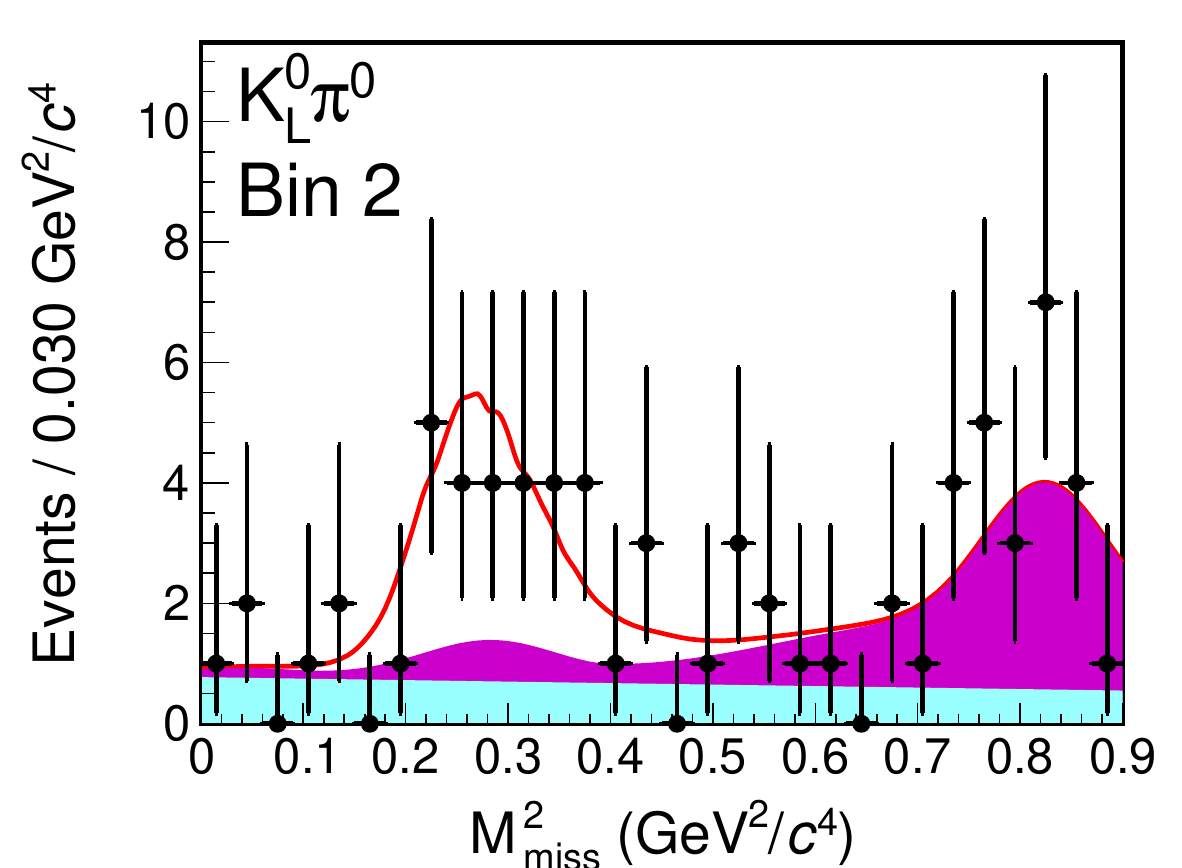}
    \includegraphics[width=0.495\textwidth,trim={0 5.0cm 0 0},clip=true]{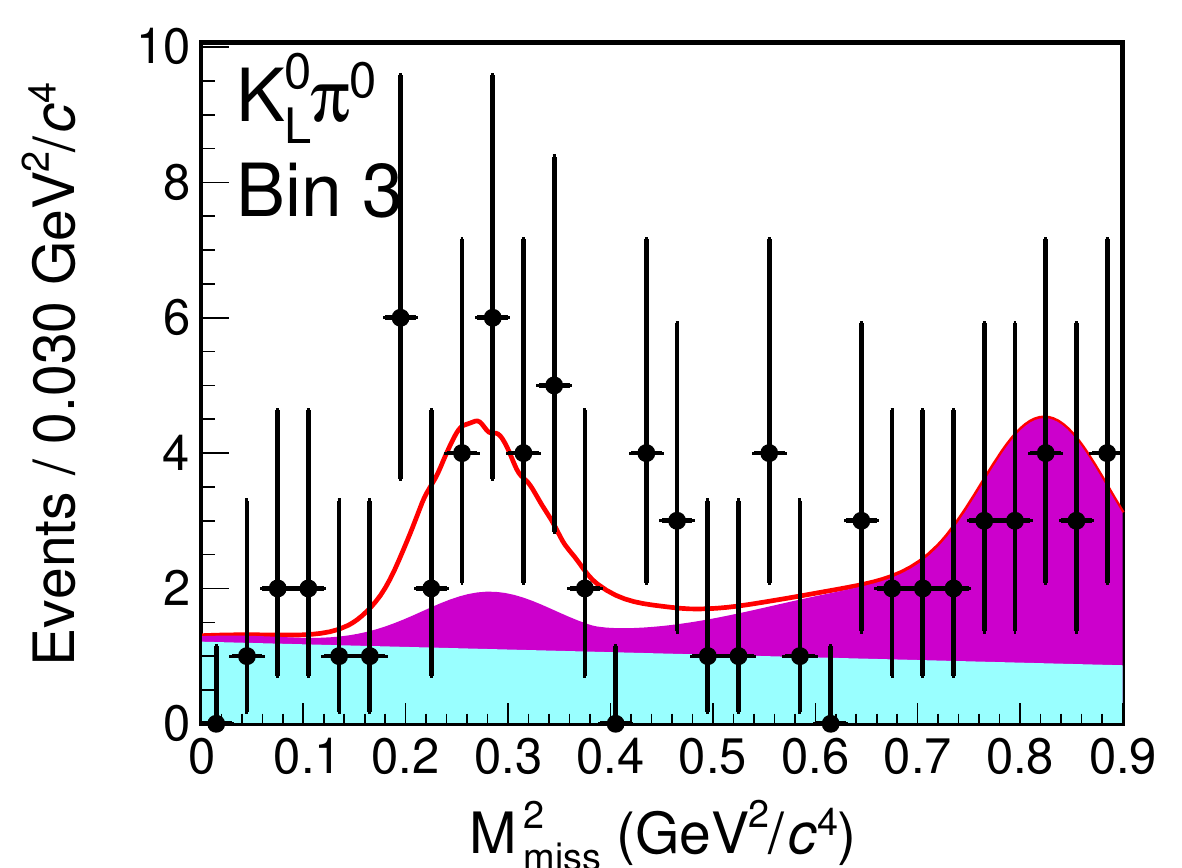}
    \includegraphics[width=0.495\textwidth,trim={0 5.0cm 0 0},clip=true]{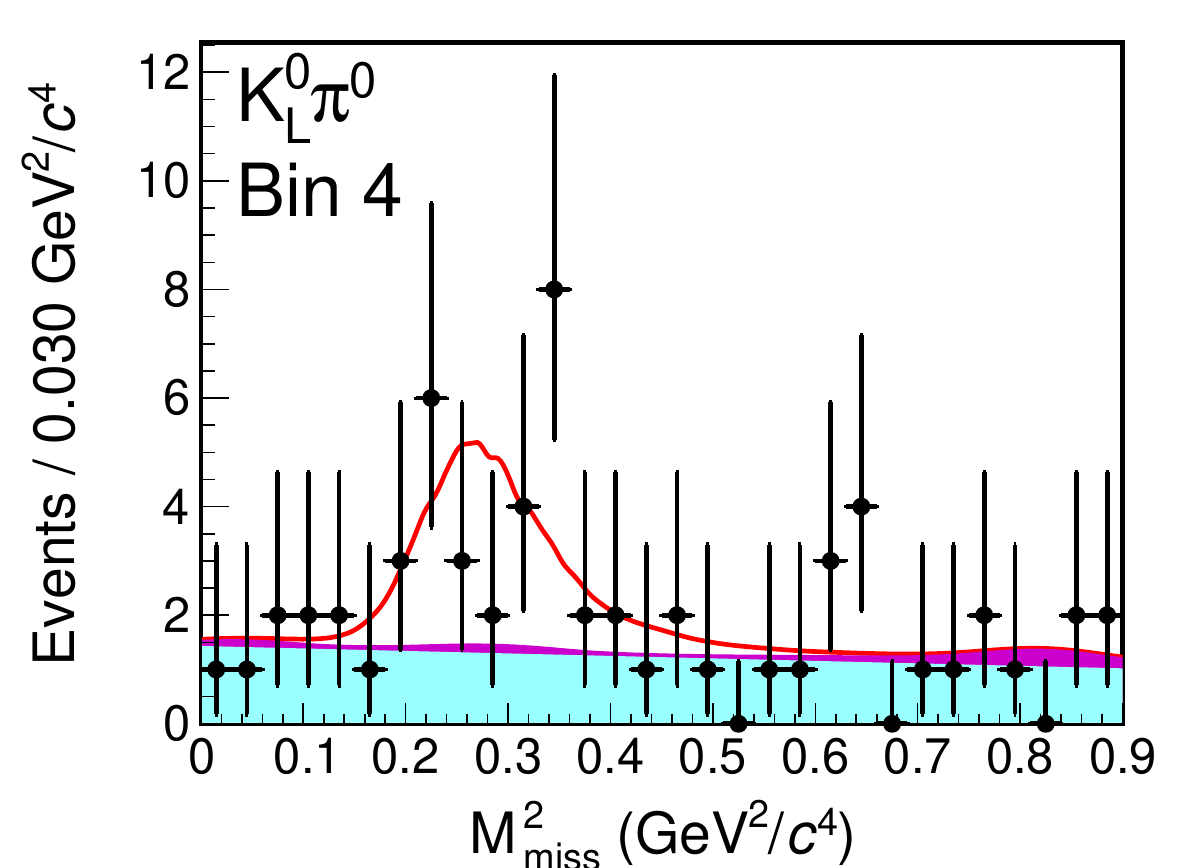}
    \caption{$M_{\rm miss}^2$ distributions of the signal mode in $D\to K^+K^-\pi^+\pi^-$ versus $D\to K_L^0\pi^0$ events. Data points are shown in black with error bars and the red curve is the fit result. The solid blue shape is the combinatorial background. The magenta, stacked on top of the blue, is the peaking background.}
    \label{figure:DT_MBC_KLpi0}
\end{figure*}

The fit projections of the $D\to K^+K^-$ and $D\to K_S^0\pi^0$ tags are shown in Figs.~\ref{figure:DT_MBC_KK} and \ref{figure:DT_MBC_KSpi0}, respectively, while that of the partially reconstructed tag mode $D\to K_L^0\pi^0$ is shown in Fig.~\ref{figure:DT_MBC_KLpi0}. Comparing these to Figs.~\ref{figure:DT_MBC_Kpi_minus} and \ref{figure:DT_MBC_Kpi_plus}, interesting quantum correlations can be seen qualitatively. In particular, in the $D\to K^+K^-$ and $D\to K_L^0\pi^0$ tags, which are $C\!P$-even decay modes, the yields in Figs.~\ref{figure:DT_MBC_KK} and \ref{figure:DT_MBC_KLpi0} are larger in bins $1$ and $4$, compared to bins $2$ and $3$. In comparison, for $D\to K_S^0\pi^0$, which is a $C\!P$-odd tag, the behavior is in the opposite sense. This observation suggests that in bins $1$ and $4$, the decay $D\to K^+K^-\pi^+\pi^-$ is predominantly $C\!P$-odd, and it is therefore enhanced when tagged with a $C\!P$-even decay due to quantum correlations, or suppressed by $C\!P$-odd tags. Similarly, in bins $2$ and $3$ it may be inferred that the behaviour is predominantly $C\!P$-even. This conclusion agrees with the quantitative findings presented in Sect.~\ref{section:Phase_space_binned_strong_phase_measurement}.

Finally, for the tag modes $D\to K_{S, L}^0\pi^+\pi^-$, which have mixed $C\!P$ content, the events are split into bins on both the tag and the signal side. The bin numbers are labelled with a pair of indices $(i, j)$, where $i$ ($j$) refers to the signal- (tag-) side bin number. The tag modes are analysed in a $2 \times 8$ binning scheme. However, since bin $(i, -j)$ in the event is equivalent to bin $(-i, j)$, these bins are merged such that there are $2\times4$ bins on the signal side and $8$ bins on the tag side, making $2\times4\times8 = 64$ bin combinations in total.

The bin yields of the fully reconstructed $D \to K_S^0\pi^+\pi^-$ tag mode, the $D \to K_S^0\pi^+\pi^-$ tag with partially reconstructed $D \to K^+K^-\pi^+\pi^-$ and the partially reconstructed $D \to K_L^0\pi^+\pi^-$ tag are listed in Tables~\ref{table:Double_tag_yields_KSpipi}, \ref{table:Double_tag_yields_KSpipiPartReco} and \ref{table:Double_tag_yields_KLpipi}, respectively.

\begin{table*}[htb!]
    \centering
    \caption{DT yields of the $D \to K_S^0\pi^+\pi^-$ tag. The uncertainties are statistical only.}
    \label{table:Double_tag_yields_KSpipi}
    {\renewcommand{\arraystretch}{1.5}
    \begin{tabular}{ccccccccc}
        \hline
        Tag bin $\rightarrow$ & $1$ & $2$ & $3$ & $4$ & $5$ & $6$ & $7$ & $8$ \\
        Signal bin $\downarrow$    &     &     &     &     &     &     &     & \\
        \hline
        $-4$ & $9.6^{+3.7}_{-3.1}$       & $9.2^{+3.6}_{-2.9}$       & $5.8^{+2.7}_{-2.0}$       & $1.6^{+1.8}_{-1.1}$       & $3.7^{+2.4}_{-1.7}$       & $1.0^{+1.3}_{-0.7}$       & $0.5^{+1.4}_{-0.5}$       & $7.2^{+3.2}_{-2.5}$       \\
        $-3$ & $7.6^{+3.5}_{-2.9}$       & $8.1^{+3.3}_{-2.7}$       & $6.1^{+3.3}_{-2.6}$       & $8.6^{+3.4}_{-2.7}$       & $21.0^{+5.2}_{-4.6}$      & $9.0^{+3.4}_{-2.7}$       & $4.9^{+2.6}_{-1.9}$       & $6.7^{+3.1}_{-2.4}$       \\
        $-2$ & $9.6^{+3.8}_{-3.1}$       & $2.0^{+2.1}_{-1.4}$       & $6.4^{+3.4}_{-2.7}$       & $11.9^{+3.8}_{-3.1}$      & $19.0^{+5.1}_{-4.4}$      & $4.8^{+2.9}_{-2.2}$       & $2.4^{+2.2}_{-1.5}$       & $7.1^{+3.2}_{-2.5}$       \\
        $-1$ & $8.0^{+3.3}_{-2.6}$       & $3.9^{+2.3}_{-1.7}$       & $2.0^{+1.9}_{-1.2}$       & $1.0^{+1.3}_{-0.7}$       & $3.9^{+2.3}_{-1.7}$       & $4.3^{+2.6}_{-1.9}$       & $4.8^{+2.9}_{-2.2}$       & $4.3^{+2.8}_{-2.1}$       \\
        $\phantom{-}1$  & $17.1^{+4.7}_{-4.1}$      & $10.8^{+3.6}_{-2.9}$      & $6.4^{+3.1}_{-2.4}$       & $2.9^{+2.0}_{-1.4}$       & $7.8^{+3.1}_{-2.4}$       & $4.0^{+2.8}_{-2.0}$       & $7.6^{+3.3}_{-2.6}$       & $4.9^{+3.0}_{-2.2}$       \\
        $\phantom{-}2$  & $15.9^{+4.8}_{-4.1}$      & $5.7^{+2.9}_{-2.2}$       & $13.9^{+4.2}_{-3.6}$      & $9.6^{+3.6}_{-2.9}$       & $20.7^{+4.9}_{-4.3}$      & $7.0^{+3.0}_{-2.3}$       & $12.3^{+4.2}_{-3.5}$      & $22.5^{+5.3}_{-4.6}$      \\
        $\phantom{-}3$  & $21.5^{+5.2}_{-4.5}$      & $10.3^{+3.9}_{-3.2}$      & $10.6^{+3.7}_{-3.0}$      & $4.0^{+2.3}_{-1.7}$       & $31.0^{+6.0}_{-5.3}$      & $16.9^{+4.4}_{-3.8}$      & $19.9^{+4.8}_{-4.1}$      & $21.1^{+5.1}_{-4.4}$      \\
        $\phantom{-}4$  & $17.7^{+4.8}_{-4.1}$      & $6.9^{+2.9}_{-2.3}$       & $7.8^{+3.1}_{-2.5}$       & $0.7^{+1.3}_{-0.7}$       & $2.6^{+2.0}_{-1.4}$       & $0.7^{+1.4}_{-0.7}$       & $5.8^{+2.8}_{-2.1}$       & $13.7^{+4.0}_{-3.3}$      \\
        \hline
    \end{tabular}}
\end{table*}

\begin{table*}[htb!]
    \centering
    \caption{DT yields of the $D \to K_S^0\pi^+\pi^-$ tag with partially reconstructed $D \to K^+K^-\pi^+\pi^-$. The uncertainties are statistical only.}
    \label{table:Double_tag_yields_KSpipiPartReco}
    {\renewcommand{\arraystretch}{1.5}
    \begin{tabular}{ccccccccc}
        \hline
        Tag bin $\rightarrow$ & $1$ & $2$ & $3$ & $4$ & $5$ & $6$ & $7$ & $8$ \\
        Signal bin $\downarrow$    &     &     &     &     &     &     &     & \\
        \hline
        $-4$ & $5.3^{+4.0}_{-3.2}$       & $7.7^{+4.1}_{-3.3}$       & $3.0^{+3.1}_{-2.4}$       & $0.0^{+1.4}_{-0.0}$       & $0.0^{+1.3}_{-0.0}$       & $3.2^{+2.5}_{-1.8}$       & $0.0^{+1.7}_{-0.0}$       & $6.8^{+3.5}_{-2.8}$       \\
        $-3$ & $11.3^{+5.7}_{-4.9}$      & $6.5^{+3.4}_{-2.7}$       & $6.5^{+3.9}_{-3.2}$       & $11.1^{+4.2}_{-3.5}$      & $39.0^{+7.8}_{-7.0}$      & $7.2^{+3.9}_{-3.1}$       & $0.3^{+3.2}_{-0.3}$       & $4.9^{+4.1}_{-3.3}$       \\
        $-2$ & $10.6^{+5.2}_{-4.5}$      & $2.4^{+3.1}_{-2.3}$       & $2.6^{+3.2}_{-2.3}$       & $3.6^{+2.6}_{-1.9}$       & $23.5^{+6.2}_{-5.5}$      & $1.9^{+3.1}_{-1.9}$       & $17.8^{+5.6}_{-4.8}$      & $5.6^{+4.8}_{-3.9}$       \\
        $-1$ & $7.4^{+4.8}_{-4.0}$       & $5.0^{+3.4}_{-2.6}$       & $0.0^{+1.3}_{-0.0}$       & $0.0^{+0.5}_{-0.0}$       & $0.6^{+2.2}_{-0.6}$       & $0.0^{+0.9}_{-0.0}$       & $6.7^{+3.4}_{-2.7}$       & $5.9^{+3.7}_{-3.1}$       \\
        $\phantom{-}1$  & $15.1^{+5.9}_{-5.1}$      & $8.2^{+4.1}_{-3.3}$       & $2.5^{+2.7}_{-2.0}$       & $2.3^{+2.2}_{-1.5}$       & $3.4^{+3.1}_{-2.3}$       & $2.7^{+3.0}_{-2.3}$       & $12.6^{+4.5}_{-3.8}$      & $8.2^{+4.1}_{-3.3}$       \\
        $\phantom{-}2$  & $4.8^{+4.7}_{-3.9}$       & $8.2^{+4.1}_{-3.3}$       & $6.5^{+4.3}_{-3.5}$       & $6.9^{+3.7}_{-2.9}$       & $20.8^{+5.9}_{-5.2}$      & $8.6^{+3.9}_{-3.2}$       & $10.6^{+5.0}_{-4.2}$      & $8.6^{+4.9}_{-4.1}$       \\
        $\phantom{-}3$  & $9.8^{+5.3}_{-4.5}$       & $17.4^{+5.8}_{-5.0}$      & $3.0^{+3.5}_{-2.6}$       & $5.7^{+3.6}_{-2.8}$       & $34.8^{+7.3}_{-6.6}$      & $12.8^{+4.6}_{-3.9}$      & $18.9^{+6.0}_{-5.2}$      & $21.2^{+6.0}_{-5.2}$      \\
        $\phantom{-}4$  & $10.6^{+5.2}_{-4.4}$      & $1.3^{+2.3}_{-1.3}$       & $2.1^{+2.3}_{-1.6}$       & $0.0^{+0.6}_{-0.0}$       & $0.0^{+1.3}_{-0.0}$       & $1.0^{+1.8}_{-1.0}$       & $4.1^{+2.8}_{-2.1}$       & $8.3^{+4.0}_{-3.2}$       \\
        \hline
    \end{tabular}}
\end{table*}

\begin{table*}[htb!]
    \centering
    \caption{DT yields of the $D \to K_L^0\pi^+\pi^-$ tag. The uncertainties are statistical only.}
    \label{table:Double_tag_yields_KLpipi}
    {\renewcommand{\arraystretch}{1.5}
    \begin{tabular}{ccccccccc}
        \hline
        Tag bin $\rightarrow$ & $1$ & $2$ & $3$ & $4$ & $5$ & $6$ & $7$ & $8$ \\
        Signal bin $\downarrow$    &     &     &     &     &     &     &     & \\
        \hline
        $-4$ & $10.6^{+5.0}_{-4.3}$      & $4.5^{+2.8}_{-2.1}$       & $0.2^{+2.6}_{-0.2}$       & $0.8^{+2.6}_{-0.8}$       & $4.5^{+4.3}_{-3.4}$       & $0.0^{+1.6}_{-0.0}$       & $6.7^{+3.4}_{-2.8}$       & $9.2^{+4.2}_{-3.6}$       \\
        $-3$ & $62.0^{+10.7}_{-9.9}$     & $23.9^{+5.9}_{-5.2}$      & $9.7^{+4.4}_{-3.7}$       & $1.8^{+2.1}_{-1.6}$       & $0.0^{+2.5}_{-0.0}$       & $2.1^{+3.0}_{-2.1}$       & $8.3^{+4.5}_{-3.7}$       & $27.3^{+6.4}_{-5.7}$      \\
        $-2$ & $48.8^{+9.1}_{-8.3}$      & $12.7^{+5.6}_{-4.8}$      & $3.4^{+3.5}_{-2.7}$       & $2.5^{+2.6}_{-2.0}$       & $6.5^{+3.7}_{-3.1}$       & $1.9^{+2.9}_{-1.9}$       & $11.7^{+4.7}_{-4.0}$      & $18.7^{+5.5}_{-4.8}$      \\
        $-1$ & $18.9^{+5.6}_{-5.0}$      & $11.9^{+4.4}_{-3.7}$      & $4.9^{+3.5}_{-2.8}$       & $4.6^{+3.6}_{-2.9}$       & $17.0^{+5.7}_{-4.9}$      & $3.1^{+2.9}_{-2.2}$       & $5.8^{+3.2}_{-2.6}$       & $4.7^{+3.7}_{-3.1}$       \\
        $\phantom{-}1$  & $25.9^{+6.5}_{-5.8}$      & $15.1^{+4.6}_{-4.0}$      & $8.5^{+3.7}_{-3.0}$       & $0.7^{+2.2}_{-0.7}$       & $10.4^{+4.8}_{-4.0}$      & $6.0^{+3.9}_{-3.2}$       & $16.3^{+5.0}_{-4.3}$      & $8.6^{+4.3}_{-3.6}$       \\
        $\phantom{-}2$  & $72.2^{+10.0}_{-9.4}$     & $18.3^{+5.6}_{-4.9}$      & $16.4^{+5.2}_{-4.5}$      & $0.0^{+1.7}_{-0.0}$       & $15.9^{+5.1}_{-4.4}$      & $14.7^{+4.9}_{-4.2}$      & $36.6^{+7.2}_{-6.5}$      & $58.7^{+9.2}_{-8.5}$      \\
        $\phantom{-}3$  & $94.1^{+11.5}_{-10.8}$    & $32.2^{+6.9}_{-6.2}$      & $18.3^{+5.9}_{-5.2}$      & $2.2^{+2.7}_{-2.1}$       & $6.7^{+4.0}_{-3.3}$       & $9.0^{+4.0}_{-3.4}$       & $42.4^{+7.6}_{-6.9}$      & $43.2^{+7.9}_{-7.3}$      \\
        $\phantom{-}4$  & $23.4^{+5.8}_{-5.1}$      & $6.2^{+3.8}_{-3.1}$       & $10.3^{+4.1}_{-3.4}$      & $3.0^{+2.7}_{-2.0}$       & $5.6^{+3.9}_{-3.1}$       & $5.2^{+3.2}_{-2.5}$       & $11.7^{+4.1}_{-3.4}$      & $12.4^{+4.5}_{-3.9}$      \\
        \hline
    \end{tabular}}
\end{table*}

\section{Binned strong-phase measurement}
\label{section:Phase_space_binned_strong_phase_measurement}
\noindent The yields presented in Sect.~\ref{section:Single_and_double_tag_yield_determination} can be analysed simultaneously to extract knowledge of the strong-phase differences in each bin of phase space for the $D\to K^+K^-\pi^+\pi^-$ decay, using a maximum likelihood fit based on  Eqs.~\eqref{equation:Yield_equation_flavor}-\eqref{equation:Yield_equation_SCMB}. This is referred to as the strong-phase fit.

There are in total eight $c_i$ and $s_i$ parameters of interest. Second, there are eight $K_i$ parameters, but due to the $\sum_i K_i = 1$ constraint, only seven of these are independent. To accommodate this constraint, a recursive fraction parameterisation $R_i$ is defined~\cite{LHCb-PAPER-2020-019},

\begin{equation*}
    R_i = 
    \begin{cases}
        K_i, & i = -4 \\
        K_i/\sum_{j \geq i}K_j, & -4 < i < +4,
    \end{cases}
\end{equation*}
such that $R_4 = 1$ by construction.

Furthermore, there are benefits to allowing the branching fraction of the signal mode $\mathcal{B}$ to be a free parameter in the fit as well, since the systematic uncertainties due to tracking and PID efficiencies may be absorbed into this parameter. These systematic uncertainties arise due to potentially imperfect modelling of such effects, leading to biases in the efficiencies obtained from simulation. However, in the fit strategy described, the strong-phase parameters are free of these systematic biases.

In addition, for the $D\to K_L^0\pi^+\pi^-$ tag, which is a partially reconstructed tag, the corresponding ST yield must be calculated from its branching fraction. Since this has not been measured to date, a separate and arbitrary normalisation is used for this tag.

The values of the hadronic parameters $r_D$, $R$ and $\delta_D$ for the flavor tags $D^0 \to K^-\pi^+\pi^0$ and $D^0 \to K^-\pi^+\pi^-\pi^+$, which appear in Eq.~\eqref{equation:Yield_equation_flavor}, are taken from the combination of BESIII, CLEO-c and LHCb results reported in Ref.~\cite{cite:K3piStrongPhase}. In the case of $D^0 \to K^-\pi^+$, for which $R=1$, the parameters $r^{K\pi}_D\cos(\delta^{K\pi}_D)$ and $r^{K\pi}_D\sin(\delta^{K\pi}_D)$ are free parameters in the fit, and they are Gaussian constrained to the results from Ref.~\cite{cite:deltaKpi} using the $D \to K_{S, L}^0\pi^+\pi^-$ tags.  This approach brings negligible degradation in the sensitivity to the $D \to K^-K^+\pi^+\pi^-$ strong-phase parameters and allows the analysis to return information on the $D^0 \to K^-\pi^+$ decay. In total, therefore, there are 19 fit parameters, which are denoted $p_\alpha$, with $\alpha = 1, 2, ..., 19$.

The efficiency matrices in Eqs.~\eqref{equation:Yield_equation_flavor}-\eqref{equation:Yield_equation_SCMB} are taken from simulation. The diagonal elements of a given efficiency matrix, which are interpreted as bin efficiencies, are similar but also have variations between bins of about $1\%$-$2\%$. The off-diagonal elements, which describe bin migration, are all less than $15\%$.

The values of $K_i$, $c_i$ and $s_i$ for the $D \to K_{S, L}^0\pi^+\pi^-$ modes are taken from the combination of BESIII and CLEO results in Ref.~\cite{cite:KSpipiStrongPhase}. The $C\!P$-even fraction $F_+$ of $D \to \pi^+\pi^-\pi^0$ is taken from Ref.~\cite{cite:pipipi0_BESIII} and that for $D \to K^0_S \pi^+\pi^-\pi^0$ from Ref.~\cite{cite:KSpipipi0_BESIII}. 

If it is assumed that DT yields have uncertainties that have a Gaussian distribution, the likelihood is defined as
\begin{equation}
    \mathcal{L} = \sum_{i, j}(V^{-1})_{ij}\big(N_i - \hat{N_i}(p_\alpha)\big)\big(N_j - \hat{N_j}(p_\alpha)\big),
    \label{equation:Likelihood}
\end{equation}
where $V_{ij} = \rho_{ij}\sigma_i\sigma_j$ is the covariance matrix of the DT yields, defined in terms of the correlation matrix $\rho_{ij}$ and the DT yield uncertainties $\sigma_i$. The indices $i$ and $j$ run over all binned DT yields. $N_i$ are the measured DT yields, while $\hat{N}_i$ denotes the predicted DT yield for a given set of $19$ parameters $p_\alpha$, which are varied in the fit.

However, for the flavor and $C\!P$ tags, the DT yields from Tables~\ref{table:Double_tag_yields_flavor} and \ref{table:Double_tag_yields_CP} have non-negligible differences between positive and negative uncertainties, which must be accounted for. A procedure from Ref.~\cite{cite:Barlow} is applied, where the covariance matrix $V_{ij}$ is linearly extrapolated from the asymmetric uncertainties,
\begin{linenomath}
    \begin{align}
        V_{ij} =& \rho_{ij}\sigma_i(N_i - \hat{N_i})\sigma_j(N_j - \hat{N_j}), \nonumber \\
        \sigma_i(N_i - \hat{N_i}) =& \sqrt{\sigma_+\sigma_- + (N_i - \hat{N_i})(\sigma_- - \sigma_+)},\label{equation:Barlow_CovMatrix}
    \end{align}
\end{linenomath}
where $\sigma_\pm$ are the positive and negative uncertainties in Tables~\ref{table:Double_tag_yields_flavor}-\ref{table:Double_tag_yields_KLpipi}.

To cross check this procedure, pseudo experiments are performed where simulated data sets are generated and fitted using the nominal fit model. The strong-phase parameters used to generate the data sets are calculated from the amplitude model in Ref.~\cite{LHCb-PAPER-2018-041}. This exercise shows that using the expression in Eq.~\eqref{equation:Barlow_CovMatrix} in the likelihood function reduces fit biases significantly. Unfortunately, in the limiting case where $\sigma_-$ is very small, Eq.~\eqref{equation:Barlow_CovMatrix} can be undefined due to a negative number under the square root. This occurs for several bins in Tables~\ref{table:Double_tag_yields_KSpipi}-\ref{table:Double_tag_yields_KLpipi} and it causes a significant fraction of the pseudo-experiment fits to fail.

Therefore, for the $D \to K_{S, L}^0\pi^+\pi^-$ tag modes, the full likelihood function, from which the values in  Tables~\ref{table:Double_tag_yields_KSpipi}-\ref{table:Double_tag_yields_KLpipi} are obtained, is used directly in the strong-phase fit. This approach reduces the number of failed pseudo experiments to a negligible level, and the resulting fit biases are also much smaller. For the $c_i$, $s_i$ and $\mathcal{B}$ parameters, the fit biases are around $1\%$-$15\%$ of the statistical uncertainty and are corrected for after the strong-phase fit.

\begin{table}[htb]
    \centering
    \caption{Results from the strong-phase fit, after bias corrections. The first uncertainty is statistical and the second systematic.}
    \label{table:Strong_phase_fit_results_after_bias_corrections}
    \begin{tabular}{cc}
        \hline
        Variable                            & Fit result \\
        \hline
        $c_1$                               & $-0.22 \pm 0.08 \pm 0.01$ \\
        $c_2$                               & $0.79 \pm 0.04 \pm 0.01$ \\
        $c_3$                               & $0.862 \pm 0.029 \pm 0.008$ \\
        $c_4$                               & $-0.39 \pm 0.08 \pm 0.01$ \\
        $s_1$                               & $-0.47 \pm 0.22 \pm 0.04$ \\
        $s_2$                               & $-0.17 \pm 0.16 \pm 0.04$ \\
        $s_3$                               & $0.26 \pm 0.14 \pm 0.02$ \\
        $s_4$                               & $0.52 \pm 0.24 \pm 0.04$ \\
        $R_{-4}$                            & $0.0871 \pm 0.0027 \pm 0.0038$ \\
        $R_{-3}$                            & $0.294 \pm 0.005 \pm 0.006$ \\
        $R_{-2}$                            & $0.400 \pm 0.007 \pm 0.003$ \\
        $R_{-1}$                            & $0.252 \pm 0.007 \pm 0.012$ \\
        $R_1$                               & $0.119 \pm 0.006 \pm 0.006$ \\
        $R_2$                               & $0.405 \pm 0.012 \pm 0.002$ \\
        $R_3$                               & $0.818 \pm 0.010 \pm 0.010$ \\
        ${\mathcal{B}}(KK\pi\pi)$                & $(2.863 \pm 0.028 \pm 0.045)\times 10^{-3}$ \\
        \hline
    \end{tabular}
\end{table}

\begin{figure*}[htb]
    \centering
    \includegraphics[width=0.75\textwidth]{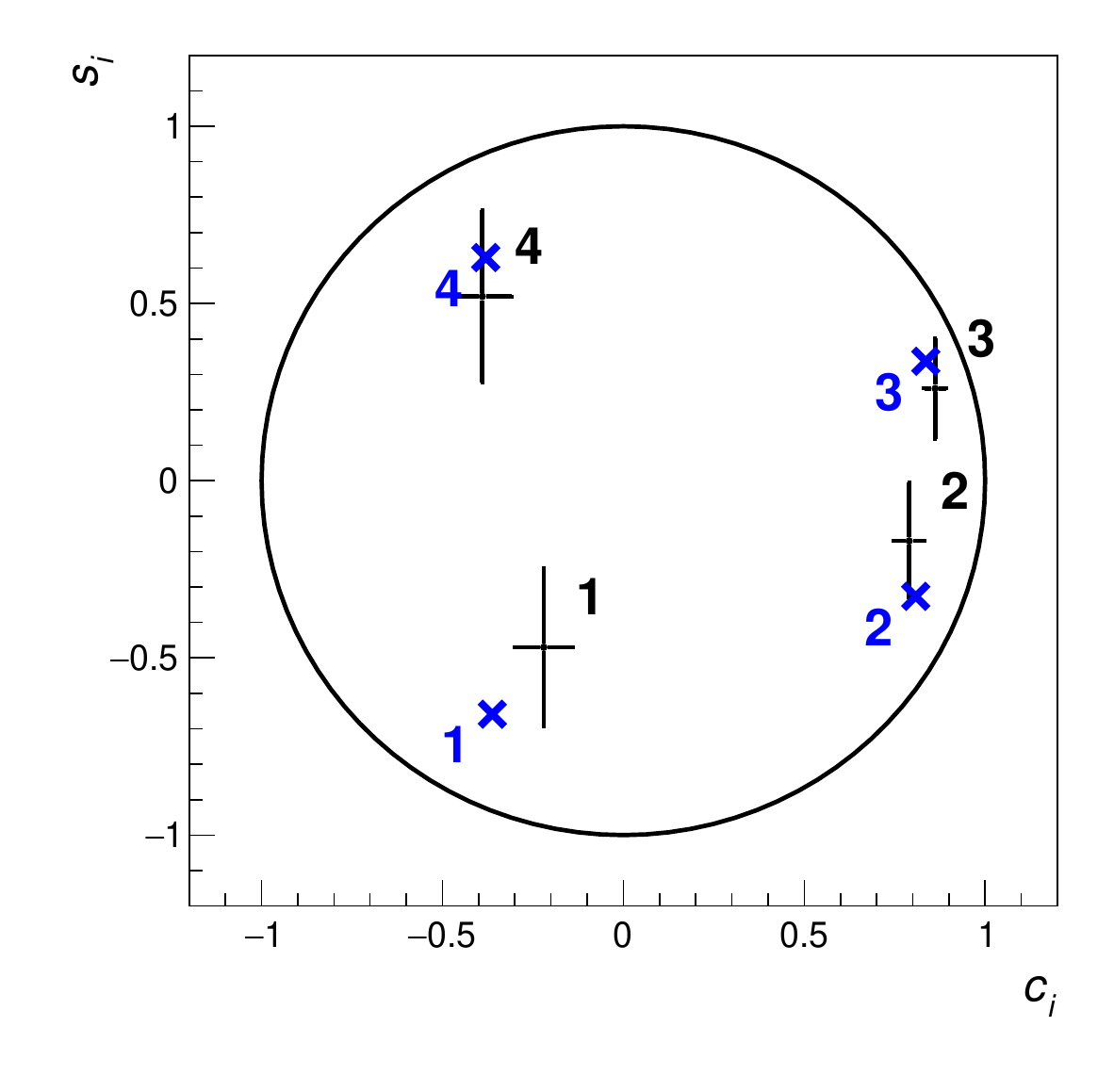}
    \caption{The fit results for $(c_i, s_i)$ in each phase-space bin, with error bars corresponding to $68\%$ confidence intervals. Only the statistical uncertainties are considered. The model predictions are shown as blue diagonal crosses, labelled with their bin numbers. The bin numbers are shown above and to the right of the measurements, and below and to the left of the model predictions.}
    \label{figure:StrongPhase_fit_cisi}
\end{figure*}

The complete results of the strong-phase fit are shown in Table~\ref{table:Strong_phase_fit_results_after_bias_corrections}. The assignment of the systematic uncertainties, which are given here, is discussed in Sect.~\ref{sec:syst}. The fit results of $(c_i, s_i)$, corresponding to a $68\%$ confidence interval, are plotted in Fig.~\ref{figure:StrongPhase_fit_cisi} for each bin. The results for each bin are significantly displaced from the origin, indicating that the binning scheme from Ref.~\cite{LHCb-PAPER-2022-037} effectively identifies regions with similar strong-phase differences. The model predictions, using the model in Ref.~\cite{LHCb-PAPER-2018-041}, are also shown as crosses, and a good agreement is found between the model and the fitted parameters.

The Gaussian-constrained fit results are $r_D^{K\pi}\cos(\delta_D^{K\pi}) = -0.0501 \pm 0.0084 \pm 0.0005$ and $r_D^{K\pi}\sin(\delta_D^{K\pi}) = -0.0091 \pm 0.0136 \pm 0.0006$, with a correlation of $0.02$. This may be compared with the external constraint from Ref.~\cite{cite:deltaKpi}, $r_D^{K\pi}\cos(\delta_D^{K\pi}) = -0.0562 \pm 0.0096$ and $r_D^{K\pi}\sin(\delta_D^{K\pi}) = -0.011 \pm 0.014$. A small improvement is seen.

\begin{figure*}[htb]
    \centering
    \includegraphics[width=0.75\textwidth]{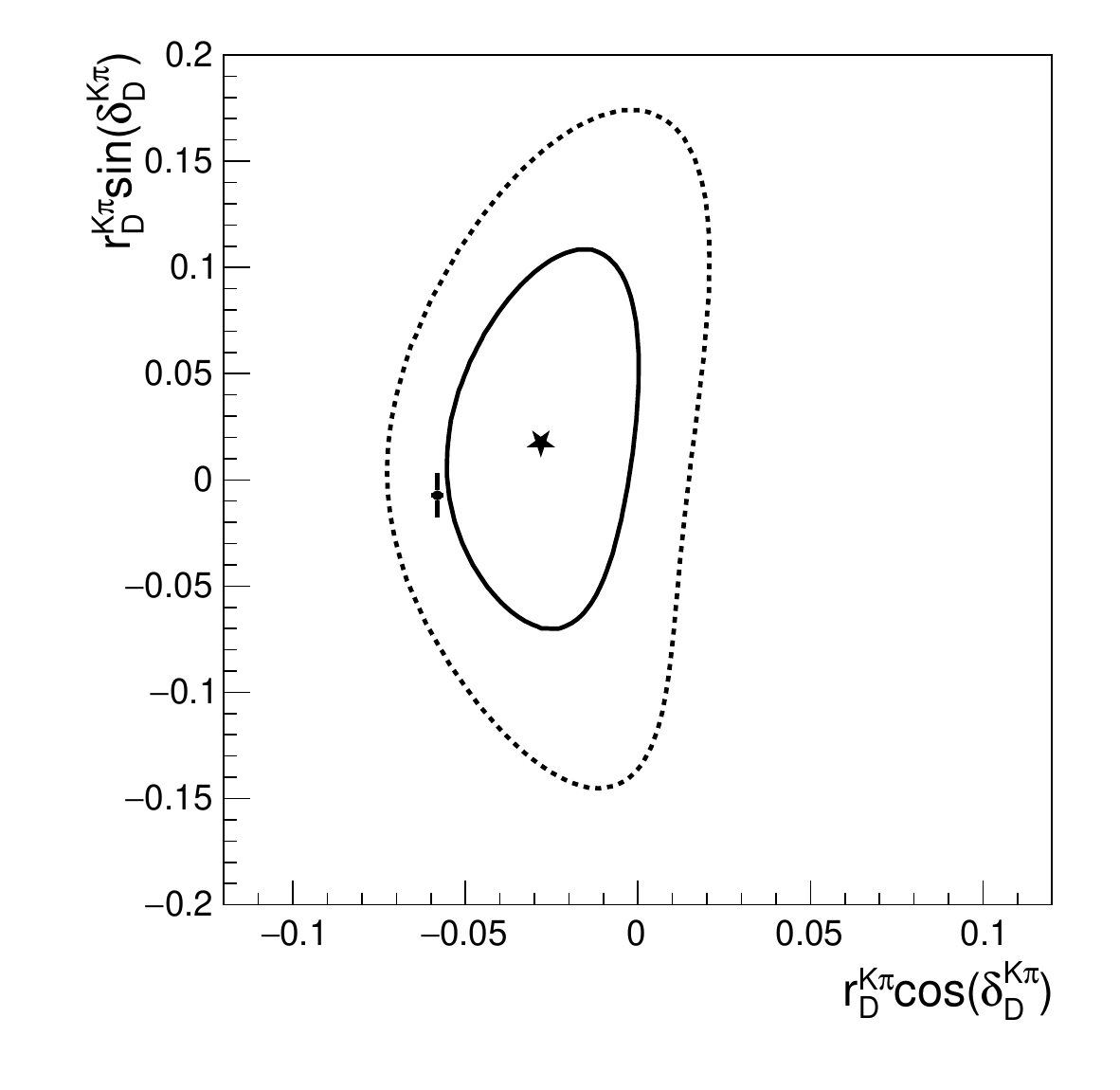}
    \caption{The (solid) $\Delta\chi^2 = 2.30$ and (dashed) $\Delta\chi^2 = 6.18$ contours of $(r^{K\pi}_D\cos(\delta^{K\pi}_D), r^{K\pi}_D\sin(\delta^{K\pi}_D))$. The data point with error bars is the current average from HFLAV~\cite{cite:HFLAV2021} and the black asterisk is the fitted value.}
    \label{figure:StrongPhase_fit_DeltaKpi}
\end{figure*}

An alternative fit where $(r^{K\pi}_D\cos(\delta^{K\pi}_D))$ and $(r^{K\pi}_D\sin(\delta^{K\pi}_D))$ are allowed to vary freely is performed, and their $\Delta\chi^2 = 2.30$ and $\Delta\chi^2 = 6.18$ contours are shown in Fig.~\ref{figure:StrongPhase_fit_DeltaKpi}. Under conditions in which Wilks' theorem applies~\cite{cite:Wilks}, these contours correspond to $68.3\%$ and $95.4\%$ confidence regions. However, due to insufficient sensitivity, this is found to not be the case, and a study with pseudo experiments shows that these contours underestimate the confidence levels. The polar angle of the fitted result in Fig.~\ref{figure:StrongPhase_fit_DeltaKpi} should be compared with the current world average from HFLAV, allowing for $C\!P$ violation, which is taken from Section 10.1 in Ref.~\cite{cite:HFLAV2021}. The results are found to be consistent.

It is interesting to also compare the fitted values of $R_i$, which are related to the fractional bin yields $K_i$. The amplitude model~\cite{LHCb-PAPER-2018-041} is expected to describe these parameters well, and from comparison in Table~\ref{table:Ri_comparison} excellent agreement is found.

\begin{table}[htb]
    \centering
    \caption{Comparison of $R_i$ between fit results and model predictions~\cite{LHCb-PAPER-2018-041}.}
    \label{table:Ri_comparison}
    \begin{tabular}{ccc}
        \hline
        Variable   & Fit result        & Model prediction \\
        \hline
        $R_{-4}$   & $0.0876 \pm 0.0027$ & $0.086$    \\
        $R_{-3}$   & $0.294 \pm 0.005$ & $0.297$    \\
        $R_{-2}$   & $0.399 \pm 0.007$ & $0.398$    \\
        $R_{-1}$   & $0.252 \pm 0.007$ & $0.267$    \\
        $R_1$      & $0.119 \pm 0.006$ & $0.110$    \\
        $R_2$      & $0.406 \pm 0.012$ & $0.401$    \\
        $R_3$      & $0.815 \pm 0.010$ & $0.833$    \\
        \hline
    \end{tabular}
\end{table}

From Eq.~\eqref{equation:ci}, it can be shown that $2F_+ - 1$ is the amplitude-averaged cosine of the strong-phase difference, integrated over the whole phase space. From the fit results, a weighted average of $c_i$, using $\sqrt{K_i\bar{K_i}}$ as weights, is performed and the $C\!P$-even fraction is found to be $F_+ = 0.754 \pm 0.010 \pm 0.008$.

The fitted value of the $D^0\to K^+K^-\pi^+\pi^-$ branching fraction ${\mathcal{B}}(D^0\to K^+K^-\pi^+\pi^-) = (2.863 \pm 0.028 \pm 0.045)\times10^{-3}$, more than 3$\,\sigma$ higher than the value reported in the PDG, $\mathcal{B} = (2.47 \pm 0.11)\times10^{-3}$~\cite{pdg}, and is significantly more precise. The result derives from an absolute measurement, in contrast to previous studies, in which the branching fraction was determined relative to that of other decays.

\section{Systematic uncertainties}
\label{sec:syst}
\noindent Systematic uncertainties are assigned on the measurements of the strong-phase parameters and on the branching fraction. Most sources of systematic uncertainties are due to uncertainties on external parameters and parameters taken from simulation, which are fixed in the strong-phase fit. Unless specified otherwise, the uncertainties are propagated to the strong-phase measurements by repeating the strong-phase fit, each time smearing the parameters according to their uncertainties and correlation matrices. The resulting sample covariance of the $19$ free parameters of the strong-phase fit are assigned as the systematic uncertainties.

For the fully reconstructed tags, the uncertainties on the ST yields are the statistical uncertainties from the fits described in Sect.~\ref{section:Single_and_double_tag_yield_determination} and listed in Table~\ref{table:Single_tag_yields_efficiencies}. For the partially reconstructed tags, the uncertainties originate from the knowledge of the branching fractions, which are a few percent, and $N_{D\bar{D}}$  (the contribution from the knowledge of the effective ST efficiency is considered with those of the other efficiencies, discussed below). Therefore, the relative uncertainties on ST yield of $D \to K_L^0\pi^0$ and $D^0 \to K^-e^+\nu_e$ in Table~\ref{table:Single_tag_yields_efficiencies} are larger than those of fully reconstructed tags. The uncertainty of the ST yield of $D\to K_L^0\pi^+\pi^-$ is not considered since it only affects the arbitrary normalisation of this mode.

For the $D \to K_{S, L}^0\pi^+\pi^-$, $D \to \pi^+\pi^-\pi^0$ and $D \to K_S^0\pi^+\pi^-\pi^0$ tag modes, there are uncertainties associated with the external strong-phase parameters that are fixed in the strong-phase fit. These uncertainties are taken from Refs.~\cite{cite:KSpipiStrongPhase,cite:pipipi0_CPfraction,cite:KSpipipi0_BESIII}, respectively.

There are uncertainties associated with the efficiencies and efficiency matrices obtained from simulation due to the finite sample sizes. For each tag, the uncertainty on the ST efficiency is assumed to follow a binomial distribution, using the fraction of events that is reconstructed. For the efficiency matrices in DT yields, each matrix element is assumed to follow a binomial distribution.

In the fit of ST and DT yields, there are uncertainties associated with the size of the peaking backgrounds. These uncertainties arise from the knowledge of the branching fractions and from the finite sample sizes of the simulation used to obtain the relative efficiencies. In the DT fits, there is also an additional uncertainty on the correction factors that account for quantum correlations when estimating the peaking-background size. This uncertainty comes from the knowledge of the $C\!P$ content of the signal and tag modes, as well as the $C\!P$ content of the background decays that contaminate the DT selection. The strong-phase information of these decays is obtained from previous studies performed with quantum-correlated $D\bar{D}$ decays~\cite{cite:pipipi0_BESIII,cite:KSpipiStrongPhase,cite:cisi4pi}. To propagate these uncertainties to the strong-phase measurement, the fits in Sects.~\ref{section:Single_and_double_tag_yield_determination} and \ref{section:Phase_space_binned_strong_phase_measurement} are repeated many times. In each iteration, the peaking backgrounds are smeared according to their uncertainties, and the resulting spread and correlation between parameters of the strong-phase fit is taken as the systematic uncertainty in the ST and DT fit.

To assess the systematic uncertainty due to the $K_S^0$ veto in the $D\to K^+K^-\pi^+\pi^-$ selection, which may affect the strong-phase differences, the amplitude model~\cite{LHCb-PAPER-2018-041} is used to calculate the strong-phase parameters with and without the veto. The difference between the results is assigned as the systematic uncertainty.

Finally, the systematic uncertainty due to tracking and PID efficiencies must be assigned to the branching fraction $\mathcal{B}$. The discrepancy in tracking and PID efficiencies between data and simulation for charged kaons and pions are studied in bins of momentum using calibration samples. The systematic uncertainty is evaluated with a weighted sum over momentum bins, using the momentum distribution of the charged kaons and pions from the $D\to K^+K^-\pi^+\pi^-$ decay. For each charged kaon (pion) track, a systematic uncertainty of $0.4\%$ ($0.5\%$) is assigned for the tracking efficiency, and a systematic uncertainty of $0.4\%$ ($0.1\%$) is assigned for the PID efficiency. The uncertainties of the two kaons, or the two pions, are assumed to be fully correlated and can be added together. The uncertainties due to tracking and PID efficiencies are assumed to be uncorrelated, and it is also assumed that the uncertainties for the kaons and pions are also uncorrelated, as different control samples are used in the determination of these efficiencies. Therefore, these contributions can be added in quadrature, and the combined systematic uncertainty for the tracking and PID efficiencies of $1.5\%$ is assigned. It is important to note that the strong-phase parameters are not affected by systematic uncertainties due to tracking or PID efficiencies since these mainly cancel in the ratio of ST and DT efficiencies in Eqs.~\eqref{equation:Yield_equation_flavor}-\eqref{equation:Yield_equation_SCMB}, and any residual uncertainty is considered to be negligible.

The uncertainties of $c_i$, $s_i$ and $\mathcal{B}$ are summarised in Table~\ref{table:Systematic_uncertainties}, where they are summed in quadrature. In the case of the strong-phase parameters, the total systematic uncertainties are all much smaller than the statistical uncertainties. Similarly, the uncertainties of $R_i$ are shown in Table~\ref{table:Systematic_uncertainties_Ri}.

\begin{table*}[htb]
    \centering
    \caption{Summary of systematic uncertainties on $c_i$, $s_i$ and $\mathcal{B}$. Each entry is multiplied by $10^3$, with the exception of $\mathcal{B}$, which is multiplied by $10^5$.}
    \label{table:Systematic_uncertainties}
    \begin{tabular}{cccccccccc}
        \hline
        & $\quad\mathcal{B}\quad$ & $\quad c_1\quad$ & $\quad c_2\quad$ & $\quad c_3\quad$ & $\quad c_4\quad$ & $\quad s_1\quad$ & $\quad s_2\quad$ & $\quad s_3\quad$ & $\quad s_4\quad$ \\
        \hline
        ST yield                    & $0.1$ & $0.6$ & $0.3$ & $0.2$ & $0.6$ & $0.0$ & $0.1$ & $0.0$ & $0.1$ \\
        $K_L^0\pi^0$ ST yield       & $0.1$ & $3.7$ & $3.4$ & $3.4$ & $3.9$ & $0.2$ & $0.0$ & $0.2$ & $0.4$ \\
        $K^-e^+\nu_e$ ST yield      & $0.6$ & $0.5$ & $0.2$ & $0.2$ & $0.5$ & $0.0$ & $0.0$ & $0.0$ & $0.1$ \\
        External strong phases      & $0.2$ & $3.9$ & $3.1$ & $2.4$ & $4.5$ & $39.0$ & $38.9$ & $21.5$ & $34.6$ \\
        Finite MC size              & $0.4$ & $2.9$ & $1.5$ & $1.2$ & $3.2$ & $6.0$ & $4.5$ & $4.1$ & $6.8$ \\
        ST and DT fit               & $0.3$ & $2.3$ & $3.6$ & $4.7$ & $1.9$ & $1.2$ & $1.5$ & $2.5$ & $1.8$ \\
        $K_S^0$ veto                & $0.0$ & $0.2$ & $4.3$ & $1.9$ & $5.8$ & $0.4$ & $6.2$ & $1.7$ & $2.8$ \\
        Tracking and PID efficiency & $4.4$ & $0.0$ & $0.0$ & $0.0$ & $0.0$ & $0.0$ & $0.0$ & $0.0$ & $0.0$ \\
        \hline
        Total systematic            & $4.5$ & $6.6$ & $7.4$ & $6.7$ & $9.1$ & $39.5$ & $39.7$ & $22.1$ & $35.4$ \\
        \hline
        Statistical                 & $2.8$ & $76.8$ & $36.9$ & $29.2$ & $82.1$ & $220.1$ & $157.6$ & $142.2$ & $238.3$ \\
        \hline
    \end{tabular}
\end{table*}

\begin{table*}[htb]
    \centering
    \caption{Summary of systematic uncertainties on $R_i$, $r^{K\pi}_D\cos(\delta^{K\pi}_D)$ and $r^{K\pi}_D\sin(\delta^{K\pi}_D)$. Each entry is multiplied by $10^3$.}
    \label{table:Systematic_uncertainties_Ri}
    \begin{tabular}{cccccccccc}
        \hline
        & $\quad R_{-4}\quad$ & $\quad R_{-3}\quad$ & $\quad R_{-2}\quad$ & $\quad R_{-1}\quad$ & $\quad R_1\quad$ & $\quad R_2\quad$ & $\quad R_3\quad$ & $r_D\cos(\delta_D)$ & $r_D\sin(\delta_D)$ \\
        \hline
        ST yield                    & $0.0$ & $0.0$ & $0.0$ & $0.0$ & $0.0$ & $0.0$ & $0.0$ & $0.1$ & $0.0$ \\
        $K_L^0\pi^0$ ST yield       & $0.0$ & $0.0$ & $0.1$ & $0.0$ & $0.0$ & $0.0$ & $0.1$ & $0.0$ & $0.0$ \\
        $K^- e^+\nu_e$ ST yield     & $0.0$ & $0.0$ & $0.1$ & $0.0$ & $0.0$ & $0.0$ & $0.1$ & $0.3$ & $0.0$ \\
        External strong phases      & $0.1$ & $0.1$ & $0.1$ & $0.2$ & $0.1$ & $0.2$ & $0.2$ & $0.1$ & $0.4$ \\
        Finite MC size              & $0.3$ & $0.6$ & $0.8$ & $0.8$ & $0.7$ & $1.4$ & $1.2$ & $0.4$ & $0.4$ \\
        ST and DT fit               & $0.1$ & $0.5$ & $0.7$ & $0.4$ & $0.3$ & $0.7$ & $0.6$ & $0.2$ & $0.1$ \\
        $K_S^0$ veto                & $3.8$ & $5.5$ & $3.0$ & $11.6$ & $6.3$ & $0.4$ & $9.8$ & $0.0$ & $0.0$ \\
        Tracking and PID efficiency & $0.0$ & $0.0$ & $0.0$ & $0.0$ & $0.0$ & $0.0$ & $0.0$ & $0.0$ & $0.0$ \\
        \hline
        Total systematic            & $3.8$ & $5.6$ & $3.2$ & $11.7$ & $6.4$ & $1.7$ & $9.9$ & $0.5$ & $0.6$ \\
        \hline
        Statistical                 & $2.7$ & $5.2$ & $6.9$ & $7.1$ & $6.3$ & $11.8$ & $10.5$ & $8.4$ & $13.6$ \\
        \hline
    \end{tabular}
\end{table*}

\section{Impact on the measurement of the CKM angle \texorpdfstring{\boldmath{$\gamma$}}{gamma}}
\noindent To assess the impact of the measured strong-phase parameters on the determination of $\gamma$, a study with pseudo experiments is performed. Initially, the bin yields and correlation matrix of the $B^\pm\to [K^+K^-\pi^+\pi^-]_Dh^\pm$ from Appendix A in Ref.~\cite{LHCb-PAPER-2022-037} are used to generate $1000$ simulated data sets, each one of a similar size to that used in the measurement of LHCb reported in Ref.~\cite{LHCb-PAPER-2022-037}. All the yields are assumed to have Gaussian-distributed uncertainties. These data sets are fitted and the mean statistical uncertainty is found to be $12.7^\circ$.

Since the systematic uncertainty of $\gamma$ from the strong-phase inputs may be of similar size to the statistical uncertainty, a suitable strategy for combining the current measurement with that in Ref.~\cite{LHCb-PAPER-2022-037} is to perform a simultaneous maximum-likelihood fit. To assess the combined uncertainty on $\gamma$, the pseudo experiments of the strong-phase fit described in Section ~\ref{section:Phase_space_binned_strong_phase_measurement} are combined with the pseudo experiments of the $B^\pm\to Dh^\pm$ bin yields.

The simultaneous fit of these combined pseudo experiments shows that the resulting strong-phase parameters and the value of $\gamma$ are consistent with those obtained from separate fits, but the uncertainty on $\gamma$ now includes both contributions from the statistical uncertainty of the $B^\pm\to Dh^\pm$ bin yields, and the uncertainties from the strong-phase parameters. The combined uncertainty is found to be $16.2^\circ$, which implies a systematic uncertainty of $10.1^\circ$ on $\gamma$ due to the strong-phase inputs.

\section{Summary and outlook}
\noindent A first measurement of the  strong-phase difference between $D^0$ and $\bar{D^0}\to K^+K^-\pi^+\pi^-$ decays has been performed in bins of phase space, using a binning scheme with $2\times4$ bins, optimised for the measurement of the angle $\gamma$ of the Unitarity Triangle. The strong-phase parameters $c_i$ and $s_i$ are similar to the model-predicted parameters used in Ref.~\cite{LHCb-PAPER-2022-037}, but this analysis is the first model-independent measurement of these parameters in phase-space bins. The results are
\begin{linenomath}
    \begin{align*}
        c_1 = -0.22\phantom{0} \pm 0.08\phantom{0} \pm 0.01\phantom{0}, \quad s_1 = -0.47 \pm 0.22 \pm 0.04, \\
        c_2 = \phantom{-}0.79\phantom{0} \pm 0.04\phantom{0} \pm 0.01\phantom{0}, \quad s_2 = -0.17 \pm 0.16 \pm 0.04, \\
        c_3 = \phantom{-}0.862 \pm 0.029 \pm 0.008, \quad s_3 = \phantom{-}0.26 \pm 0.14 \pm 0.02, \\
        c_4 = -0.39\phantom{0} \pm 0.08\phantom{0} \pm 0.01\phantom{0}, \quad s_4 = \phantom{-}0.52 \pm 0.24 \pm 0.04, 
    \end{align*}
\end{linenomath}
where the first uncertainty is statistical and the second is systematic. The associated correlation matrices are listed in Appendix~\ref{appendix:Correlation_matrices}. In all cases $c_i^2 + s_i^2$ is significantly different from zero, indicating that the binning scheme from Ref.~\cite{LHCb-PAPER-2022-037} is effective in identifying regions with similar strong-phase differences. These results will be valuable in enabling a model-independent measurement of $\gamma$ in $B^\pm \to DK^\pm$, $D^0\to K^+K^-\pi^+\pi^-$ decays at LHCb and Belle~II, and in studies of mixing and $C\!P$ violation in the $D^0 \bar{D{}^0}$ system. The expected uncertainty from the strong-phase parameters in such a measurement is around  $10^\circ$, for a $B$-meson sample of the size accumulated by LHCb in Runs 1 and 2 of the LHC. In addition, this analysis demonstrates that the strong-phase fit can be extended to give sensitivity to the $D \to K^-\pi^+$ parameters $r^{K\pi}_D$ and $\delta^{K\pi}_D$.

From the measured values of $c_i$ and $R_i$, the $C\!P$-even fraction is determined to be $F_+ = 0.754 \pm 0.010 \pm 0.008$, which is more precise than the result reported in Ref.~\cite{cite:KKpipi_FPlus} due to the larger data set and the inclusion of additional tag modes. 

Finally, the branching fraction of $D^0\to K^+K^-\pi^+\pi^-$ has been measured to be 
\begin{equation*}
    {\mathcal{B}}(D^0\to K^+K^-\pi^+\pi^-) = (2.863 \pm 0.028 \pm 0.045)\times10^{-3}, 
\end{equation*}
where the first uncertainty is statistical and the second is systematic.  This result more than 3$\sigma$ higher than the average of previous measurements~\cite{pdg}, and is around a factor of two more precise.

\appendix
\section{Correlation matrices}
\label{appendix:Correlation_matrices}
Tables~\ref{table:Statistical_correlations} and \ref{table:Systematic_correlations} contain the correlations associated with the $c_i$ and $s_i$ parameters.

\begin{table*}[htb]
    \centering
    \caption{The correlation matrix associated with the statistical uncertainties.}
    \label{table:Statistical_correlations}
    \begin{tabular}{cccccccc}
        \hline
        $\quad c_1\quad$ & $\quad c_2\quad$ & $\quad c_3\quad$ & $\quad c_4\quad$ & $\quad s_1\quad$ & $\quad s_2\quad$ & $\quad s_3\quad$ & $\quad s_4\quad$ \\
        \hline
        $\phantom{-}1.00$ & $-0.04$ & $\phantom{-}0.01$ & $-0.10$ & $-0.01$ & $\phantom{-}0.00$ & $\phantom{-}0.00$ & $\phantom{-}0.00$ \\
        $-0.04$ & $\phantom{-}1.00$ & $-0.08$ & $\phantom{-}0.01$ & $\phantom{-}0.00$ & $-0.01$ & $\phantom{-}0.00$ & $\phantom{-}0.00$ \\
        $\phantom{-}0.01$ & $-0.08$ & $\phantom{-}1.00$ & $-0.03$ & $\phantom{-}0.00$ & $\phantom{-}0.00$ & $\phantom{-}0.00$ & $\phantom{-}0.00$ \\
        $-0.10$ & $\phantom{-}0.01$ & $-0.03$ & $\phantom{-}1.00$ & $\phantom{-}0.00$ & $\phantom{-}0.00$ & $\phantom{-}0.00$ & $\phantom{-}0.03$ \\
        $-0.01$ & $\phantom{-}0.00$ & $\phantom{-}0.00$ & $\phantom{-}0.00$ & $\phantom{-}1.00$ & $-0.06$ & $-0.01$ & $\phantom{-}0.00$ \\
        $\phantom{-}0.00$ & $-0.01$ & $\phantom{-}0.00$ & $\phantom{-}0.00$ & $-0.06$ & $\phantom{-}1.00$ & $\phantom{-}0.00$ & $\phantom{-}0.00$ \\
        $\phantom{-}0.00$ & $\phantom{-}0.00$ & $\phantom{-}0.00$ & $\phantom{-}0.00$ & $-0.01$ & $\phantom{-}0.00$ & $\phantom{-}1.00$ & $-0.05$ \\
        $\phantom{-}0.00$ & $\phantom{-}0.00$ & $\phantom{-}0.00$ & $\phantom{-}0.03$ & $\phantom{-}0.00$ & $\phantom{-}0.00$ & $-0.05$ & $\phantom{-}1.00$ \\
        \hline
    \end{tabular}
\end{table*}

\begin{table*}[htb]
    \centering
    \caption{The correlation matrix associated with the systematic uncertainties.}
    \label{table:Systematic_correlations}
    \begin{tabular}{cccccccc}
        \hline
        $\quad c_1\quad$ & $\quad c_2\quad$ & $\quad c_3\quad$ & $\quad c_4\quad$ & $\quad s_1\quad$ & $\quad s_2\quad$ & $\quad s_3\quad$ & $\quad s_4\quad$ \\
        \hline
        $\phantom{-}1.00$ & $\phantom{-}0.29$ & $\phantom{-}0.27$ & $\phantom{-}0.38$ & $-0.03$ & $\phantom{-}0.04$ & $\phantom{-}0.03$ & $\phantom{-}0.03$ \\
        $\phantom{-}0.29$ & $\phantom{-}1.00$ & $\phantom{-}0.65$ & $\phantom{-}0.57$ & $-0.01$ & $\phantom{-}0.10$ & $-0.02$ & $\phantom{-}0.06$ \\
        $\phantom{-}0.27$ & $\phantom{-}0.65$ & $\phantom{-}1.00$ & $\phantom{-}0.40$ & $\phantom{-}0.00$ & $\phantom{-}0.05$ & $\phantom{-}0.00$ & $\phantom{-}0.04$ \\
        $\phantom{-}0.38$ & $\phantom{-}0.57$ & $\phantom{-}0.40$ & $\phantom{-}1.00$ & $\phantom{-}0.01$ & $\phantom{-}0.11$ & $-0.09$ & $\phantom{-}0.15$ \\
        $-0.03$ & $-0.01$ & $\phantom{-}0.00$ & $\phantom{-}0.01$ & $\phantom{-}1.00$ & $-0.35$ & $-0.56$ & $-0.41$ \\
        $\phantom{-}0.04$ & $\phantom{-}0.10$ & $\phantom{-}0.05$ & $\phantom{-}0.11$ & $-0.35$ & $\phantom{-}1.00$ & $\phantom{-}0.34$ & $\phantom{-}0.19$ \\
        $\phantom{-}0.03$ & $-0.02$ & $\phantom{-}0.00$ & $-0.09$ & $-0.56$ & $\phantom{-}0.34$ & $\phantom{-}1.00$ & $\phantom{-}0.46$ \\
        $\phantom{-}0.03$ & $\phantom{-}0.06$ & $\phantom{-}0.04$ & $\phantom{-}0.15$ & $-0.41$ & $\phantom{-}0.19$ & $\phantom{-}0.46$ & $\phantom{-}1.00$ \\
        \hline
    \end{tabular}
\end{table*}

\textbf{Acknowledgement}

The BESIII Collaboration thanks the staff of BEPCII and the IHEP computing center for their strong support. This work is supported in part by National Key R\&D Program of China under Contracts Nos. 2023YFA1606000, 2020YFA0406300, 2020YFA0406400; National Natural Science Foundation of China (NSFC) under Contracts Nos. 11635010, 11735014, 11935015, 11935016, 11935018, 12025502, 12035009, 12035013, 12061131003, 12192260, 12192261, 12192262, 12192263, 12192264, 12192265, 12221005, 12225509, 12235017, 12361141819; the Chinese Academy of Sciences (CAS) Large-Scale Scientific Facility Program; the CAS Center for Excellence in Particle Physics (CCEPP); Joint Large-Scale Scientific Facility Funds of the NSFC and CAS under Contract No. U1832207; 100 Talents Program of CAS; The Institute of Nuclear and Particle Physics (INPAC) and Shanghai Key Laboratory for Particle Physics and Cosmology; German Research Foundation DFG under Contracts Nos. FOR5327, GRK 2149; Istituto Nazionale di Fisica Nucleare, Italy; Knut and Alice Wallenberg Foundation under Contracts Nos. 2021.0174, 2021.0299; Ministry of Development of Turkey under Contract No. DPT2006K-120470; National Research Foundation of Korea under Contract No. NRF-2022R1A2C1092335; National Science and Technology fund of Mongolia; National Science Research and Innovation Fund (NSRF) via the Program Management Unit for Human Resources \& Institutional Development, Research and Innovation of Thailand under Contracts Nos. B16F640076, B50G670107; Polish National Science Centre under Contract No. 2019/35/O/ST2/02907; Swedish Research Council under Contract No. 2019.04595; The Swedish Foundation for International Cooperation in Research and Higher Education under Contract No. CH2018-7756; U. S. Department of Energy under Contract No. DE-FG02-05ER41374

\clearpage
\bibliographystyle{apsrev4-1}
\bibliography{References}

\begin{thebibliography}{57}%
\makeatletter
\providecommand \@ifxundefined [1]{%
 \@ifx{#1\undefined}
}%
\providecommand \@ifnum [1]{%
 \ifnum #1\expandafter \@firstoftwo
 \else \expandafter \@secondoftwo
 \fi
}%
\providecommand \@ifx [1]{%
 \ifx #1\expandafter \@firstoftwo
 \else \expandafter \@secondoftwo
 \fi
}%
\providecommand \natexlab [1]{#1}%
\providecommand \enquote  [1]{``#1''}%
\providecommand \bibnamefont  [1]{#1}%
\providecommand \bibfnamefont [1]{#1}%
\providecommand \citenamefont [1]{#1}%
\providecommand \href@noop [0]{\@secondoftwo}%
\providecommand \href [0]{\begingroup \@sanitize@url \@href}%
\providecommand \@href[1]{\@@startlink{#1}\@@href}%
\providecommand \@@href[1]{\endgroup#1\@@endlink}%
\providecommand \@sanitize@url [0]{\catcode `\\12\catcode `\$12\catcode
  `\&12\catcode `\#12\catcode `\^12\catcode `\_12\catcode `\%12\relax}%
\providecommand \@@startlink[1]{}%
\providecommand \@@endlink[0]{}%
\providecommand \url  [0]{\begingroup\@sanitize@url \@url }%
\providecommand \@url [1]{\endgroup\@href {#1}{\urlprefix }}%
\providecommand \urlprefix  [0]{URL }%
\providecommand \Eprint [0]{\href }%
\providecommand \doibase [0]{http://dx.doi.org/}%
\providecommand \selectlanguage [0]{\@gobble}%
\providecommand \bibinfo  [0]{\@secondoftwo}%
\providecommand \bibfield  [0]{\@secondoftwo}%
\providecommand \translation [1]{[#1]}%
\providecommand \BibitemOpen [0]{}%
\providecommand \bibitemStop [0]{}%
\providecommand \bibitemNoStop [0]{.\EOS\space}%
\providecommand \EOS [0]{\spacefactor3000\relax}%
\providecommand \BibitemShut  [1]{\csname bibitem#1\endcsname}%
\let\auto@bib@innerbib\@empty
\bibitem [{\citenamefont {Cabibbo}(1963)}]{Cabibbo:1963yz}%
  \BibitemOpen
  \bibfield  {author} {\bibinfo {author} {\bibfnamefont {N.}~\bibnamefont
  {Cabibbo}},\ }\href {\doibase 10.1103/PhysRevLett.10.531} {\bibfield
  {journal} {\bibinfo  {journal} {Phys. Rev. Lett.}\ }\textbf {\bibinfo
  {volume} {10}},\ \bibinfo {pages} {531} (\bibinfo {year} {1963})}\BibitemShut
  {NoStop}%
\bibitem [{\citenamefont {Kobayashi}\ and\ \citenamefont
  {Maskawa}(1973)}]{Kobayashi:1973fv}%
  \BibitemOpen
  \bibfield  {author} {\bibinfo {author} {\bibfnamefont {M.}~\bibnamefont
  {Kobayashi}}\ and\ \bibinfo {author} {\bibfnamefont {T.}~\bibnamefont
  {Maskawa}},\ }\href {\doibase 10.1143/PTP.49.652} {\bibfield  {journal}
  {\bibinfo  {journal} {Prog. Theor. Phys.}\ }\textbf {\bibinfo {volume}
  {49}},\ \bibinfo {pages} {652} (\bibinfo {year} {1973})}\BibitemShut
  {NoStop}%
\bibitem [{\citenamefont {Workman}\ \emph {et~al.}(2022)\citenamefont {Workman}
  \emph {et~al.}}]{pdg}%
  \BibitemOpen
  \bibfield  {author} {\bibinfo {author} {\bibfnamefont {R.~L.}\ \bibnamefont
  {Workman}} \emph {et~al.} (\bibinfo {collaboration} {Particle Data Group}),\
  }\href {\doibase 10.1093/ptep/ptac097} {\bibfield  {journal} {\bibinfo
  {journal} {PTEP}\ }\textbf {\bibinfo {volume} {2022}},\ \bibinfo {pages}
  {083C01} (\bibinfo {year} {2022})}\BibitemShut {NoStop}%
\bibitem [{\citenamefont {Brod}\ and\ \citenamefont
  {Zupan}(2014)}]{cite:BrodZupan}%
  \BibitemOpen
  \bibfield  {author} {\bibinfo {author} {\bibfnamefont {J.}~\bibnamefont
  {Brod}}\ and\ \bibinfo {author} {\bibfnamefont {J.}~\bibnamefont {Zupan}},\
  }\href {\doibase 10.1007/JHEP01(2014)051} {\bibfield  {journal} {\bibinfo
  {journal} {JHEP}\ }\textbf {\bibinfo {volume} {01}},\ \bibinfo {pages} {051}
  (\bibinfo {year} {2014})},\ \Eprint {http://arxiv.org/abs/1308.5663}
  {arXiv:1308.5663 [hep-ph]} \BibitemShut {NoStop}%
\bibitem [{\citenamefont {Bondar}\ and\ \citenamefont
  {Poluektov}(2006)}]{BondarPoluektov2006}%
  \BibitemOpen
  \bibfield  {author} {\bibinfo {author} {\bibfnamefont {A.}~\bibnamefont
  {Bondar}}\ and\ \bibinfo {author} {\bibfnamefont {A.}~\bibnamefont
  {Poluektov}},\ }\href {\doibase 10.1140/epjc/s2006-02590-x} {\bibfield
  {journal} {\bibinfo  {journal} {Eur. Phys. J. C}\ }\textbf {\bibinfo {volume}
  {47}},\ \bibinfo {pages} {347} (\bibinfo {year} {2006})},\ \Eprint
  {http://arxiv.org/abs/hep-ph/0510246} {arXiv:hep-ph/0510246 [hep-ph]}
  \BibitemShut {NoStop}%
\bibitem [{\citenamefont {Bondar}\ and\ \citenamefont
  {Poluektov}(2008)}]{BondarPoluektov2008}%
  \BibitemOpen
  \bibfield  {author} {\bibinfo {author} {\bibfnamefont {A.}~\bibnamefont
  {Bondar}}\ and\ \bibinfo {author} {\bibfnamefont {A.}~\bibnamefont
  {Poluektov}},\ }\href {\doibase 10.1140/epjc/s10052-008-0600-z} {\bibfield
  {journal} {\bibinfo  {journal} {Eur. Phys. J. C}\ }\textbf {\bibinfo {volume}
  {55}},\ \bibinfo {pages} {51} (\bibinfo {year} {2008})},\ \Eprint
  {http://arxiv.org/abs/0801.0840} {arXiv:0801.0840 [hep-ex]} \BibitemShut
  {NoStop}%
\bibitem [{\citenamefont {Giri}\ \emph {et~al.}(2003)\citenamefont {Giri},
  \citenamefont {Grossman}, \citenamefont {Soffer},\ and\ \citenamefont
  {Zupan}}]{GiriGrossmanSofferZupan}%
  \BibitemOpen
  \bibfield  {author} {\bibinfo {author} {\bibfnamefont {A.}~\bibnamefont
  {Giri}}, \bibinfo {author} {\bibfnamefont {Y.}~\bibnamefont {Grossman}},
  \bibinfo {author} {\bibfnamefont {A.}~\bibnamefont {Soffer}}, \ and\ \bibinfo
  {author} {\bibfnamefont {J.}~\bibnamefont {Zupan}},\ }\href {\doibase
  10.1103/PhysRevD.68.054018} {\bibfield  {journal} {\bibinfo  {journal} {Phys.
  Rev. D}\ }\textbf {\bibinfo {volume} {68}},\ \bibinfo {pages} {054018}
  (\bibinfo {year} {2003})},\ \Eprint {http://arxiv.org/abs/hep-ph/0303187}
  {arXiv:hep-ph/0303187 [hep-ph]} \BibitemShut {NoStop}%
\bibitem [{\citenamefont {Ablikim}\ \emph
  {et~al.}(2020{\natexlab{a}})\citenamefont {Ablikim} \emph
  {et~al.}}]{cite:KSpipiStrongPhase}%
  \BibitemOpen
  \bibfield  {author} {\bibinfo {author} {\bibfnamefont {M.}~\bibnamefont
  {Ablikim}} \emph {et~al.} (\bibinfo {collaboration} {BESIII Collaboration}),\
  }\href {\doibase 10.1103/PhysRevD.101.112002} {\bibfield  {journal} {\bibinfo
   {journal} {Phys. Rev. D}\ }\textbf {\bibinfo {volume} {101}},\ \bibinfo
  {pages} {112002} (\bibinfo {year} {2020}{\natexlab{a}})},\ \Eprint
  {http://arxiv.org/abs/2003.00091} {arXiv:2003.00091 [hep-ex]} \BibitemShut
  {NoStop}%
\bibitem [{\citenamefont {Libby}\ \emph {et~al.}(2010)\citenamefont {Libby}
  \emph {et~al.}}]{cite:CLEOcisiKSpipi}%
  \BibitemOpen
  \bibfield  {author} {\bibinfo {author} {\bibfnamefont {J.}~\bibnamefont
  {Libby}} \emph {et~al.} (\bibinfo {collaboration} {CLEO Collaboration}),\
  }\href {\doibase 10.1103/PhysRevD.82.112006} {\bibfield  {journal} {\bibinfo
  {journal} {Phys. Rev. D}\ }\textbf {\bibinfo {volume} {82}},\ \bibinfo
  {pages} {112006} (\bibinfo {year} {2010})},\ \Eprint
  {http://arxiv.org/abs/1010.2817} {arXiv:1010.2817 [hep-ex]} \BibitemShut
  {NoStop}%
\bibitem [{\citenamefont {Ablikim}\ \emph
  {et~al.}(2020{\natexlab{b}})\citenamefont {Ablikim} \emph
  {et~al.}}]{cite:cisiKSKK}%
  \BibitemOpen
  \bibfield  {author} {\bibinfo {author} {\bibfnamefont {M.}~\bibnamefont
  {Ablikim}} \emph {et~al.} (\bibinfo {collaboration} {BESIII Collaboration}),\
  }\href {\doibase 10.1103/PhysRevD.102.052008} {\bibfield  {journal} {\bibinfo
   {journal} {Phys. Rev. D}\ }\textbf {\bibinfo {volume} {102}},\ \bibinfo
  {pages} {052008} (\bibinfo {year} {2020}{\natexlab{b}})},\ \Eprint
  {http://arxiv.org/abs/2007.07959} {arXiv:2007.07959 [hep-ex]} \BibitemShut
  {NoStop}%
\bibitem [{\citenamefont {Aaij}\ \emph {et~al.}(2021)\citenamefont {Aaij} \emph
  {et~al.}}]{LHCb-PAPER-2020-019}%
  \BibitemOpen
  \bibfield  {author} {\bibinfo {author} {\bibfnamefont {R.}~\bibnamefont
  {Aaij}} \emph {et~al.} (\bibinfo {collaboration} {LHCb Collaboration}),\
  }\href {\doibase 10.1007/JHEP02(2021)169} {\bibfield  {journal} {\bibinfo
  {journal} {JHEP}\ }\textbf {\bibinfo {volume} {02}},\ \bibinfo {pages} {169}
  (\bibinfo {year} {2021})},\ \Eprint {http://arxiv.org/abs/2010.08483}
  {arXiv:2010.08483 [hep-ex]} \BibitemShut {NoStop}%
\bibitem [{\citenamefont {Aaij}\ \emph {et~al.}(2016)\citenamefont {Aaij} \emph
  {et~al.}}]{LHCb-PAPER-2016-007}%
  \BibitemOpen
  \bibfield  {author} {\bibinfo {author} {\bibfnamefont {R.}~\bibnamefont
  {Aaij}} \emph {et~al.} (\bibinfo {collaboration} {LHCb Collaboration}),\
  }\href {\doibase 10.1007/JHEP08(2016)137} {\bibfield  {journal} {\bibinfo
  {journal} {JHEP}\ }\textbf {\bibinfo {volume} {08}},\ \bibinfo {pages} {137}
  (\bibinfo {year} {2016})},\ \Eprint {http://arxiv.org/abs/1605.01082}
  {arXiv:1605.01082 [hep-ex]} \BibitemShut {NoStop}%
\bibitem [{\citenamefont {Abudin\'en}\ \emph {et~al.}(2022)\citenamefont
  {Abudin\'en} \emph {et~al.}}]{cite:Belle_gamma}%
  \BibitemOpen
  \bibfield  {author} {\bibinfo {author} {\bibfnamefont {F.}~\bibnamefont
  {Abudin\'en}} \emph {et~al.} (\bibinfo {collaboration} {Belle and Belle II
  Collaborations}),\ }\href {\doibase 10.1007/JHEP02(2022)063} {\bibfield
  {journal} {\bibinfo  {journal} {JHEP}\ }\textbf {\bibinfo {volume} {02}},\
  \bibinfo {pages} {063} (\bibinfo {year} {2022})},\ \bibinfo {note} {[Erratum:
  JHEP 12, 034 (2022)]},\ \Eprint {http://arxiv.org/abs/2110.12125}
  {arXiv:2110.12125 [hep-ex]} \BibitemShut {NoStop}%
\bibitem [{\citenamefont {Harnew}\ \emph {et~al.}(2018)\citenamefont {Harnew},
  \citenamefont {Naik}, \citenamefont {Prouve}, \citenamefont {Rademacker},\
  and\ \citenamefont {Asner}}]{cite:cisi4pi}%
  \BibitemOpen
  \bibfield  {author} {\bibinfo {author} {\bibfnamefont {S.}~\bibnamefont
  {Harnew}}, \bibinfo {author} {\bibfnamefont {P.}~\bibnamefont {Naik}},
  \bibinfo {author} {\bibfnamefont {C.}~\bibnamefont {Prouve}}, \bibinfo
  {author} {\bibfnamefont {J.}~\bibnamefont {Rademacker}}, \ and\ \bibinfo
  {author} {\bibfnamefont {D.}~\bibnamefont {Asner}},\ }\href {\doibase
  10.1007/JHEP01(2018)144} {\bibfield  {journal} {\bibinfo  {journal} {JHEP}\
  }\textbf {\bibinfo {volume} {01}},\ \bibinfo {pages} {144} (\bibinfo {year}
  {2018})},\ \Eprint {http://arxiv.org/abs/1709.03467} {arXiv:1709.03467
  [hep-ex]} \BibitemShut {NoStop}%
\bibitem [{\citenamefont {Ablikim}\ \emph
  {et~al.}(2024{\natexlab{a}})\citenamefont {Ablikim} \emph
  {et~al.}}]{cite:cisi4pi_BESIII}%
  \BibitemOpen
  \bibfield  {author} {\bibinfo {author} {\bibfnamefont {M.}~\bibnamefont
  {Ablikim}} \emph {et~al.} (\bibinfo {collaboration} {BESIII}),\ }\href
  {\doibase 10.1103/PhysRevD.110.112008} {\bibfield  {journal} {\bibinfo
  {journal} {Phys. Rev. D}\ }\textbf {\bibinfo {volume} {110}},\ \bibinfo
  {pages} {112008} (\bibinfo {year} {2024}{\natexlab{a}})},\ \Eprint
  {http://arxiv.org/abs/2408.16279} {arXiv:2408.16279 [hep-ex]} \BibitemShut
  {NoStop}%
\bibitem [{\citenamefont {Resmi}\ \emph {et~al.}(2018)\citenamefont {Resmi},
  \citenamefont {Libby}, \citenamefont {Malde},\ and\ \citenamefont
  {Wilkinson}}]{cite:KSpipipi0_CLEOc}%
  \BibitemOpen
  \bibfield  {author} {\bibinfo {author} {\bibfnamefont {P.~K.}\ \bibnamefont
  {Resmi}}, \bibinfo {author} {\bibfnamefont {J.}~\bibnamefont {Libby}},
  \bibinfo {author} {\bibfnamefont {S.}~\bibnamefont {Malde}}, \ and\ \bibinfo
  {author} {\bibfnamefont {G.}~\bibnamefont {Wilkinson}},\ }\href {\doibase
  10.1007/JHEP01(2018)082} {\bibfield  {journal} {\bibinfo  {journal} {JHEP}\
  }\textbf {\bibinfo {volume} {01}},\ \bibinfo {pages} {082} (\bibinfo {year}
  {2018})},\ \Eprint {http://arxiv.org/abs/1710.10086} {arXiv:1710.10086
  [hep-ex]} \BibitemShut {NoStop}%
\bibitem [{\citenamefont {Resmi}\ \emph {et~al.}(2019)\citenamefont {Resmi}
  \emph {et~al.}}]{Belle:2019uav}%
  \BibitemOpen
  \bibfield  {author} {\bibinfo {author} {\bibfnamefont {P.~K.}\ \bibnamefont
  {Resmi}} \emph {et~al.} (\bibinfo {collaboration} {Belle Collaboration}),\
  }\href {\doibase 10.1007/JHEP10(2019)178} {\bibfield  {journal} {\bibinfo
  {journal} {JHEP}\ }\textbf {\bibinfo {volume} {10}},\ \bibinfo {pages} {178}
  (\bibinfo {year} {2019})},\ \Eprint {http://arxiv.org/abs/1908.09499}
  {arXiv:1908.09499 [hep-ex]} \BibitemShut {NoStop}%
\bibitem [{\citenamefont {Rademacker}\ and\ \citenamefont
  {Wilkinson}(2007)}]{cite:Rademacker_Wilkinson_KKpipi}%
  \BibitemOpen
  \bibfield  {author} {\bibinfo {author} {\bibfnamefont {J.}~\bibnamefont
  {Rademacker}}\ and\ \bibinfo {author} {\bibfnamefont {G.}~\bibnamefont
  {Wilkinson}},\ }\href {\doibase 10.1016/j.physletb.2007.01.071} {\bibfield
  {journal} {\bibinfo  {journal} {Phys. Lett. B}\ }\textbf {\bibinfo {volume}
  {647}},\ \bibinfo {pages} {400} (\bibinfo {year} {2007})},\ \Eprint
  {http://arxiv.org/abs/hep-ph/0611272} {arXiv:hep-ph/0611272} \BibitemShut
  {NoStop}%
\bibitem [{\citenamefont {Aaij}\ \emph {et~al.}(2023)\citenamefont {Aaij} \emph
  {et~al.}}]{LHCb-PAPER-2022-037}%
  \BibitemOpen
  \bibfield  {author} {\bibinfo {author} {\bibfnamefont {R.}~\bibnamefont
  {Aaij}} \emph {et~al.} (\bibinfo {collaboration} {LHCb Collaboration}),\
  }\href {\doibase 10.1140/epjc/s10052-023-11560-5} {\bibfield  {journal}
  {\bibinfo  {journal} {Eur. Phys. J. C}\ }\textbf {\bibinfo {volume} {83}},\
  \bibinfo {pages} {547} (\bibinfo {year} {2023})},\ \Eprint
  {http://arxiv.org/abs/2301.10328} {arXiv:2301.10328 [hep-ex]} \BibitemShut
  {NoStop}%
\bibitem [{\citenamefont {Ablikim}\ \emph {et~al.}(2013)\citenamefont {Ablikim}
  \emph {et~al.}}]{cite:IntegratedLuminosity_2010_2011}%
  \BibitemOpen
  \bibfield  {author} {\bibinfo {author} {\bibfnamefont {M.}~\bibnamefont
  {Ablikim}} \emph {et~al.} (\bibinfo {collaboration} {BESIII Collaboration}),\
  }\href {\doibase 10.1088/1674-1137/37/12/123001} {\bibfield  {journal}
  {\bibinfo  {journal} {Chin. Phys. C}\ }\textbf {\bibinfo {volume} {37}},\
  \bibinfo {pages} {123001} (\bibinfo {year} {2013})},\ \Eprint
  {http://arxiv.org/abs/1307.2022} {arXiv:1307.2022 [hep-ex]} \BibitemShut
  {NoStop}%
\bibitem [{\citenamefont {Ablikim}\ \emph {et~al.}(2016)\citenamefont {Ablikim}
  \emph {et~al.}}]{cite:NDD_corr}%
  \BibitemOpen
  \bibfield  {author} {\bibinfo {author} {\bibfnamefont {M.}~\bibnamefont
  {Ablikim}} \emph {et~al.} (\bibinfo {collaboration} {BESIII Collaboration}),\
  }\href {\doibase 10.1016/j.physletb.2015.11.043} {\bibfield  {journal}
  {\bibinfo  {journal} {Phys. Lett. B}\ }\textbf {\bibinfo {volume} {753}},\
  \bibinfo {pages} {629} (\bibinfo {year} {2016})},\ \bibinfo {note} {[Erratum:
  Phys.Lett.B 812, 135982 (2021)]},\ \Eprint {http://arxiv.org/abs/1507.08188}
  {arXiv:1507.08188 [hep-ex]} \BibitemShut {NoStop}%
\bibitem [{\citenamefont {Ablikim}\ \emph
  {et~al.}(2023{\natexlab{a}})\citenamefont {Ablikim} \emph
  {et~al.}}]{cite:KKpipi_FPlus}%
  \BibitemOpen
  \bibfield  {author} {\bibinfo {author} {\bibfnamefont {M.}~\bibnamefont
  {Ablikim}} \emph {et~al.} (\bibinfo {collaboration} {BESIII Collaboration}),\
  }\href {\doibase 10.1103/PhysRevD.107.032009} {\bibfield  {journal} {\bibinfo
   {journal} {Phys. Rev. D}\ }\textbf {\bibinfo {volume} {107}},\ \bibinfo
  {pages} {032009} (\bibinfo {year} {2023}{\natexlab{a}})},\ \Eprint
  {http://arxiv.org/abs/2212.06489} {arXiv:2212.06489 [hep-ex]} \BibitemShut
  {NoStop}%
\bibitem [{\citenamefont {Nayak}\ \emph {et~al.}(2015)\citenamefont {Nayak},
  \citenamefont {Libby}, \citenamefont {Malde}, \citenamefont {Thomas},
  \citenamefont {Wilkinson}, \citenamefont {Briere}, \citenamefont {Naik},
  \citenamefont {Gershon},\ and\ \citenamefont
  {Bonvicini}}]{cite:Firstpipipi0}%
  \BibitemOpen
  \bibfield  {author} {\bibinfo {author} {\bibfnamefont {M.}~\bibnamefont
  {Nayak}}, \bibinfo {author} {\bibfnamefont {J.}~\bibnamefont {Libby}},
  \bibinfo {author} {\bibfnamefont {S.}~\bibnamefont {Malde}}, \bibinfo
  {author} {\bibfnamefont {C.}~\bibnamefont {Thomas}}, \bibinfo {author}
  {\bibfnamefont {G.}~\bibnamefont {Wilkinson}}, \bibinfo {author}
  {\bibfnamefont {R.~A.}\ \bibnamefont {Briere}}, \bibinfo {author}
  {\bibfnamefont {P.}~\bibnamefont {Naik}}, \bibinfo {author} {\bibfnamefont
  {T.}~\bibnamefont {Gershon}}, \ and\ \bibinfo {author} {\bibfnamefont
  {G.}~\bibnamefont {Bonvicini}},\ }\href {\doibase
  10.1016/j.physletb.2014.11.022} {\bibfield  {journal} {\bibinfo  {journal}
  {Phys. Lett. B}\ }\textbf {\bibinfo {volume} {740}},\ \bibinfo {pages} {1}
  (\bibinfo {year} {2015})},\ \Eprint {http://arxiv.org/abs/1410.3964}
  {arXiv:1410.3964 [hep-ex]} \BibitemShut {NoStop}%
\bibitem [{\citenamefont {Di~Canto}\ \emph {et~al.}(2019)\citenamefont
  {Di~Canto}, \citenamefont {Garra~Tic\'o}, \citenamefont {Gershon},
  \citenamefont {Jurik}, \citenamefont {Martinelli}, \citenamefont {Pila\v{r}},
  \citenamefont {Stahl},\ and\ \citenamefont {Tonelli}}]{cite:BinFlip}%
  \BibitemOpen
  \bibfield  {author} {\bibinfo {author} {\bibfnamefont {A.}~\bibnamefont
  {Di~Canto}}, \bibinfo {author} {\bibfnamefont {J.}~\bibnamefont
  {Garra~Tic\'o}}, \bibinfo {author} {\bibfnamefont {T.}~\bibnamefont
  {Gershon}}, \bibinfo {author} {\bibfnamefont {N.}~\bibnamefont {Jurik}},
  \bibinfo {author} {\bibfnamefont {M.}~\bibnamefont {Martinelli}}, \bibinfo
  {author} {\bibfnamefont {T.}~\bibnamefont {Pila\v{r}}}, \bibinfo {author}
  {\bibfnamefont {S.}~\bibnamefont {Stahl}}, \ and\ \bibinfo {author}
  {\bibfnamefont {D.}~\bibnamefont {Tonelli}},\ }\href {\doibase
  10.1103/PhysRevD.99.012007} {\bibfield  {journal} {\bibinfo  {journal} {Phys.
  Rev. D}\ }\textbf {\bibinfo {volume} {99}},\ \bibinfo {pages} {012007}
  (\bibinfo {year} {2019})},\ \Eprint {http://arxiv.org/abs/1811.01032}
  {arXiv:1811.01032 [hep-ex]} \BibitemShut {NoStop}%
\bibitem [{\citenamefont {Ablikim}\ \emph {et~al.}(2018)\citenamefont {Ablikim}
  \emph {et~al.}}]{cite:NDD}%
  \BibitemOpen
  \bibfield  {author} {\bibinfo {author} {\bibfnamefont {M.}~\bibnamefont
  {Ablikim}} \emph {et~al.} (\bibinfo {collaboration} {BESIII Collaboration}),\
  }\href {\doibase 10.1088/1674-1137/42/8/083001} {\bibfield  {journal}
  {\bibinfo  {journal} {Chin. Phys. C}\ }\textbf {\bibinfo {volume} {42}},\
  \bibinfo {pages} {083001} (\bibinfo {year} {2018})}\BibitemShut {NoStop}%
\bibitem [{\citenamefont {Ablikim}\ \emph
  {et~al.}(2024{\natexlab{b}})\citenamefont {Ablikim} \emph
  {et~al.}}]{cite:NewLuminosity}%
  \BibitemOpen
  \bibfield  {author} {\bibinfo {author} {\bibfnamefont {M.}~\bibnamefont
  {Ablikim}} \emph {et~al.} (\bibinfo {collaboration} {BESIII Collaboration}),\
  }\href {\doibase 10.1088/1674-1137/ad70a0} {\bibfield  {journal} {\bibinfo
  {journal} {Chin. Phys. C}\ }\textbf {\bibinfo {volume} {48}},\ \bibinfo
  {pages} {123001} (\bibinfo {year} {2024}{\natexlab{b}})},\ \Eprint
  {http://arxiv.org/abs/2406.05827} {arXiv:2406.05827 [hep-ex]} \BibitemShut
  {NoStop}%
\bibitem [{\citenamefont {Aaij}\ \emph {et~al.}(2019)\citenamefont {Aaij} \emph
  {et~al.}}]{LHCb-PAPER-2018-041}%
  \BibitemOpen
  \bibfield  {author} {\bibinfo {author} {\bibfnamefont {R.}~\bibnamefont
  {Aaij}} \emph {et~al.} (\bibinfo {collaboration} {LHCb Collaboration}),\
  }\href {\doibase 10.1007/JHEP02(2019)126} {\bibfield  {journal} {\bibinfo
  {journal} {JHEP}\ }\textbf {\bibinfo {volume} {02}},\ \bibinfo {pages} {126}
  (\bibinfo {year} {2019})},\ \Eprint {http://arxiv.org/abs/1811.08304}
  {arXiv:1811.08304 [hep-ex]} \BibitemShut {NoStop}%
\bibitem [{\citenamefont {Baltrusaitis}\ \emph {et~al.}(1986)\citenamefont
  {Baltrusaitis} \emph {et~al.}}]{cite:DT_method}%
  \BibitemOpen
  \bibfield  {author} {\bibinfo {author} {\bibfnamefont {R.~M.}\ \bibnamefont
  {Baltrusaitis}} \emph {et~al.} (\bibinfo {collaboration} {MARK-III
  Collaboration}),\ }\href {\doibase 10.1103/PhysRevLett.56.2140} {\bibfield
  {journal} {\bibinfo  {journal} {Phys. Rev. Lett.}\ }\textbf {\bibinfo
  {volume} {56}},\ \bibinfo {pages} {2140} (\bibinfo {year}
  {1986})}\BibitemShut {NoStop}%
\bibitem [{\citenamefont {Atwood}\ and\ \citenamefont
  {Soni}(2003)}]{Atwood:2003mj}%
  \BibitemOpen
  \bibfield  {author} {\bibinfo {author} {\bibfnamefont {D.}~\bibnamefont
  {Atwood}}\ and\ \bibinfo {author} {\bibfnamefont {A.}~\bibnamefont {Soni}},\
  }\href {\doibase 10.1103/PhysRevD.68.033003} {\bibfield  {journal} {\bibinfo
  {journal} {Phys. Rev. D}\ }\textbf {\bibinfo {volume} {68}},\ \bibinfo
  {pages} {033003} (\bibinfo {year} {2003})},\ \Eprint
  {http://arxiv.org/abs/hep-ph/0304085} {arXiv:hep-ph/0304085 [hep-ph]}
  \BibitemShut {NoStop}%
\bibitem [{\citenamefont {Harnew}\ and\ \citenamefont
  {Rademacker}(2014)}]{Harnew:2013wea}%
  \BibitemOpen
  \bibfield  {author} {\bibinfo {author} {\bibfnamefont {S.}~\bibnamefont
  {Harnew}}\ and\ \bibinfo {author} {\bibfnamefont {J.}~\bibnamefont
  {Rademacker}},\ }\href {\doibase 10.1016/j.physletb.2013.11.065} {\bibfield
  {journal} {\bibinfo  {journal} {Phys. Lett. B}\ }\textbf {\bibinfo {volume}
  {728}},\ \bibinfo {pages} {296} (\bibinfo {year} {2014})},\ \Eprint
  {http://arxiv.org/abs/1309.0134} {arXiv:1309.0134 [hep-ph]} \BibitemShut
  {NoStop}%
\bibitem [{\citenamefont {Ablikim}\ \emph {et~al.}(2021)\citenamefont {Ablikim}
  \emph {et~al.}}]{cite:K3piStrongPhase}%
  \BibitemOpen
  \bibfield  {author} {\bibinfo {author} {\bibfnamefont {M.}~\bibnamefont
  {Ablikim}} \emph {et~al.} (\bibinfo {collaboration} {BESIII Collaboration}),\
  }\href {\doibase 10.1007/JHEP05(2021)164} {\bibfield  {journal} {\bibinfo
  {journal} {JHEP}\ }\textbf {\bibinfo {volume} {05}},\ \bibinfo {pages} {164}
  (\bibinfo {year} {2021})},\ \Eprint {http://arxiv.org/abs/2103.05988}
  {arXiv:2103.05988 [hep-ex]} \BibitemShut {NoStop}%
\bibitem [{\citenamefont {Ablikim}\ \emph {et~al.}(2022)\citenamefont {Ablikim}
  \emph {et~al.}}]{cite:deltaKpi}%
  \BibitemOpen
  \bibfield  {author} {\bibinfo {author} {\bibfnamefont {M.}~\bibnamefont
  {Ablikim}} \emph {et~al.} (\bibinfo {collaboration} {BESIII Collaboration}),\
  }\href {\doibase 10.1140/epjc/s10052-022-10872-2} {\bibfield  {journal}
  {\bibinfo  {journal} {Eur. Phys. J. C}\ }\textbf {\bibinfo {volume} {82}},\
  \bibinfo {pages} {1009} (\bibinfo {year} {2022})},\ \Eprint
  {http://arxiv.org/abs/2208.09402} {arXiv:2208.09402 [hep-ex]} \BibitemShut
  {NoStop}%
\bibitem [{\citenamefont {Malde}\ \emph {et~al.}(2015)\citenamefont {Malde},
  \citenamefont {Thomas}, \citenamefont {Wilkinson}, \citenamefont {Naik},
  \citenamefont {Prouve}, \citenamefont {Rademacker}, \citenamefont {Libby},
  \citenamefont {Nayak}, \citenamefont {Gershon},\ and\ \citenamefont
  {Briere}}]{cite:pipipi0_CPfraction}%
  \BibitemOpen
  \bibfield  {author} {\bibinfo {author} {\bibfnamefont {S.}~\bibnamefont
  {Malde}}, \bibinfo {author} {\bibfnamefont {C.}~\bibnamefont {Thomas}},
  \bibinfo {author} {\bibfnamefont {G.}~\bibnamefont {Wilkinson}}, \bibinfo
  {author} {\bibfnamefont {P.}~\bibnamefont {Naik}}, \bibinfo {author}
  {\bibfnamefont {C.}~\bibnamefont {Prouve}}, \bibinfo {author} {\bibfnamefont
  {J.}~\bibnamefont {Rademacker}}, \bibinfo {author} {\bibfnamefont
  {J.}~\bibnamefont {Libby}}, \bibinfo {author} {\bibfnamefont
  {M.}~\bibnamefont {Nayak}}, \bibinfo {author} {\bibfnamefont
  {T.}~\bibnamefont {Gershon}}, \ and\ \bibinfo {author} {\bibfnamefont
  {R.~A.}\ \bibnamefont {Briere}},\ }\href {\doibase
  10.1016/j.physletb.2015.05.043} {\bibfield  {journal} {\bibinfo  {journal}
  {Phys. Lett. B}\ }\textbf {\bibinfo {volume} {747}},\ \bibinfo {pages} {9}
  (\bibinfo {year} {2015})},\ \Eprint {http://arxiv.org/abs/1504.05878}
  {arXiv:1504.05878 [hep-ex]} \BibitemShut {NoStop}%
\bibitem [{\citenamefont {Ablikim}\ \emph
  {et~al.}(2024{\natexlab{c}})\citenamefont {Ablikim} \emph
  {et~al.}}]{cite:pipipi0_BESIII}%
  \BibitemOpen
  \bibfield  {author} {\bibinfo {author} {\bibfnamefont {M.}~\bibnamefont
  {Ablikim}} \emph {et~al.} (\bibinfo {collaboration} {BESIII Collaboration}),\
  }\href@noop {} {\  (\bibinfo {year} {2024}{\natexlab{c}})},\ \Eprint
  {http://arxiv.org/abs/2409.07197} {arXiv:2409.07197 [hep-ex]} \BibitemShut
  {NoStop}%
\bibitem [{\citenamefont {Ablikim}\ \emph
  {et~al.}(2023{\natexlab{b}})\citenamefont {Ablikim} \emph
  {et~al.}}]{cite:KSpipipi0_BESIII}%
  \BibitemOpen
  \bibfield  {author} {\bibinfo {author} {\bibfnamefont {M.}~\bibnamefont
  {Ablikim}} \emph {et~al.} (\bibinfo {collaboration} {BESIII Collaboration}),\
  }\href {\doibase 10.1103/PhysRevD.108.032003} {\bibfield  {journal} {\bibinfo
   {journal} {Phys. Rev. D}\ }\textbf {\bibinfo {volume} {108}},\ \bibinfo
  {pages} {032003} (\bibinfo {year} {2023}{\natexlab{b}})},\ \Eprint
  {http://arxiv.org/abs/2305.03975} {arXiv:2305.03975 [hep-ex]} \BibitemShut
  {NoStop}%
\bibitem [{\citenamefont {Asner}\ and\ \citenamefont
  {Sun}(2006)}]{cite:AsnerSun}%
  \BibitemOpen
  \bibfield  {author} {\bibinfo {author} {\bibfnamefont {D.~M.}\ \bibnamefont
  {Asner}}\ and\ \bibinfo {author} {\bibfnamefont {W.~M.}\ \bibnamefont
  {Sun}},\ }\href {\doibase 10.1103/PhysRevD.73.034024} {\bibfield  {journal}
  {\bibinfo  {journal} {Phys. Rev. D}\ }\textbf {\bibinfo {volume} {73}},\
  \bibinfo {pages} {034024} (\bibinfo {year} {2006})},\ \bibinfo {note}
  {[Erratum: Phys.Rev.D 77, 019901 (2008)]},\ \Eprint
  {http://arxiv.org/abs/hep-ph/0507238} {arXiv:hep-ph/0507238} \BibitemShut
  {NoStop}%
\bibitem [{\citenamefont {Amhis}\ \emph {et~al.}(2023)\citenamefont {Amhis}
  \emph {et~al.}}]{cite:HFLAV2021}%
  \BibitemOpen
  \bibfield  {author} {\bibinfo {author} {\bibfnamefont {Y.~S.}\ \bibnamefont
  {Amhis}} \emph {et~al.} (\bibinfo {collaboration} {Heavy Flavor Averaging
  Group, HFLAV}),\ }\href {\doibase 10.1103/PhysRevD.107.052008} {\bibfield
  {journal} {\bibinfo  {journal} {Phys. Rev. D}\ }\textbf {\bibinfo {volume}
  {107}},\ \bibinfo {pages} {052008} (\bibinfo {year} {2023})},\ \Eprint
  {http://arxiv.org/abs/2206.07501} {arXiv:2206.07501 [hep-ex]} \BibitemShut
  {NoStop}%
\bibitem [{\citenamefont {Ablikim}\ \emph {et~al.}(2010)\citenamefont {Ablikim}
  \emph {et~al.}}]{Ablikim:2009aa}%
  \BibitemOpen
  \bibfield  {author} {\bibinfo {author} {\bibfnamefont {M.}~\bibnamefont
  {Ablikim}} \emph {et~al.} (\bibinfo {collaboration} {BESIII Collaboration}),\
  }\href {\doibase 10.1016/j.nima.2009.12.050} {\bibfield  {journal} {\bibinfo
  {journal} {Nucl. Instrum. Meth. A}\ }\textbf {\bibinfo {volume} {614}},\
  \bibinfo {pages} {345} (\bibinfo {year} {2010})},\ \Eprint
  {http://arxiv.org/abs/0911.4960} {arXiv:0911.4960 [physics.ins-det]}
  \BibitemShut {NoStop}%
\bibitem [{\citenamefont {Yu}\ \emph {et~al.}(2016)\citenamefont {Yu} \emph
  {et~al.}}]{Yu:2016cof}%
  \BibitemOpen
  \bibfield  {author} {\bibinfo {author} {\bibfnamefont {C.}~\bibnamefont {Yu}}
  \emph {et~al.},\ }in\ \href {\doibase 10.18429/JACoW-IPAC2016-TUYA01} {\emph
  {\bibinfo {booktitle} {{7th International Particle Accelerator
  Conference}}}}\ (\bibinfo {year} {2016})\ p.\ \bibinfo {pages}
  {TUYA01}\BibitemShut {NoStop}%
\bibitem [{\citenamefont {Ablikim}\ \emph
  {et~al.}(2020{\natexlab{c}})\citenamefont {Ablikim} \emph
  {et~al.}}]{Ablikim:2019hff}%
  \BibitemOpen
  \bibfield  {author} {\bibinfo {author} {\bibfnamefont {M.}~\bibnamefont
  {Ablikim}} \emph {et~al.} (\bibinfo {collaboration} {BESIII Collaboration}),\
  }\href {\doibase 10.1088/1674-1137/44/4/040001} {\bibfield  {journal}
  {\bibinfo  {journal} {Chin. Phys. C}\ }\textbf {\bibinfo {volume} {44}},\
  \bibinfo {pages} {040001} (\bibinfo {year} {2020}{\natexlab{c}})},\ \Eprint
  {http://arxiv.org/abs/1912.05983} {arXiv:1912.05983 [hep-ex]} \BibitemShut
  {NoStop}%
\bibitem [{\citenamefont {Lu}\ \emph {et~al.}(2020)\citenamefont {Lu},
  \citenamefont {Xiao},\ and\ \citenamefont {Ji}}]{EcmsMea}%
  \BibitemOpen
  \bibfield  {author} {\bibinfo {author} {\bibfnamefont {J.}~\bibnamefont
  {Lu}}, \bibinfo {author} {\bibfnamefont {Y.}~\bibnamefont {Xiao}}, \ and\
  \bibinfo {author} {\bibfnamefont {X.}~\bibnamefont {Ji}},\ }\href {\doibase
  10.1007/s41605-020-00188-8} {\bibfield  {journal} {\bibinfo  {journal}
  {Radiat. Detect. Technol. Methods}\ }\textbf {\bibinfo {volume} {4}},\
  \bibinfo {pages} {337} (\bibinfo {year} {2020})}\BibitemShut {NoStop}%
\bibitem [{\citenamefont {Zhang}\ \emph {et~al.}(2022)\citenamefont {Zhang}
  \emph {et~al.}}]{EventFilter}%
  \BibitemOpen
  \bibfield  {author} {\bibinfo {author} {\bibfnamefont {J.-W.}\ \bibnamefont
  {Zhang}} \emph {et~al.},\ }\href {\doibase 10.1007/s41605-022-00331-7}
  {\bibfield  {journal} {\bibinfo  {journal} {Radiat. Detect. Technol.
  Methods}\ }\textbf {\bibinfo {volume} {6}},\ \bibinfo {pages} {289} (\bibinfo
  {year} {2022})}\BibitemShut {NoStop}%
\bibitem [{\citenamefont {Li}\ \emph {et~al.}(2017)\citenamefont {Li} \emph
  {et~al.}}]{etof1}%
  \BibitemOpen
  \bibfield  {author} {\bibinfo {author} {\bibfnamefont {X.}~\bibnamefont {Li}}
  \emph {et~al.},\ }\href {\doibase 10.1007/s41605-017-0014-2} {\bibfield
  {journal} {\bibinfo  {journal} {Radiat. Detect. Technol. Methods}\ }\textbf
  {\bibinfo {volume} {1}},\ \bibinfo {pages} {13} (\bibinfo {year}
  {2017})}\BibitemShut {NoStop}%
\bibitem [{\citenamefont {Guo}\ \emph {et~al.}(2017)\citenamefont {Guo} \emph
  {et~al.}}]{etof2}%
  \BibitemOpen
  \bibfield  {author} {\bibinfo {author} {\bibfnamefont {Y.-X.}\ \bibnamefont
  {Guo}} \emph {et~al.},\ }\href {\doibase 10.1007/s41605-017-0012-4}
  {\bibfield  {journal} {\bibinfo  {journal} {Radiat. Detect. Technol.
  Methods}\ }\textbf {\bibinfo {volume} {1}},\ \bibinfo {pages} {15} (\bibinfo
  {year} {2017})}\BibitemShut {NoStop}%
\bibitem [{\citenamefont {Cao}\ \emph {et~al.}(2020)\citenamefont {Cao} \emph
  {et~al.}}]{etof3}%
  \BibitemOpen
  \bibfield  {author} {\bibinfo {author} {\bibfnamefont {P.}~\bibnamefont
  {Cao}} \emph {et~al.},\ }\href {\doibase 10.1016/j.nima.2019.163053}
  {\bibfield  {journal} {\bibinfo  {journal} {Nucl. Instrum. Meth. A}\ }\textbf
  {\bibinfo {volume} {953}},\ \bibinfo {pages} {163053} (\bibinfo {year}
  {2020})}\BibitemShut {NoStop}%
\bibitem [{\citenamefont {Agostinelli}\ \emph {et~al.}(2003)\citenamefont
  {Agostinelli} \emph {et~al.}}]{geant4}%
  \BibitemOpen
  \bibfield  {author} {\bibinfo {author} {\bibfnamefont {S.}~\bibnamefont
  {Agostinelli}} \emph {et~al.} (\bibinfo {collaboration} {GEANT4
  Collaboration}),\ }\href {\doibase 10.1016/S0168-9002(03)01368-8} {\bibfield
  {journal} {\bibinfo  {journal} {Nucl. Instrum. Meth. A}\ }\textbf {\bibinfo
  {volume} {506}},\ \bibinfo {pages} {250} (\bibinfo {year}
  {2003})}\BibitemShut {NoStop}%
\bibitem [{\citenamefont {Jadach}\ \emph {et~al.}(2000)\citenamefont {Jadach},
  \citenamefont {Ward},\ and\ \citenamefont {Was}}]{ref:kkmc}%
  \BibitemOpen
  \bibfield  {author} {\bibinfo {author} {\bibfnamefont {S.}~\bibnamefont
  {Jadach}}, \bibinfo {author} {\bibfnamefont {B.}~\bibnamefont {Ward}}, \ and\
  \bibinfo {author} {\bibfnamefont {Z.}~\bibnamefont {Was}},\ }\href {\doibase
  10.1016/S0010-4655(00)00048-5} {\bibfield  {journal} {\bibinfo  {journal}
  {Comput. Phys. Commun.}\ }\textbf {\bibinfo {volume} {130}},\ \bibinfo
  {pages} {260} (\bibinfo {year} {2000})}\BibitemShut {NoStop}%
\bibitem [{\citenamefont {Lange}(2001)}]{Lange:2001uf}%
  \BibitemOpen
  \bibfield  {author} {\bibinfo {author} {\bibfnamefont {D.}~\bibnamefont
  {Lange}},\ }\href {\doibase 10.1016/S0168-9002(01)00089-4} {\bibfield
  {journal} {\bibinfo  {journal} {Nucl. Instrum. Meth. A}\ }\textbf {\bibinfo
  {volume} {462}},\ \bibinfo {pages} {152} (\bibinfo {year}
  {2001})}\BibitemShut {NoStop}%
\bibitem [{\citenamefont {Ping}(2008)}]{Ping:2008zz}%
  \BibitemOpen
  \bibfield  {author} {\bibinfo {author} {\bibfnamefont {R.~G.}\ \bibnamefont
  {Ping}},\ }\href {\doibase 10.1088/1674-1137/32/8/001} {\bibfield  {journal}
  {\bibinfo  {journal} {Chin. Phys. C}\ }\textbf {\bibinfo {volume} {32}},\
  \bibinfo {pages} {599} (\bibinfo {year} {2008})}\BibitemShut {NoStop}%
\bibitem [{\citenamefont {Yang}\ \emph {et~al.}(2014)\citenamefont {Yang},
  \citenamefont {Ping},\ and\ \citenamefont {Chen}}]{YANGRui-Ling:61301}%
  \BibitemOpen
  \bibfield  {author} {\bibinfo {author} {\bibfnamefont {R.-L.}\ \bibnamefont
  {Yang}}, \bibinfo {author} {\bibfnamefont {R.-G.}\ \bibnamefont {Ping}}, \
  and\ \bibinfo {author} {\bibfnamefont {H.}~\bibnamefont {Chen}},\ }\href
  {\doibase 10.1088/0256-307X/31/6/061301} {\bibfield  {journal} {\bibinfo
  {journal} {Chin. Phys. Lett.}\ }\textbf {\bibinfo {volume} {31}},\ \bibinfo
  {pages} {061301} (\bibinfo {year} {2014})}\BibitemShut {NoStop}%
\bibitem [{\citenamefont {Chen}\ \emph {et~al.}(2000)\citenamefont {Chen},
  \citenamefont {Huang}, \citenamefont {Qi}, \citenamefont {Zhang},\ and\
  \citenamefont {Zhu}}]{Chen:2000tv}%
  \BibitemOpen
  \bibfield  {author} {\bibinfo {author} {\bibfnamefont {J.~C.}\ \bibnamefont
  {Chen}}, \bibinfo {author} {\bibfnamefont {G.~S.}\ \bibnamefont {Huang}},
  \bibinfo {author} {\bibfnamefont {X.~R.}\ \bibnamefont {Qi}}, \bibinfo
  {author} {\bibfnamefont {D.~H.}\ \bibnamefont {Zhang}}, \ and\ \bibinfo
  {author} {\bibfnamefont {Y.~S.}\ \bibnamefont {Zhu}},\ }\href {\doibase
  10.1103/PhysRevD.62.034003} {\bibfield  {journal} {\bibinfo  {journal} {Phys.
  Rev. D}\ }\textbf {\bibinfo {volume} {62}},\ \bibinfo {pages} {034003}
  (\bibinfo {year} {2000})}\BibitemShut {NoStop}%
\bibitem [{\citenamefont {Richter-Was}(1993)}]{RICHTERWAS1993163}%
  \BibitemOpen
  \bibfield  {author} {\bibinfo {author} {\bibfnamefont {E.}~\bibnamefont
  {Richter-Was}},\ }\href {\doibase 10.1016/0370-2693(93)90062-M} {\bibfield
  {journal} {\bibinfo  {journal} {Phys. Lett. B}\ }\textbf {\bibinfo {volume}
  {303}},\ \bibinfo {pages} {163} (\bibinfo {year} {1993})}\BibitemShut
  {NoStop}%
\bibitem [{\citenamefont {Ablikim}\ \emph {et~al.}(2015)\citenamefont {Ablikim}
  \emph {et~al.}}]{cite:KeNu_selection}%
  \BibitemOpen
  \bibfield  {author} {\bibinfo {author} {\bibfnamefont {M.}~\bibnamefont
  {Ablikim}} \emph {et~al.} (\bibinfo {collaboration} {BESIII Collaboration}),\
  }\href {\doibase 10.1103/PhysRevD.92.072012} {\bibfield  {journal} {\bibinfo
  {journal} {Phys. Rev. D}\ }\textbf {\bibinfo {volume} {92}},\ \bibinfo
  {pages} {072012} (\bibinfo {year} {2015})},\ \Eprint
  {http://arxiv.org/abs/1508.07560} {arXiv:1508.07560 [hep-ex]} \BibitemShut
  {NoStop}%
\bibitem [{\citenamefont {Albrecht}\ \emph {et~al.}(1990)\citenamefont
  {Albrecht} \emph {et~al.}}]{cite:Argus}%
  \BibitemOpen
  \bibfield  {author} {\bibinfo {author} {\bibfnamefont {H.}~\bibnamefont
  {Albrecht}} \emph {et~al.} (\bibinfo {collaboration} {ARGUS Collaboration}),\
  }\href {\doibase 10.1016/0370-2693(90)91293-K} {\bibfield  {journal}
  {\bibinfo  {journal} {Phys. Lett. B}\ }\textbf {\bibinfo {volume} {241}},\
  \bibinfo {pages} {278} (\bibinfo {year} {1990})}\BibitemShut {NoStop}%
\bibitem [{\citenamefont {Tat}\ \emph {et~al.}(2022)\citenamefont {Tat},
  \citenamefont {Wilkinson},\ and\ \citenamefont
  {Malde}}]{cite:KKpipi_BinningScheme}%
  \BibitemOpen
  \bibfield  {author} {\bibinfo {author} {\bibfnamefont {M.}~\bibnamefont
  {Tat}}, \bibinfo {author} {\bibfnamefont {G.}~\bibnamefont {Wilkinson}}, \
  and\ \bibinfo {author} {\bibfnamefont {S.}~\bibnamefont {Malde}},\ }\href
  {\doibase 10.5281/zenodo.6940031} {\enquote {\bibinfo {title}
  {{D0-\textgreater KKpipi binning scheme}},}\ } (\bibinfo {year} {2022}),\
  \bibinfo {note} {{DOI}:10.5281/zenodo.6940031}\BibitemShut {NoStop}%
\bibitem [{\citenamefont {Barlow}(2004)}]{cite:Barlow}%
  \BibitemOpen
  \bibfield  {author} {\bibinfo {author} {\bibfnamefont {R.}~\bibnamefont
  {Barlow}},\ }in\ \href@noop {} {\emph {\bibinfo {booktitle} {{PHYSTAT (2005):
  Statistical Problems in Particle Physics, Astrophysics and Cosmology}}}}\
  (\bibinfo {year} {2004})\ p.~\bibinfo {pages} {56},\ \Eprint
  {http://arxiv.org/abs/physics/0406120} {arXiv:physics/0406120} \BibitemShut
  {NoStop}%
\bibitem [{\citenamefont {Wilks}(1938)}]{cite:Wilks}%
  \BibitemOpen
  \bibfield  {author} {\bibinfo {author} {\bibfnamefont {S.~S.}\ \bibnamefont
  {Wilks}},\ }\href {\doibase 10.1214/aoms/1177732360} {\bibfield  {journal}
  {\bibinfo  {journal} {Ann. Math.Stat.}\ }\textbf {\bibinfo {volume} {9}},\
  \bibinfo {pages} {60} (\bibinfo {year} {1938})}\BibitemShut {NoStop}%
\end{thebibliography}%

\end{document}